\documentclass{aa}  
\usepackage{graphicx}
\usepackage{txfonts}
\usepackage{hyperref}
\usepackage{makecell}
\usepackage{subfiles}
\usepackage{siunitx}
\usepackage[normalem]{ulem}
\usepackage{threeparttable}
\usepackage{booktabs}
\usepackage{multirow}
\DeclareMathAlphabet{\mathcal}{OMS}{cmsy}{m}{n}
\newcommand\teff{\ensuremath{T_{\textup{eff}}}\xspace}
\newcommand\logg{\ensuremath{\log g}\xspace}
\newcommand\muHz{\ensuremath{\mu \mathrm{Hz}}\xspace}
\newcommand\numax{\ensuremath{\nu_{\textup{max}}}\xspace}
\newcommand\dnu{\ensuremath{\Delta\nu}\xspace}

\newcommand\FeH{\ensuremath{[\mathrm{Fe}/\mathrm{H}]}\xspace}

\newcommand\Nstars{749\xspace}
\newcommand\Ninit{753\xspace}

\newcommand\gaia{\textit{Gaia}\xspace}
\newcommand\kepler{\textit{Kepler}\xspace}

\begin{document} 
    \title{Granulation signatures as seen by \kepler short-cadence data}
    \subtitle{I. A decoupling between granulation and oscillation timescales for dwarfs}

   \author{J. R. Larsen\inst{1}\fnmsep\thanks{E-mail: jensrl@phys.au.dk}
            \and
            M. S. Lundkvist\inst{1}
            \and
            M. B. Nielsen\inst{2}
            \and
            G. R. Davies\inst{2}
            \and 
            Y. Zhou\inst{3,1}
            \and
            M. N. Lund\inst{1} 
        }
        
   \institute{Stellar Astrophysics Centre (SAC), Department of Physics and Astronomy, Aarhus University, Ny Munkegade 120, 8000 Aarhus C, Denmark
            \and
              School of Physics and Astronomy, University of Birmingham, Edgbaston B15 2TT, United Kingdom 
            \and
              Rosseland Centre for Solar Physics, Institute of Theoretical Astrophysics, University of Oslo, P.O. Box 1029, Blindern, NO-0315 Oslo, Norway
            }
   \date{Received 19 December, 2025; Accepted 9 February, 2026}

  \abstract
   {Granulation is the observable surface signature of convection in the envelopes of low-mass stars, forming the background in stellar power spectra. While well-studied in evolved giants, granulation on the main-sequence has received less attention.}
   {We aim to study and characterise granulation signatures of main-sequence and subgiant stars, extending previous studies of giants to provide a continuous physical picture across evolutionary stages.}
   {We analyse \Ninit\ \kepler\ short-cadence stars using a Bayesian nested-sampling framework to evaluate three background descriptions and compare model preferences. This yields full posterior distributions for all parameters, enabling robust comparisons across a diverse stellar sample.}
   {No universal preference between background models is found, thus an a priori choice is not justified. Assuming a Gaussian oscillation envelope, \numax\ estimates become sensitive to model misspecification, with the resulting systematics exceeding the formal uncertainties. The envelope width scales with \numax\ across models and shows a dependence on effective temperature. Total granulation amplitudes in dwarfs broadly follow giant-based scalings, however a decoupling appears between the timescale of the primary granulation and the oscillations for main-sequence stars cooler than the Sun. The prolonged granulation timescale is reproduced by 3D hydrodynamical simulations of a K-dwarf, driven by reduced convective velocities resulting from a more efficient convective energy transport in denser envelopes.}   
   {Our study represents the most extensive Bayesian background modelling of \kepler\ short-cadence stars to date and reveals a decoupling between granulation and oscillation timescales in K-dwarfs. The prolonged granulation timescale increases the frequency separation to the oscillation excess, potentially aiding seismic detectability, while the reduced convective velocities may influence the excitation of stellar oscillations and relate to the low amplitudes observed in cool dwarfs. Finally, we contribute a dataset linking granulation, oscillations, and stellar parameters, establishing a foundation for future investigations into their interdependence across the Hertzsprung–Russell diagram.}
       
   \titlerunning{A decoupling between granulation and oscillation timescales for dwarfs}  
   \maketitle
   
\section{Introduction}\label{sec:Intro}
Stellar granulation is the photometric signature of convection in the outer layers of stars with convective envelopes. Hot plasma rises toward the photosphere, cools, and sinks back into the stellar interior, producing a dynamic pattern of bright granules and darker intergranular lanes. These motions occur on characteristic timescales set by the fundamental stellar properties, reflecting the interplay between gravity, temperature, and composition in the outer layers, resulting in the introduction of a stochastic signal in photometric time series. In the frequency domain, granulation manifests as a background in the power density spectrum (PDS) that decays with increasing frequency and is often modelled by Harvey-like functions \citep{Harvey85}. The high precision and long baselines of space-based missions such as \kepler \citep{Borucki10} have made it possible to measure granulation parameters for large numbers of stars with a wide range of fundamental properties.

Granulation has been extensively characterised in the Sun (e.g. \citealt{Karoff13}), where high signal-to-noise data allow detailed modelling of the temporal and spatial properties of convection. In evolved stars, \citet{Kallinger2014} analysed thousands of \kepler red giants, establishing empirical scaling relations between granulation parameters and the global asteroseismic quantity known as the frequency of maximum oscillation power, \numax. Working purely in the time domain, \citet{Diaz22} evaluated the autocorrelation time of the Legacy stars \citep{Lund2017}, pushing towards studying the granulation of main-sequence (MS) stars. In doing so, they found that the scaling laws roughly agree with those of \citet{Kallinger2014}. Parallel theoretical and numerical work, notably 3D radiative hydrodynamical simulations of stellar atmospheres, has provided physical justification for those scaling relations and explored their dependence on metallicity, surface gravity and convection prescription (\citealt{Samadi13b,Zhou21}). Yet, the stellar samples where detailed granulation studies have been performed primarily consists of more evolved stars on the late-subgiant (SGB) and red-giant branch (RGB).

Extending granulation studies to MS and less-evolved SGB stars is essential for establishing how surface convection scales across different stellar regimes. Whereas current empirical scaling relations are largely informed by evolved stars, the behaviour of granulation in less-evolved stars is not as well characterised. MS and SGB stars probe a broad range of temperatures, surface gravities, and Mach numbers, providing an ideal setting to examine whether the empirical relations derived from red giants remain valid when applied to hotter, denser stellar envelopes. By analysing the granulation signatures in frequency space for the largest sample of MS and SGB stars to date, this work bridges the observational gap between dwarfs and giants and offers new constraints on how granulation properties evolve with stellar structure. This calibration has immediate significance not only for convection theory, but also for asteroseismic applications -- such as improved background modelling for upcoming missions (e.g. PLATO; \citealt{PLATO24}) and a refined understanding of seismic detectability in cool MS stars -- and more reliable noise characterisation in precision exoplanetary studies. 

\citet{Larsen2025b} introduced a Bayesian nested-sampling framework for evaluating competing granulation background models using 3D radiative hydrodynamical simulations, and demonstrated its potential through limited application to stellar observations. Their analysis showed that the accuracy and robustness of comparisons between granulation background models might be obscured by the commonly adopted Gaussian envelope model for the oscillation excess that sits atop the background profile. Further development was therefore required before reliable application of the framework to large, heterogeneous observational datasets. In this work, we extend the application of the framework to a catalogue of \kepler short-cadence stars by \citet{Sayeed25}, spanning a wide range of evolutionary states, observing durations, and signal-to-noise ratios. 

This work first presents the stellar sample studied in Sect.~\ref{sec:Sample}. The further developments to the framework of \citet{Larsen2025b} are outlined in Sect.~\ref{sec:Framework}, which also details the methodology underlying this study. In Sect.~\ref{sec:ModPref}, we investigate the model preferences and sensitivities across the sample, before studying in detail the scaling of the granulation parameters for MS and SGB stars in Sect.~\ref{sec:GranScaling}. In doing so, we uncover what appears to be a previously unreported decoupling between granulation and oscillation timescales -- a result that motivates a dedicated extension of the sample in Sect.~\ref{sec:Decoupling} with additional K-dwarfs observed by the Transiting Exoplanet Survey Satellite (TESS; \citealt{Ricker14}) and theoretical considerations using both 1D and 3D K-dwarf models. Finally, in Sect.~\ref{sec:Conclusion} we make our concluding remarks and outline potential future implications of our findings.

\section{\kepler short-cadence sample}\label{sec:Sample}

\begin{figure}[t]
    \resizebox{\hsize}{!}{\includegraphics[width=\linewidth]{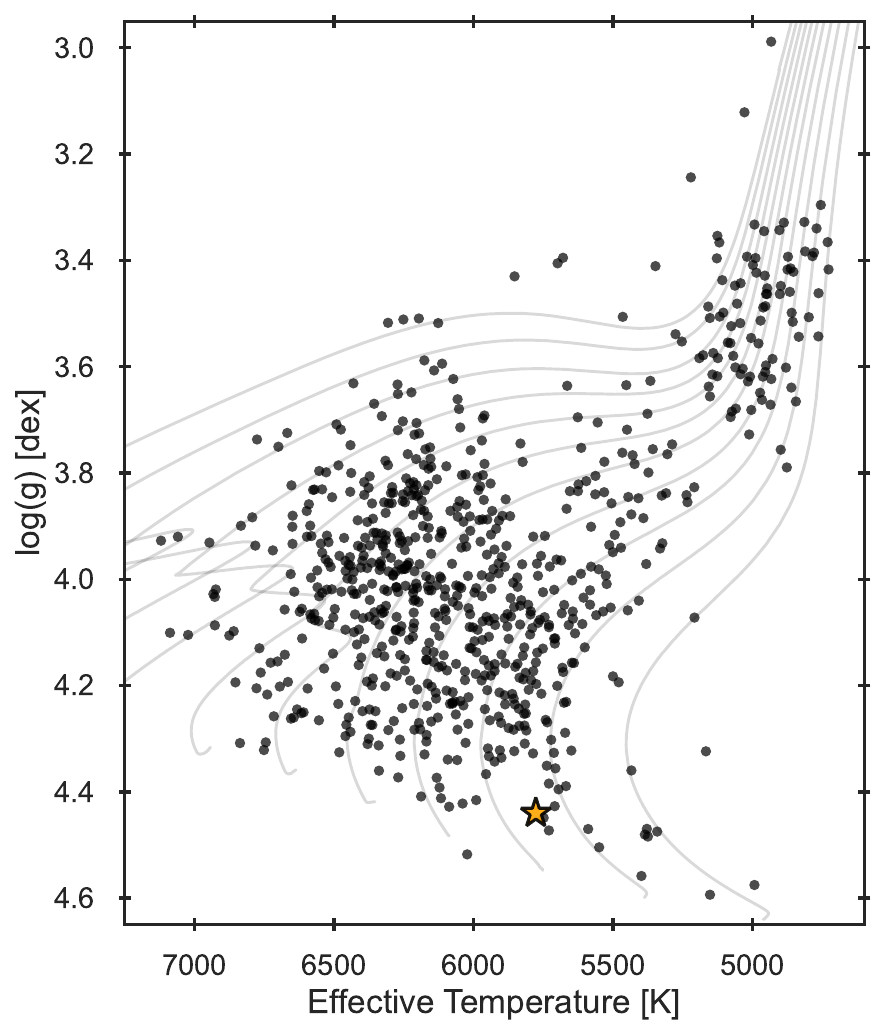}}
    \vspace*{-6mm}
    \centering
    \caption{Kiel diagram of the 733 stars with available effective temperatures and surface gravities, taken from the sample after the sorting in Sect.~\ref{sec:Sample}. The \teff values are retrieved from Table 3 of \citet{Sayeed25}, while the seismic \logg was calculated using the asteroseismic scaling relations with this \teff and the \dnu estimate from \citet{Sayeed25}. Simple stellar evolution tracks of solar metallicity and a range of masses are overplotted to guide the eye. The solar location is indicated by the yellow star symbol.}
    \label{fig:KielSample}
    \vspace*{-2mm}
\end{figure}

The sample used in this work was drawn from the catalogue of \citet{Sayeed25}. The catalogue consists of all known solar-like oscillators observed in short-cadence mode by \kepler\ and numbers a total of 765 stars -- primarily sourced from the Legacy \citep{Lund2017}, KAGES \citep{Aguirre15_KAGES}, and APOKASC \citep{Serenelli17} catalogues. Figure~\ref{fig:KielSample} shows the distribution of these stars in a Kiel diagram. The sample is dominated by MS and SGB stars, but also includes a modest number of stars on the lower RGB. Importantly, there are several G and K dwarfs present situated in the vicinity of and lower on the MS than the Sun, respectively. 

\begin{table*}
\renewcommand{\arraystretch}{2.3}
\centering
\caption{Background models used in this work, presented in an adapted version of Table~1 from \citet{Larsen2025b}.}
\begin{tabular}{lcccc}
\hline
\multicolumn{1}{c}{\thead{Model}} & \multicolumn{1}{c}{\thead{Functional form}} & \multicolumn{1}{c}{\thead{Free parameters}} &  \multicolumn{1}{c}{\thead{No. of components}} & \multicolumn{1}{c}{\thead{Reference}} \\
\hline
J & {\Large $\frac{a^2}{1+(\nu/b)^{l} + (\nu/d)^{k}}$} & $a,b,d,l,k$ & Hybrid & \citet{Lundkvist21} \\
H & {\Large $\frac{a^2/b}{1+(\nu/b)^{l}}+\frac{c^2/d}{1+(\nu/d)^{k}}$} & $a,b,c,d,l,k$ & 2 &\citet{Kallinger2014} \\
T & {\Large $\frac{a^2/b}{1+(\nu/b)^{l}}+\frac{c^2/d}{1+(\nu/d)^{k}} + \frac{e^2/f}{1+(\nu/f)^{m}}$} & $a,b,c,d,e,f,l,k,m$ & 3 & \citet{Larsen2025b} \vspace*{1mm}
\\
\hline
\end{tabular}
\tablefoot{The model name, functional form, summarised free parameters, number of components, and reference is given by the table. The parameters $a$, $c$, and $e$ are amplitudes in ppm, while $b$, $d$ and $f$ are the associated characteristic frequencies in \muHz, respectively. The exponents of the characteristic frequencies are denoted $l$, $k$ and $m$.}
\label{tab:Models}
\vspace*{-2mm}
\end{table*}

This catalogue provides ample grounds for further development and testing of the framework presented in \citet{Larsen2025b} -- see Sect.~\ref{sec:Framework} for details -- as it contains a large number of stars spanning a broad range of evolutionary stages across the MS, SGB, and the lower RGB; whereas previous similar studies focused largely on evolved stars observed by \kepler in long cadence \citep{Kallinger2014}. Furthermore, for most stars in the catalogue a seismic characterisation exists. This means that the global asteroseismic parameters -- \numax and the large frequency separation \dnu -- are largely available, as well as seismically determined masses, radii and ages for the majority of the sample. In short, the catalogue studied in this work consists of some of the most well-characterised dwarf stars observed by \kepler.

For the entire catalogue of \citet{Sayeed25}, we retrieved the raw \kepler data for all targets and processed them with the KASOC filter \citep{Handberg14} to ensure a homogeneous treatment of the data. The corresponding power density spectra were then computed following the approach of \citet{Handberg11}. Known \kepler systematics were removed based on the list provided in Table~5 of the \kepler Data Characteristics Handbook \citep{KeplerHandbook16}. To mitigate the influence of strong rotational variability, we omitted PDS data below 10~\muHz, corresponding to rotational signatures with periods longer than approximately 1.15 days. For our sample of \kepler short-cadence stars -- composed primarily of low-mass MS and SGB stars (see Fig.~\ref{fig:KielSample}) -- rotation periods shorter than about 1.15~days are unlikely \citep{Santos24}. While some harmonics of the rotational peaks may extend into the low-frequency regime, their contribution to the overall power budget is minor, and the activity component of our background models effectively absorbs their influence on the granulation signal.

The sample spans stars with \kepler observations ranging from less than a single quarter ($\lesssim$30 days) to over three years. Some of the more evolved stars with short time series may have longer-duration long-cadence data available, which could improve PDS quality. Nevertheless, we restricted this analysis to the short-cadence data to maintain consistency. Applying the \citet{Larsen2025b} framework to a broad range of stars including lower-quality cases -- and as we were focusing on granulation rather than pulsation features -- provided a stringent test of the method and supported the further developments outlined in Sect.~\ref{sec:Framework}.

Before moving on, we note that the catalogue of \citet{Sayeed25} contained a few stars not suitable for our work. The specific stars which were removed from the sample and the associated reason(s) can be found in Table~\ref{tab:removals}. As will become clear in Sect.~\ref{sec:Framework}, we use an estimate of the observed $\numax$ for our setup and prior definitions. We took the following steps to obtain them, listed in order of priority:
\begin{enumerate}
    \item \numax directly from Table~3 of \citet{Sayeed25} taken from literature estimates.
    \item \numax from source papers if applicable \citep{Aguirre15_KAGES,Lund2017,Serenelli17}.
    \item \numax estimated by pySYD \citep{Chontos21} from Table~3 of \citet{Sayeed25}.
    \item Recover \dnu and \teff estimates from Tables~3 and 4 of \citet{Sayeed25}, respectively, and calculate \numax using the asteroseismic scaling relation (Eq.~\ref{Eq:numaxscal}).
\end{enumerate}
Through the steps above, we recovered a \numax estimate for all stars in the sample. However, as we were dealing with short-cadence \kepler data and considered stars on the MS, SGB and lower RGB, we chose to discard the star if the estimated $\numax<100$ \muHz, indicating an evolved giant star. In summary, the discarded stars in Table~\ref{tab:removals} number 12 in total and contain both those deemed unsuitable and those below the adopted \numax cut-off. This resulted in a reduced sample of \Ninit stars for our studies.

\section{Framework and methodology}\label{sec:Framework}
The foundation of the framework for performing the background model inference was developed and described in \citet{Larsen2025b}. It is a Bayesian setup based on nested sampling using the inference algorithm \texttt{Dynesty} \citep{Speagle20}, which allows simultaneous estimation of the posterior probability distributions and the Bayesian evidence, $\mathcal{Z}$. For details on the main body of the framework we refer to \citet{Larsen2025b}. Specifically, their Sect.~2.2 outlines the complete description of stellar power spectra including how apodisation \citep{Chaplin11b}, stellar activity, oscillations, and white noise is accounted for (see also Appendix~\ref{App:Priors}). Herein, we describe the further developments enabling extensive application to the \kepler short-cadence sample. The developments are described in Sect.~\ref{subsec:corrinf} and Appendix~\ref{App:peakbogging}, and also concern refined priors (see Appendix~\ref{App:ModSelectDiscuss} and \ref{App:Priors}). The background models considered in this work are those concluded in \citet{Larsen2025b} to have merit and are seen in Table~\ref{tab:Models}: a hybrid model with a single amplitude and two characteristic frequencies (J), a two-component Harvey model (H) and a three-component Harvey model (T). 

In \citet{Larsen2025b}, the framework was applied primarily to 3D hydrodynamical simulations of convection, but subsequently extended to two real stars: the solar analogue KIC8006161 (Doris) and the Sun. These stars were of exceptionally high quality and signal-to-noise ratios. Applying the framework widely to our sample of \Ninit stars requires handling cases where the PDS components are less distinct, necessitating further developments of the framework. The setup we developed encodes physically motivated connections between the various components of power density spectra and is outlined in Sect.~\ref{subsec:corrinf}. Moreover, \citet{Larsen2025b} noted that the conclusions drawn on the background model preference may depend on the implemented model for the power excess due to the presence of stellar oscillations. To investigate this, we present extensive tests of a new approach dubbed `peakbogging' to account for the oscillation excess, briefly summarised in Sect.~\ref{subsec:PeakbogIntro} and outlined in detail throughout Appendix~\ref{App:peakbogging}. Both the traditional treatment using a Gaussian envelope and peakbogging utilise the developments in Sect.~\ref{subsec:corrinf}.

The log-likelihood $\ln{\mathcal{L}}$ used throughout this work describes independent frequency bins in the PDS combined with a standard $\chi^2$ probability distribution. In the present work, the likelihood defined in \citet{Larsen2025b} is slightly modified to allow for binning of the PDS by introducing the factor $s$ as the number of datapoints per $i$'th bin into the expression for $\chi^2$ (see e.g., \citealt{Appourchaux04}; \citealt{Handberg11}; \citealt{Lundkvist21}), such that 
\begin{align}
    \ln\mathcal{L} &= \ln p(D|\theta,M) = \sum_i \ln\left(f(D_i,\theta,M_i) \right), \label{eq:loglike}\\
    f(D_i,\theta,M_i) &= \frac{s^{s-1}}{(s-1)!} \frac{D_i^{s-1}}{M_i(\theta)^s}\exp\left(-\frac{sD_i}{M_i(\theta)}\right) \label{eq:chi2} \ .
\end{align}
As in the original framework of \citet{Larsen2025b}, $D$ denotes the observed data (power), $\theta$ is the model parameters and $M$ an assumed model predicting a power $M_i(\theta)$ for a given frequency bin.

\subsection{A correlated inference setup}\label{subsec:corrinf}
It is well known that $\numax$ is closely linked to both the amplitudes and characteristic timescales of granulation, as all three quantities reflect the properties of the underlying convective motions \citep[e.g.][]{Kjeldsen11,Mathur11,Samadi13b,Kallinger2014,Diaz22}. This physical connection implies that these parameters are correlated and unlikely to vary independently during our inference: a star with a low $\numax$ must also exhibit granulation with correspondingly larger amplitudes and longer timescales, while a high $\numax$ demands the opposite. 

In our framework, this correlation provides a natural way to let the PDS as a whole guide the sampling. When we consider the entire PDS -- spanning several orders of magnitude in both frequency and power -- the inference is naturally sensitive to any combination of $\numax$ and granulation parameters that fails to reproduce the overall behaviour of the spectrum. For instance, a MS-like $\numax$ paired with granulation amplitudes of several hundred parts-per-million and timescales of only tens of microhertz would contradict the expected behaviour. During the sampling, any tentative move toward such mismatched parameter combinations should therefore result in a disfavoured likelihood given the data, effectively steering the sampler away from unphysical regions of parameter space.

To implement this, we developed a correlated inference setup for the granulation parameters for which scaling relations with \numax were derived by \citet{Kallinger2014} -- which is the amplitude, $a$, and timescale, $b$, of the first granulation component and the timescale, $d$, of the second. This means that instead of freely sampling the granulation parameters, we calculate them based on the sampled \numax. In turn, we control the strength of this correlation by adding a scatter parameter to be inferred:
\begin{align}
    a \ [\text{ppm}] &= 3382\numax^{-0.609} \sigma_a , \\
    b \ [\muHz] &= 0.317\numax^{0.970} \sigma_b, \\
    d \ [\muHz] &= 0.948\numax^{0.992} \sigma_d  .
\end{align}
The power-law coefficients come from \citet{Kallinger2014} and \numax is a free parameter during the inference. The scatter parameters, $\sigma_{a,b,d}$, describe the potential scatter of the scaling relations and are also freely sampled. Priors forcing the scatter parameters close to $1$ imposes a very tight correlation, while allowing wider variation from $1$ lessens the correlation between granulation and \numax. Inspecting Fig.~8 of \citet{Kallinger2014}, we see that the scatter for their RGB sample is quite low and roughly symmetric in the log-parameters; $\sim$10-15\% for the timescales and $\sim$15-30\% for the amplitude. These considerations are carefully taken into account when defining the prior ranges for $\sigma_{a,b,d}$ in Appendix~\ref{App:Priors}. 

\begin{figure}[h!]
    \resizebox{\hsize}{!}{\includegraphics[width=\linewidth]{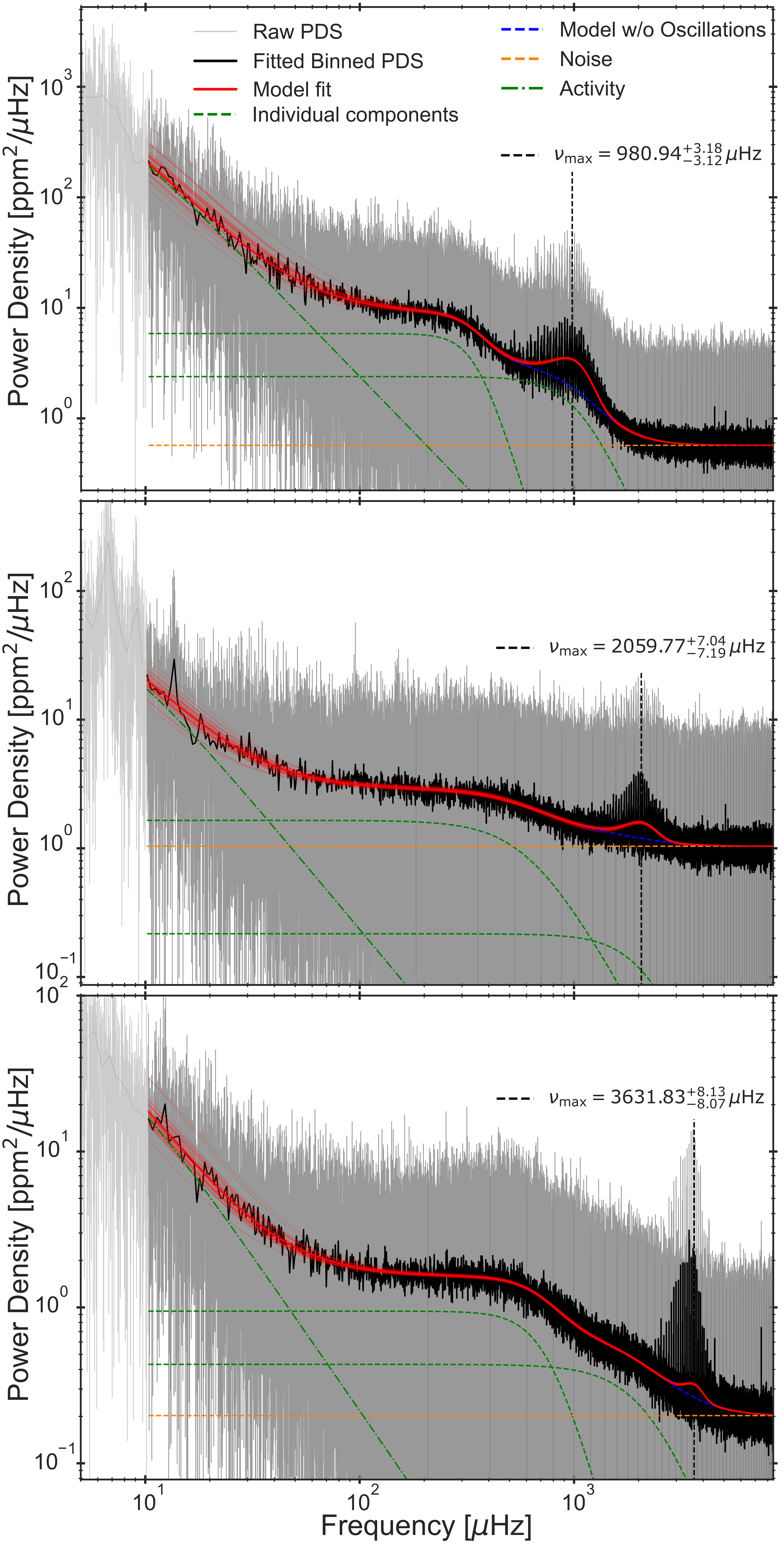}}
    \centering
    \caption{Power density spectra with overlaid results of the background model inference using model H (see Table~\ref{tab:Models}) when binning to $0.5$\muHz resolution for three stars: KIC6679371 (top), KIC8866102 (middle) and KIC8006161 (bottom). The unbinned PDS is shown in grey with the binned version overplotted in black. The model is plotted in red using the median of the obtained posteriors for each fit parameter. Additionally, 50 randomly drawn samples from the posteriors are used to replot the model to indicate the scatter. The individual granulation components are plotted as dashed green profiles. The fitted value of \numax is given in each panel and indicated by the vertical dashed black line, while the noise is shown by the horizontal dashed orange line. The activity component is the dash-dotted green line. The model without the influence of the Gaussian oscillation excess is plotted as the dashed blue profile, visible underneath the oscillation excess.}
    \label{fig:PDSComp}
    \vspace*{-3mm}
\end{figure}

\subsection{Studying granulation for a large sample}\label{subsec:InfSampBinning}
Our sample of \Ninit stars exhibits substantial diversity. Our focus is on characterising the stellar granulation background -- the broad trends in the power density spectra across several orders of magnitude in frequency and power -- rather than resolving the detailed structure of the oscillation modes. This emphasis naturally motivates moderate binning of the power density spectra, providing consistent resolution across the sample and smoothing statistical fluctuations.

Binning offers several advantages. Averaging multiple independent frequency bins causes the noise distribution to converge toward a Gaussian by the central limit theorem \citep{laplace_thorie_1812}, replacing the exponential distribution characteristic of individual PDS bins \citep{Anderson1990}. This transformation improves sampling stability and effectiveness, since the sampler no longer needs to account for a highly skewed distribution. The corresponding $\chi^2$ statistics of Eq.~\ref{eq:chi2} are adjusted via the effective degrees of freedom, $s=2N$, with $N$ being the number of independent points per bin. We tested the impact of varying the binning by examining how the inferred granulation parameters respond to changes in bin size. We focused on the second granulation component, which is most sensitive to changes in resolution, as it describes the frequency range near the stellar oscillations. We applied model H (Table~\ref{tab:Models}) to three stars: KIC~6679371, KIC~8866102, and KIC~8006161 with values of $\numax \sim 1000,\ 2000,\ 3500$ \muHz, respectively. The binning was varied to provide frequency resolutions ranging from $0.025$ to $4.5$ \muHz. Variations in the inferred parameters remained below the associated $1\sigma$ uncertainties, demonstrating that the choice of binning does not meaningfully bias the measurements.

The main trade-off when binning is that low-amplitude oscillation peaks, especially in the envelope wings, may be partially smoothed. However, these features contribute negligibly to the total power budget and thus have little impact on our granulation inferences. Furthermore, as an added bonus, binning improves computational efficiency by reducing the number of data points, substantially decreasing runtime for each inference. Several stars in our sample have time series shorter than a full \kepler quarter, yielding a native frequency resolution $\sim 0.4\ \mu\mathrm{Hz}$. For stars with sufficiently long time series, we bin the PDS to a uniform resolution as close as possible to $0.5$ \muHz, while maintaining a constant number of data points per bin, $N$, to ensure consistent and robust $\chi^2$ statistics. This choice is made throughout this work and balances the need for statistical stability and computational tractability, while retaining the granulation signals of interest.

Figure~\ref{fig:PDSComp} shows the outcome of applying model H and binning to a resolution of $0.5$ \muHz for KIC 6679371, KIC 8866102, and KIC 8006161. This figure visualises the results we will obtain -- here for model H, however repeated for J and T as well -- for every sample star considered in this work using the outlined framework. We see how the two Harvey profiles describe the lower and upper granulation components at increasing frequencies, with the Gaussian envelope standing on top of the inferred background slope. By inspection of KIC~8006161 in the bottom panel, we notice that the primary granulation component at lower frequency is both prominent and very clearly separated in frequency from the oscillation excess. Through the study of similar stars this trend will be assessed more thoroughly in Sect.~\ref{sec:Decoupling}, but already here one may notice that this wide a separation in frequency is not readily apparent for the other two more evolved stars in Fig.~\ref{fig:PDSComp}. The choice to remove the PDS contributions below 10 \muHz in order to avoid the influence of strong rotational peaks only plays a role for KIC~8866102 in the middle panel. However, true for all three is that this choice does not affect the remaining granulation and oscillatory parameters, as the lowest frequency regime remains adequately described by the activity component. Finally, we see that the effect of binning the PDS does indeed increase the contrast without affecting the finer details of the inference.

\subsection{Introducing `peakbogging'}\label{subsec:PeakbogIntro}
Traditionally, the oscillation excess is represented with a symmetric Gaussian envelope atop the granulation background, yet such an approach may be problematic. This can systematically misrepresent the power distribution, particularly when (i) the envelope is intrinsically asymmetric, (ii) mode visibilities are non-standard (e.g. due to inclination, mode lifetime or mixed-mode complexity), or (iii) the available data has a low signal-to-noise ratio and sparse sampling. In such cases the Gaussian envelope may trade off with the background model and misrepresent the true power distribution, thereby biasing granulation amplitudes and timescales, which might in some cases lead to degeneracies between the background and oscillation components. The issues associated with the traditional Gaussian approach also affected the results of \citet{Larsen2025b}, where it was discussed if the conclusions on model preference may depend on how the oscillation excess is accounted for. 

These limitations motivated the development and testing of the so-named `peakbogging' approach: a mixture-model likelihood setup where a flexible foreground component absorbs residual signal not described by the assumed background model, aimed at reducing the bias associated with background model choices and improving robustness when applied across diverse stellar samples. However, as is discussed in Appendix~\ref{subsec:PeakbogLimits}, certain unresolved pathologies plague peakbogging when applied to the sample studied in this work. On the other hand, it shows promise for a significant number of stars and is robust in terms of providing meaningful posteriors. Moreover, in virtually all cases peakbogging recovers an identical estimate of the total granulation amplitude (Sects.~\ref{subsec:totgranamp} and \ref{app:bog_granscal}). 

The detailed setup of peakbogging is outlined in Appendix~\ref{App:peakbogging}, and throughout the appendix we present the peakbogging results analogous to those in the main paper, while discussing the promising aspects alongside the potential issues. For the results presented in the remainder of this article, we utilised the traditional approach of treating the oscillation excess using a Gaussian envelope. We acknowledge the underlying assumptions and misrepresentations of this model in doing so.

\section{Background model preferences and sensitivities across the sample}\label{sec:ModPref}
\begin{figure}[t]
    \resizebox{\hsize}{!}{\includegraphics[width=\linewidth]{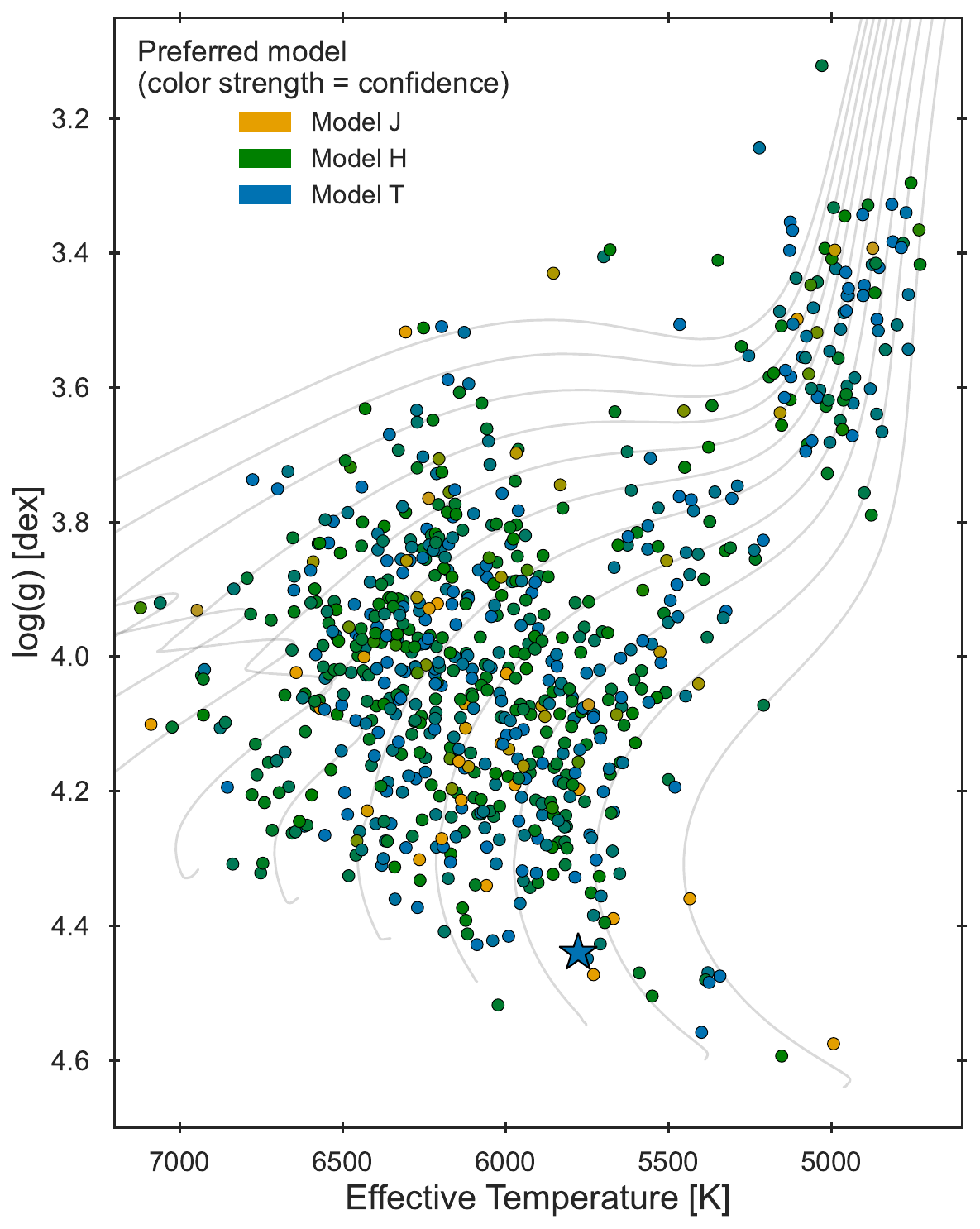}}
    \vspace*{-6.5mm}
    \centering
    \caption{Kiel diagram with colouring according to normalised evidence ratios, with model preferences as indicated by the legend. When models are comparable in their evidences the colour is blended between the two competing models. The Sun is overplotted as the enlarged star symbol at the solar location and significantly prefers model T.}
    \label{fig:Evidences_Gauss}    
    \vspace*{-4mm}
\end{figure}

By applying the granulation background inference framework to the sample outlined in Sect.~\ref{sec:Sample}, we can now begin to investigate the trends exhibited by the granulation and oscillatory components across a diverse set of MS, SGB, and lower RGB stars. Inevitably, when analysing such a large and diverse sample in detail, a small fraction of stars cannot be recovered with reliable results, typically due to a combination of short time series and low signal-to-noise ratio, contamination from spurious peaks in the PDS, or numerical convergence issues in the nested sampling procedure. To ensure that only robust results enter our analysis, we evaluate the quality of the posterior distributions obtained for all granulation parameters and retain only those that are well-sampled and statistically consistent. This conservative approach ensures that the inferred model parameters -- estimated as the median of the posterior distributions, with corresponding uncertainties from the 16th and 84th percentiles -- represent genuine features of the data rather than artefacts of a failed inference. In our case we discard the star if any of the three background model inferences fail, as proper subsequent comparisons are unable to be made. We note in passing that sometimes a star is thus discarded where a subset of the three considered background models did provide valid results. In total 4 targets failed to yield meaningful inferences and were removed. Hence, in the remainder of the paper we present results based on the remaining \Nstars stars. These stars thus provide a data sample where the complete posterior distributions for all parameters, the covariances between the parameters, and the Bayesian evidences $\mathcal{Z}$ are readily available for study herein and in future efforts.

In the following we first examine the background model preferences of the individual stars, as quantified by the Bayesian evidences obtained through nested sampling. We then assess how the choice of background model affects the inferred \numax values, before inspecting how the width of the oscillation excess changes across the evolutionary stages of the sample. Specifically for the latter two investigations in Sects.~\ref{subsec:numaxresid} and \ref{subsec:oscwidths}, we evaluated the resulting \numax estimates and Gaussian envelope widths to further sort the dataset. In cases where the oscillatory signal is unclear the Gaussian envelope model is a poor representation of the data and may perform badly. When it does, the framework attributes negligible power to the Gaussian envelope and essentially removes its contribution by narrowing it in to absorb a single noise peak. In such cases the oscillations contribute negligibly to the overall power budget, meaning that the underlying granulation background components are still well-characterised. Yet for investigations of \numax and the oscillation excess widths, such cases were removed from consideration.

\begin{figure}[t]
    \resizebox{\hsize}{!}{\includegraphics[width=\linewidth]{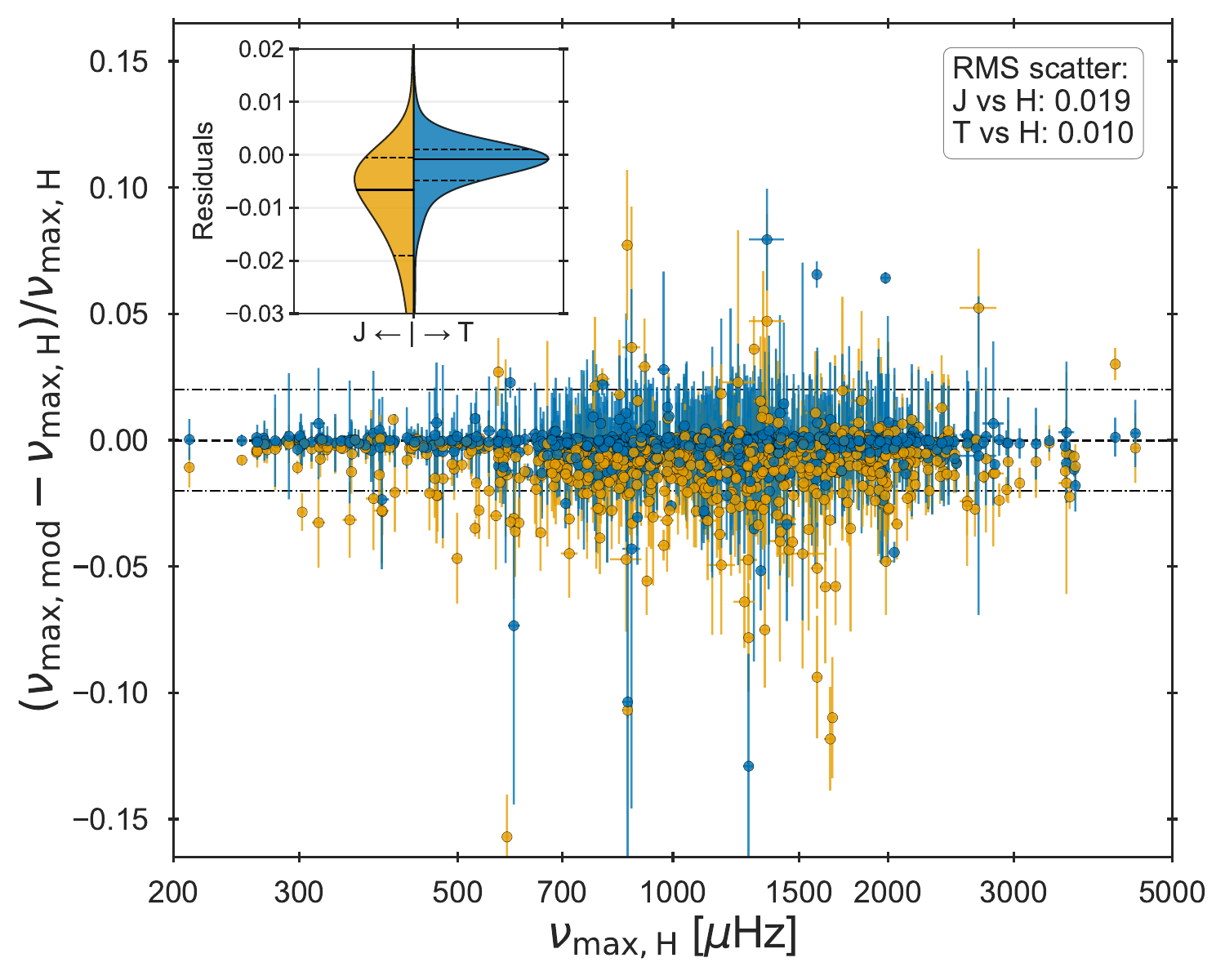}}
    \vspace*{-7mm}
    \centering
    \caption{Comparison of the \numax determination across the different models. The \numax fractional residuals of models J and T to those obtained by model H are plotted in yellow and blue, respectively. The horizontal dashed line indicates perfect agreement in \numax determinations, while the dot-dashed show the $2\%$ bounds. The RMS scatter was calculated for both cases and is provided in the inserted box in the top right. The insert shows a split violin plot of the \numax residual distributions for model J (left) and model T (right) versus model H, with medians and 16th/84th percentiles overplotted as full and dashed horizontal lines, respectively.}
    \label{fig:numaxresid}
    \vspace*{-3.5mm}
\end{figure}
\begin{figure*}[t]
    \includegraphics[width=\linewidth]{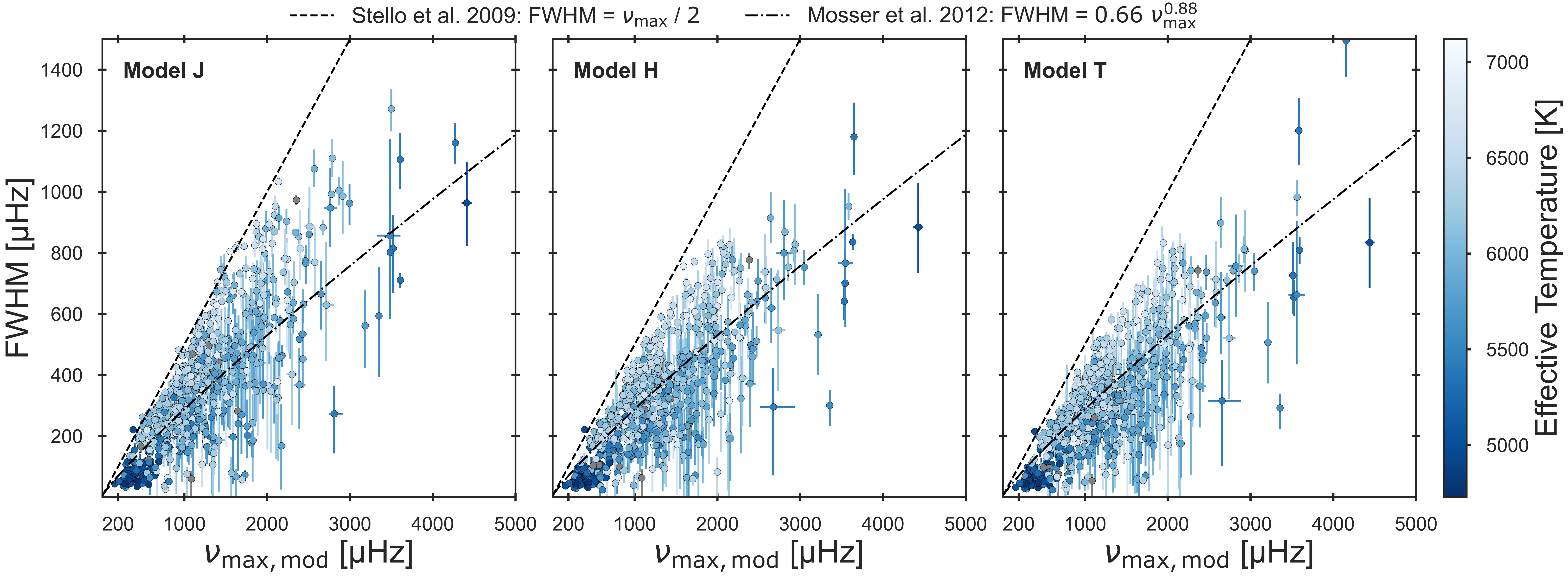}
    \centering
    \vspace*{-5mm}
    \caption{The FWHM of the Gaussian oscillation excess as a function of the determined \numax, coloured by the temperature of the star. The dashed and dot-dashed lines indicate the predictions by \citet{Stello09} and \citet{Mosser12}, respectively.}
    \label{fig:Oscwidths}
    \vspace*{-2mm}
\end{figure*}

\subsection{Background model preferences}
Previous efforts have considered how the choice of background model has implications for the resulting outcome and our conclusions in connection to observational asteroseismology (see e.g. \citealt{Sreenivas24}). Specifically, \citet{Handberg17} discussed such aspects and argue how a single choice of background model may not be suitable. They argue that fixing the background model a priori carries the assumption that the model accurately describes the granulation background for the star(s) in question -- an assumption that is hard to justify for such a diverse sample as the one studied in this work. 

By applying the three different background models of Table~\ref{tab:Models} to the entire sample, we thereby avoid this assumption. Figure~\ref{fig:Evidences_Gauss} shows that the assumption of a single background model being suitable is not justified, even in restricted regions of stellar evolution. Across the sample, models H and T -- the two- and three-component Harvey models, respectively -- are consistently preferred over the hybrid model J. No readily apparent trends in model preference across the different evolutionary stages are seen in the figure. Hence, if the goal is to most accurately describe the granulation background for a certain dataset -- be it to study the granulation itself or correct for the background to study the oscillations -- these results indicate that one must consider various models and choose the one which best represents the data.

In \citet{Larsen2025b}, it was briefly discussed whether the conclusions regarding model preference depend on the assumption of a Gaussian envelope representing the oscillation excess. Similarly, in the present analysis, we cannot exclude the possibility that the inferred preferences depend on this assumption. We also note that Fig.~\ref{fig:Evidences_Gauss} displays evidence ratios, not posterior odds ratios, which would additionally include a factor reflecting our prior beliefs. In other words, the results shown purely indicate which model best fits the data, without accounting for physical plausibility or other prior reservations concerning the assumed background models. For instance, some readers may question the three-component model~T, whose third granulation component appears at high frequency ($\nu \gtrsim \numax$). If so, such prior beliefs should be incorporated when interpreting Fig.~\ref{fig:Evidences_Gauss}, potentially shifting the preference towards models~J or~H.

As a further consequence of relying on a Gaussian envelope, we occasionally find that the granulation components compensate for its shortcomings when describing the power around the oscillation excess. In such cases, the model effectively trades power between the granulation terms and the envelope because the latter provides only a crude approximation to the true oscillation signal. This behaviour naturally raises the question of whether the inferred model preferences are influenced by the chosen representation of the oscillation excess. To explore this, Appendix~\ref{App:subsec:bogprefs} presents an analogous analysis based on the peakbogging approach (Fig.~\ref{fig:Evidences_Peakbog}), which offers a more flexible description of the oscillation power. Although peakbogging has its own unresolved issues, Fig.~\ref{fig:Evidences_Peakbog} demonstrates that the preferred background model can change when the oscillation excess is represented differently. Consequently, as the preferred model is sensitive to the underlying assumptions, this reinforces the need to avoid a priori selection of a background model and instead choose the optimal model on a star-by-star basis \citep{Handberg17}.

\subsection{Background model sensitivity of \numax}\label{subsec:numaxresid}
In asteroseismic studies of the stellar background signal a key parameter used for subsequent analysis is \numax. Thus, the question of how \numax varies with the assumed background model naturally arises. In Fig.~\ref{fig:numaxresid} we present the fractional differences in \numax between the different models. To keep the results internally consistent, we compare to the value obtained with our framework when using the most widely applied model in literature, that is, the two-component model H, rather than the observed values from \citet{Sayeed25}.

It is readily apparent that variations between the models occur. Specifically, a slight tendency is seen for lower \numax estimates when using the hybrid model J in comparison to model H. The root-mean-square scatter is also slightly larger when the nature of the model changes from individual Harvey-like components to the hybrid model. For the majority of the sample, the variations are below the ${\sim}2\%$ level. However, some stars show larger variation up to and exceeding the ${\sim}5\%$ level when changing the assumed background model. For the comparison between models T and H, the residuals are consistent with zero within $2\sigma$ formal uncertainties for $93\%$ of stars, but for J vs H this decreases to $79\%$ (meaning $21\%$ show inconsistencies that cannot be explained by formal uncertainties alone). Hence, while the formal uncertainties -- with mean fractional values of ${\sim}1.5\%$ -- explain the scatter for the majority of the sample, there is a significant number of cases where they do not.

When using a framework such as in this work, or alternatively pipelines like pySYD \citep{Chontos21}, the internal uncertainties reported on \numax are often very small. The above recovery percentages indicate that the choice of background model introduces systematic differences that are comparable to or dominating the formal uncertainties reported on \numax.

\begin{figure*}[t]
    \includegraphics[width=\linewidth]{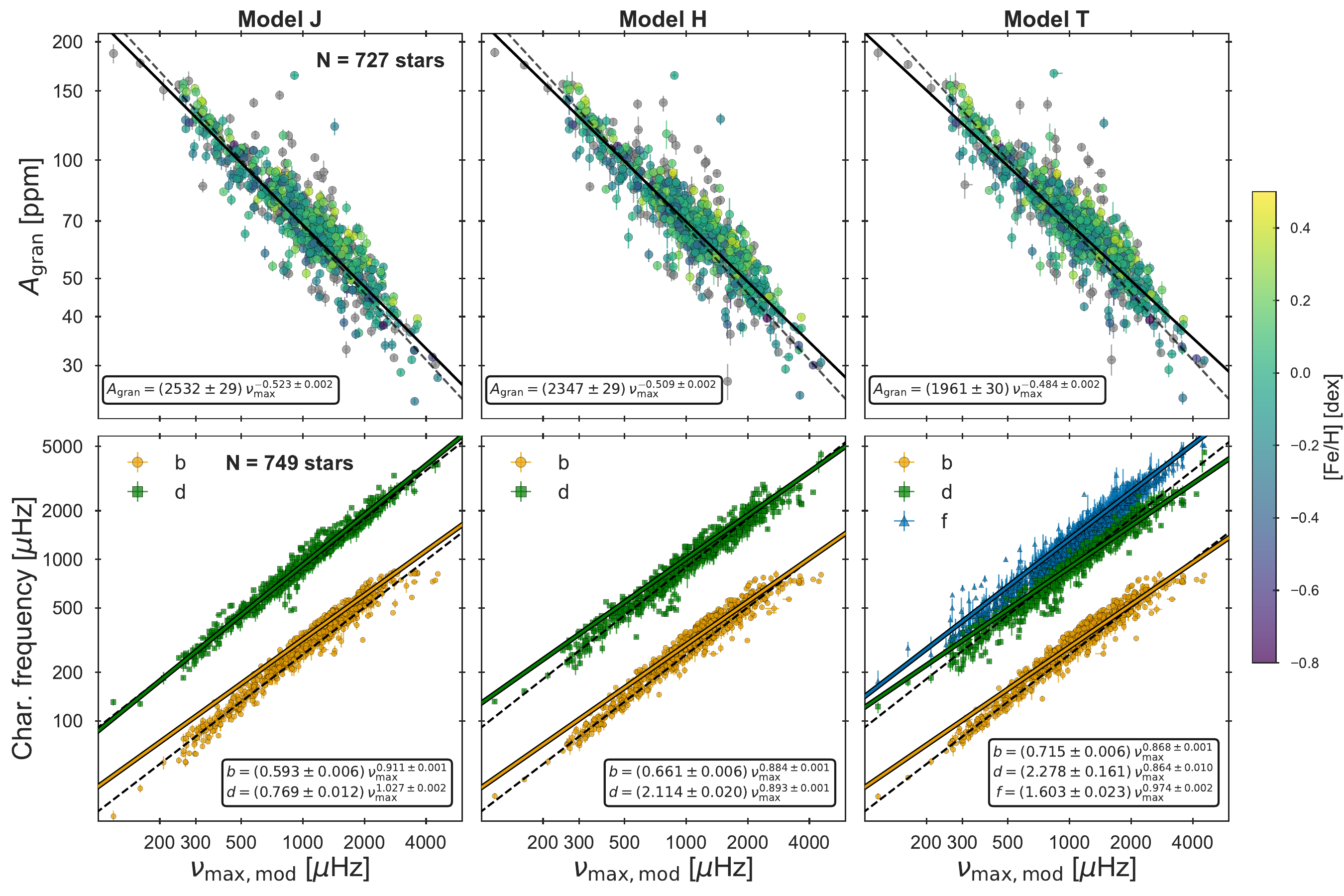}
    \centering
    \vspace*{-6mm}
    \caption{Total granulation amplitudes and characteristic frequencies as a function of \numax obtained for the three background models of Table~\ref{tab:Models}. In all panels, the dashed lines represent the corresponding scaling relation for the parameter from \citet{Kallinger2014}. \textit{Top row:} total granulation amplitudes colour-coded by the stellar metallicity \FeH for the 727 stars with available temperatures. The black lines show a power law fit to the data. \textit{Bottom row:} the characteristic frequencies of the individual granulation components for the 749 stars with consistent granulation posteriors, with colours as indicated by the legend. The corresponding coloured lines show the results of a power law fit to the data for each timescale, all given a black outline for improved readability.}
    \label{fig:GranParams}
    \vspace*{-2mm}
\end{figure*}

\subsection{Oscillation excess widths}\label{subsec:oscwidths}
Having applied the Gaussian envelope approach to the entire sample we have obtained estimates of the oscillation excess widths $\sigma$ for, to our knowledge, the largest collection of \kepler MS and SGB stars to date. How the width of the oscillation excess changes through evolution has been studied by e.g. \citet{Stello09} and \citet{Mosser12}. This sample, however, enables a future study of how the oscillations widths depend on various stellar parameters such as temperature, metallicity, or stellar masses.

Figure~\ref{fig:Oscwidths} shows the obtained full-width-half-maxima, $\text{FWHM}=2\sqrt{2\ln(2)} \sigma$, of the Gaussian envelopes obtained by the background model inferences. It can be seen how across all models, the trend suggested by \citet{Stello09} as $\numax/2$ seems to provide a rough upper boundary. Furthermore, the trend $\text{FWHM} = 0.66\numax^{0.88}$ suggested by \citet{Mosser12} follows much of the data, but significant scatter around the relation is found. For model J specifically, larger oscillation widths are found for the stars with $\numax\gtrsim 2000$ \muHz. 

Lastly, we note a trend with temperature, indicating that lower effective temperatures correspond to smaller oscillation excess widths (as also noted by \citep{Schofield19thesis}). Exploring empirical scaling relations for oscillation widths that account for temperature or other stellar parameters, and potentially include subdivisions by evolutionary stage as considered by \citet{Kim21}, would be an interesting avenue for future work using the provided dataset.

\section{Scaling of granulation parameters}\label{sec:GranScaling}
In this section we will study how the granulation behaves across the sample, similarly to how \citet{Kallinger2014} approached it for their sample of \kepler RGB stars. For brevity we restrict ourselves and only consider the total granulation amplitudes predicted by the background models, which means the combined amplitudes of the individual granulation components. Furthermore, we then study the characteristic frequencies (i.e. timescales) associated with each granulation component across the different background models of Table~\ref{tab:Models}.

\subsection{Total granulation amplitudes, $A_\mathrm{gran}$}\label{subsec:totgranamp}
The total granulation amplitude includes the bolometric correction for the \kepler\ passband, defined as $A_\mathrm{gran}^2 = C_\mathrm{bol}^2 \sum a_i^2$, where each $a_i$ denotes the normalised granulation amplitude of an individual component after accounting for apodisation, and $C_\mathrm{bol} = \left(\teff / 5934~\textup{K}\right)^{0.8}$ \citep{Michel09, Ballot11}. The top row of Fig.~\ref{fig:GranParams} shows these amplitudes for the 727 stars in our sample with available \teff measurements. The total power attributed to granulation varies only marginally among the different background models, and this stability of $A_\mathrm{gran}$ demonstrates the internal consistency of our framework.

Following the approach of \citet{Kallinger2014}, we fit a simple power law to the measured $A_\mathrm{gran}$ values as a function of \numax. The resulting scaling relations derived for $A_\mathrm{gran}$ for each background model (the fit coefficients are indicated in the figure) show the same picture: a declining amplitude with increasing \numax and an exponent similar to $-1/2$. Moreover, despite widely applying this simplistic and naive power law description to the entire sample, the resulting scaling relations qualitatively agree with those obtained by \citet{Kallinger2014}, however with a less steep slope. As \citet{Kallinger2014} focused on evolved RGB stars, while our sample is dominated by less-evolved SGB and MS stars (see Fig.~\ref{fig:KielSample}), this may indicate an evolutionary effect on the slope. However, further work would be required to confirm this statement, as the background model used by \citet{Kallinger2014} (their model F) is not identical to, and is more constrained than, any of the models considered here.

Several theoretical studies have examined how metallicity, \FeH, affects granulation amplitudes, typically predicting lower amplitudes for more metal-poor stars \citep{Corsaro17,Yu18,Diaz22}. In Fig.~\ref{fig:GranParams}, the amplitudes are colour-coded by metallicity where available. Despite recognising that the range in \FeH is modest, no systematic trend with metallicity is apparent. However, as also noted by \citet{Kallinger2014}, the amplitudes are expected to depend on stellar mass: at a fixed \numax, higher-mass stars should exhibit smaller amplitudes than lower-mass ones. This mass dependence likely contributes to the scatter seen in Fig.~\ref{fig:GranParams}, and may obscure any subtle metallicity trend that could otherwise emerge.

\begin{figure*}[t]
    \centering
    \includegraphics[width=\linewidth]{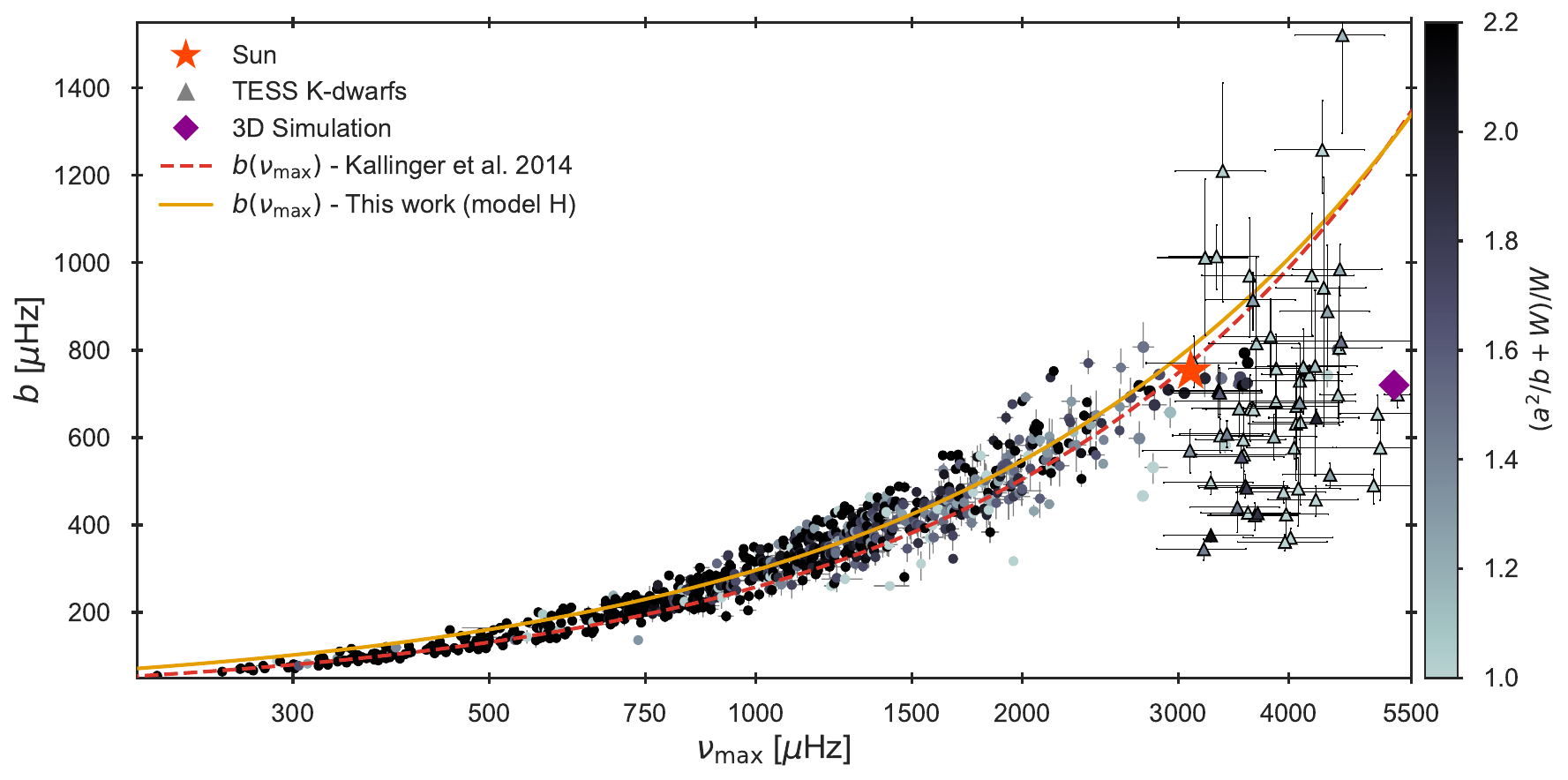}
    \vspace*{-7mm}
    \caption{The timescale of the primary granulation component against the oscillation timescale \numax, indicating a decoupling through a plateau beyond $\numax\approx 3000$ \muHz. The plotted values are those estimated by the background model preferred by the Bayesian evidence $\mathcal{Z}$. The \kepler sample is replotted from Fig.~\ref{fig:GranParams}. The TESS K-dwarf additions are shown as the triangular points with black outlines. The Sun is shown as the dark orange star symbol, obtained using a $\sim 3.15$ year time series from VIRGO \citep{Froehlich95} blue band data taken during the solar minimum between solar cycle 23 and 24. Lastly, a STAGGER \citep{Stein24} 3D hydrodynamical simulation replicating the K-dwarf $\epsilon$ Indi A is shown as the purple diamond. The colouring indicates an SNR-proxy as the ratio between the estimated granulation amplitude level and the white noise, where stars exhibiting higher degrees of contrast are darker.}
    \label{fig:TimescalePlateau}
    \vspace*{-3mm}
\end{figure*}

\subsection{The granulation timescales for dwarfs}\label{subsec:grantimescales}
A close correlation between the granulation timescale (or equivalently, the characteristic frequency, $\nu = 1/\left(2\pi\tau\right)$) and \numax\ is expected from theoretical considerations; a trend that was utilised earlier in Sect.~\ref{subsec:corrinf}. This relationship is shown in the bottom row of Fig.~\ref{fig:GranParams} for the three background models. In all cases, the characteristic frequencies increase with \numax. Interestingly, the third component of model~T also follows this trend, with a scatter comparable to that of the secondary component, which lends some credibility to its presence at high frequency. 

For the primary granulation component, however, the MS and SGB stars lie systematically above the scaling relation of \citet{Kallinger2014}. Following the same procedure as for $A_\mathrm{gran}$, we fit a simple power law to the characteristic frequencies of the first ($b$), second ($d$), and third ($f$) granulation components for each background model. The resulting fit coefficients are given in the figure panels. For the primary granulation component, clear systematic deviations from these fits indicate that, for our \kepler\ short-cadence sample, a simple power-law scaling does not adequately capture the observed behaviour. The most evolved stars in the sample at low values of \numax lie systematically below, meanwhile the MS and SGB stars hint at a non-linear trend in this log-space figure. 

The most striking deviation occurs for the primary granulation component at high \numax, from roughly the solar value ($\numax \approx 3000~\mu$Hz) upward, corresponding to MS stars cooler than the Sun. Across all background models, the characteristic frequency -- or equivalently, the granulation timescale -- appears to reach a plateau. This flattening implies that the tight correlation between granulation timescale and \numax\ breaks down, which would manifest in the stellar power spectra as an increasing frequency separation between the primary granulation component and the oscillation excess. Such behaviour was already noted for KIC8006161 in Fig.~\ref{fig:PDSComp}, which indeed lies on this plateau. The departure from this otherwise robust correlation is intriguing, as both granulation and stellar oscillations are fundamentally governed by the convective motions of the star. We further investigate this phenomenon and its implications in more detail in Sect.~\ref{sec:Decoupling}.

\section{Investigating the granulation plateau with TESS and stellar modelling}\label{sec:Decoupling}
In Sect.~\ref{subsec:grantimescales} we found an indication of a plateau in the granulation timescale for stars with a $\numax\gtrsim3000$ \muHz. We wish to investigate this surprising trend further, as it indicates a potential decoupling between granulation and oscillation timescales for cool dwarfs. The apparent plateau in Fig.~\ref{fig:GranParams} relies on a modest number of stars, as only a few in the sample display $\numax\gtrsim 3000$ \muHz. To remedy this situation, we further populate it with K-dwarfs observed by TESS \citep{Ricker14}. The target selection and data retrieval is outlined in Appendix~\ref{App:TESSKs}. 

Compared to \kepler, TESS delivers shorter time series with lower photometric contrast, which inevitably reduces the overall quality of the corresponding power spectra. For our purposes, however, the primary signal of interest is the granulation rather than the oscillations. While the latter are typically too weak to be detected reliably in TESS data of K-dwarfs (except for a few cases: \citealt{Hon24}; \citealt{Lund25}), the granulation may remain accessible. To ensure that these signals are not constrained by assumptions tailored to spectra that display clear oscillation signatures, we disable the correlated–inference configuration of Sect.~\ref{subsec:corrinf}, allowing the granulation parameters to be freely inferred. 

After application of the framework we manually evaluated all 78 stars. Those dominated by white or blue noise, displaying strong contaminants at low frequency, or miniscule granulation signals were removed. All posteriors for the stars were subsequently inspected to ensure reliable sampling and meaningful posteriors. In total, this removed 16 stars such that the TESS K-dwarf additions numbered 62, which are seen in Fig.~\ref{fig:TimescalePlateau}. The results obtained for the TESS K-dwarfs are plotted using the scaling relation \numax estimates and uncertainties, not the \numax inferred by the framework, as the oscillatory signals were undetectable.

The TESS K-dwarfs reproduce the plateau and thus lend further credibility to the decoupling between the granulation and oscillation timescales. As expected they show a larger scatter and a generally more uncertain estimation of the plateau, owing to the worse quality of the photometry. The stars showing the largest uncertainties in the estimated timescale similarly display granulation amplitudes comparable to the white noise level. Lastly, we examined the Gaia activity indexes \citep{GaiaDR3_AstParamRelease} for the stars in Fig.~\ref{fig:TimescalePlateau} to evaluate if magnetic activity could play a role in the observed scatter. However, no clear trend with the activity levels was found. While the scatter for the TESS K-dwarf additions is large, it is notable that the majority lie well below the expected scaling relation with \numax. Furthermore, they scatter to quite low values, indicating significantly longer timescales than expected.

\subsection{Signal recovery using simulated power spectra}
To validate that the framework can robustly recover granulation signals in power spectra with low signal-to-noise ratios, we performed a recovery test. We selected all \kepler and TESS targets shown in Fig.~\ref{fig:TimescalePlateau} with $\numax > 3000$\,\muHz \ -- 73 stars in total -- and simulated their power spectra using whichever background model was preferred by the evidence.

For each star, we adopted the observational parameters inferred by the framework as the underlying ‘true’ signal and added $\chi^2$-distributed noise following \citet{Gizon03}. We then generated six noise realisations per star with white-noise levels corresponding to target signal-to-noise ratios: $\text{SNR}=\text{observed primary granulation amplitude / white noise} = \{5.0,~2.0,~1.0,~0.7,~0.4,~0.2,~0.1\}$. In total, this yielded 511 simulated spectra: 259 generated using model~J, 153 using model~H, and 98 using model~T. That the simplest model (J) is preferred for the TESS data, which overall displays lower contrast between granulation and white noise levels than the \kepler data, is in line with the findings by \citet{Kallinger2014}.

We then reapplied the framework to all simulated spectra using the same background model that generated them and evaluated how well the characteristic frequency of the primary granulation component, $b$, was recovered. While all background parameters could be tested, $b$ is most relevant here, as the observed decoupling hinges on its behaviour alone. This was done using the same setup as for the TESS K-dwarfs, meaning the correlated inference of Sect.~\ref{subsec:corrinf} was disabled. Across all investigated levels of white noise, the framework recovers the true input timescale within $3\sigma$ in $94.9\%$ of cases. This demonstrates that even for power spectra dominated by noise -- as is frequently the case for the TESS sample in Fig.~\ref{fig:TimescalePlateau} -- the framework remains capable of reliably retrieving the underlying granulation timescale. This gives us confidence that the timescale plateau, and the associated decoupling it reflects, are not artefacts of noise-dominated power spectra or the methodology, but a real feature of the stellar granulation background.

\subsection{Convective energy transport of K-dwarfs} \label{sec:K-conv-1D}
The plateau in the granulation timescale can be explored further from a theoretical perspective. As a first attempt, we examine the predictions from 1D stellar models. One of the stars in our \kepler\ sample that lies on the plateau is Kepler-444 \citep{Campante15}. It has been studied in detail by \citet{Winther23}, from whom we recover the best-fitting stellar model calculated with GARSTEC \citep{Achim08}. Additionally, we also recover the model obtained from a solar calibration when using an identical setup and abundances (those of \citealt{Asplund09}). Contrasting these two models allows us to identify which differences in the outer layers distinguish the Sun -- with $\numax \approx 3090~\mu$Hz, at the onset of the plateau -- from Kepler-444, which displays $\numax\approx4400~\mu$Hz.

To reproduce the observed behaviour, some physical mechanism must act to prolong the granulation timescale. K-dwarfs possess deeper convective envelopes than G-dwarfs, but they may also differ in the efficiency with which convection transports energy. If such differences are present, they should imprint themselves on the temperature gradients in the near-surface layers.

\begin{figure}[t]
    \centering
    \resizebox{\hsize}{!}{\includegraphics[width=\linewidth]{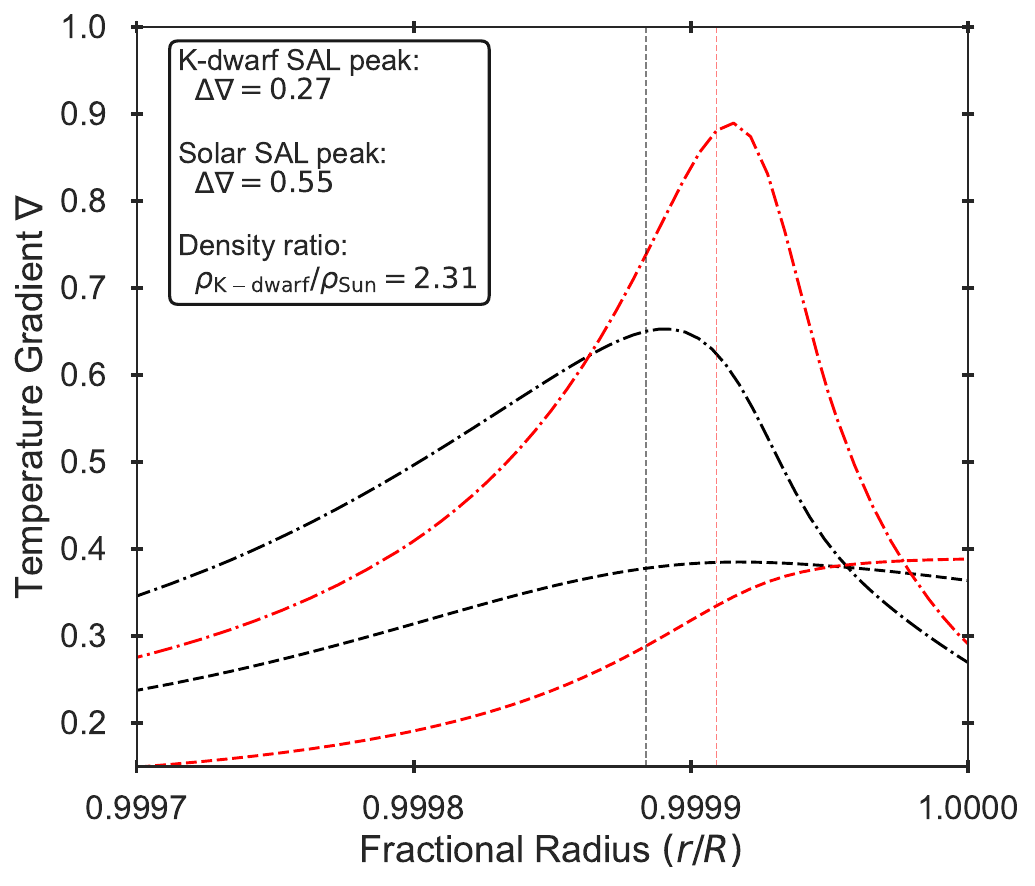}}
    \vspace*{-6mm}
    \caption{Temperature gradients in the super-adiabatic layer (SAL) of a K-dwarf (black) and solar model (red). The dashed and dot-dashed lines show the adiabatic gradient $\nabla_\mathrm{ad}$ and the structural gradient $\nabla$, respectively. The vertical dashed lines indicate the radial location of the maximum difference between the gradients, and the diagnostic inserts presents the values of this difference along with the ratio of the local density at this point between the models.}
    \label{fig:gradients}
    \vspace*{-2mm}
\end{figure}

We begin by considering the adiabatic gradient, $\nabla_\mathrm{ad}$, and the actual structural gradient, $\nabla$, in the outer envelope of the two models. These are shown in Fig.~\ref{fig:gradients} near the super-adiabatic layer \citep{Kippenhahn13} at the near-surface layers. We may compute the maximum difference between these gradients for both models,
\begin{equation}
    \Delta\nabla = \max\left(\nabla - \nabla_\mathrm{ad}\right) \ .
\end{equation}
A smaller value of $\Delta\nabla$ means the convection is less driven, which ties to the work being performed. If we consider the acceleration that a convective element experiences due to buoyancy, $a\propto g \frac{\delta \rho}{\rho}\propto g \ \Delta\nabla$ we see that it is proportional to the local gravitational acceleration and this difference in the gradients. Hence, over a displacement distance $\ell$ as defined by mixing-length theory \citep{VitenseMLT58}, the velocity of the element is, 
\begin{equation}\label{Eq:v_element}
    v^2 \propto a\ell \quad\Rightarrow\quad v \propto \sqrt{g\,\ell\,\Delta\nabla} \ .
\end{equation}
Across the thin super-adiabatic layer, the convective velocity therefore scales with the square root of the gradient difference.

Next, if we only consider the kinetic energy of the convective elements we may write the following expression for the flux transported by the convection, $F_\mathrm{conv}$, using Eq.~\ref{Eq:v_element}:
\begin{align}
    F_\mathrm{conv} &\propto \rho v^3 \propto \rho \left(\sqrt{g\ell\,\Delta\nabla}\right)^3 \label{eq:Fconv} \\
    \Rightarrow\Delta\nabla &\propto \left(\frac{F_\mathrm{conv}}{\rho(g\ell)^{3/2}}\right)^{2/3} \ . \label{eq:graddif}
\end{align}
For fixed $F_\mathrm{conv}$, $g$, and $\ell$ across the thin super-adiabatic layer, Eq.~\ref{eq:graddif} shows that an increase in the local density $\rho$ lowers $\Delta\nabla$. Indeed, the density at the position of $\Delta\nabla$ in the Kepler-444 model is higher by a factor of 2.3 and the reduction in $\Delta\nabla$ is reflected in the models (as seen in Fig.~\ref{fig:gradients}). Moreover, the local flux ratio at this same point is $\sim0.63$, showing that the required flux that must be transported is also reduced when comparing a K-dwarf to the Sun, as the former is less luminous. This immediately implies, via Eq.~\ref{eq:Fconv}, that a lower convective velocity is required to transport the required flux in a K-dwarf envelope. 

Finally, we approximate the convective turnover time in the two models. Assuming that all the energy is transported by convection -- which is a fair approximation near the super-adiabatic layer of low-mass MS stars -- the local flux, $F_\mathrm{conv} = F_r$, can be estimated as $F_r = L_r /4\pi r^2$, where $L_r$ is the luminosity at radial coordinate $r$. Using the velocity inferred from Eq.~\ref{eq:Fconv} and a mixing length $\ell = \alpha H_p$, we estimate
\begin{equation}
    \tau \approx \frac{\ell}{v} = \alpha H_p \left(\frac{\rho}{F_r}\right)^{1/3} \ .
\end{equation}
We adopt the solar-calibrated mixing-length parameter $\alpha = 1.786$ from \citet{Winther23} and evaluate the pressure scale height $H_p$ at the location of $\Delta\nabla$. Although the numerical estimates cannot be directly compared to the observed values -- owing to the assumptions and simplifications inherent in this treatment -- we can compare the ratio between them which finds $\tau_\mathrm{Kepler-444} \ / \ \tau_\odot \approx 1.08$. As the ratio is above 1, we have demonstrated that 1D stellar models naturally predict longer convective turnover times in K-dwarfs than in G-dwarfs; a behaviour primarily driven by the higher densities in their outer envelopes, and aided by their lower luminosities, leading to smaller convective velocities being necessary to transport the required flux.

\subsection{Granulation in a 3D K-dwarf simulation}\label{subsec:3Dsims}
\begin{table}
\centering
\caption{Basic properties of our 3D simulations.}
\label{tb:simu-info}
{\begin{tabular*}{\columnwidth}{@{\extracolsep{\fill}}lccc}
\toprule[2pt]
  Model name            & \texttt{t45g46m00}    & \texttt{solar}   & \texttt{t62g43m00}    
  \\
\midrule[1pt]
  \teff (K)     & $4571 \pm 7$ & $5772 \pm 16$ & $6231 \pm 14$  
  \\
  \logg (cgs)        & 4.62 & 4.438 & 4.319
  \\
  $d_{\mathrm{gran}}$  (Mm)  & 0.78 & 1.51  & 3.17
  \\
  $\hat{v}_{\mathrm{h}}$  (km/s) & 1.58    & 2.32  & 3.01
  \\
  $t_{\mathrm{gran}}$  (s)   & 248  & 326   & 527
  \\
\bottomrule[2pt]
\end{tabular*}}
\tablefoot{\texttt{t45g46m00} is set up to replicate $\epsilon$ Indi A and \texttt{t62g43m00} corresponds to an F-type MS star. As \teff is an emergent quantity of the 3D simulation that fluctuates with time, both its time-averaged mean and standard deviation are given. Symbols $d_{\mathrm{gran}}$ and $\hat{v}_{\mathrm{h}}$ denote typical granule size and representative value of horizontal velocity at the stellar surface, respectively. The granulation timescale is estimated through $t_{\mathrm{gran}} = d_{\mathrm{gran}} / (2 \hat{v}_{\mathrm{h}})$.}
\vspace*{-3mm}
\end{table}

To investigate this further we go beyond 1D stellar modelling and consider 3D hydrodynamic simulations of convection replicating the K-dwarf $\epsilon$ Indi A \citep{Campante24}. The 1D stellar models showed that owing to a higher density in the exterior of K-dwarfs, a lower velocity is required to transport the required flux via convection, likely resulting in longer convective turnover times. If this inference holds, it should be corroborated by the 3D simulations where we have access to all details concerning the velocity flows of the convective elements. 

Three simulations were calculated with the STAGGER code \citep{Stein24} using the solar abundances of \citet{Asplund09} and targeted a $\teff\approx 4580$ K and $\log(g) = 4.62$ dex, resulting in simulations of a MS star with $\numax \approx 5000$ \muHz. As done in \citet{Diaz22} the resonance modes (box modes) of the simulation domain were damped. This was done by introducing an artificial damping timescale (as the inverse $1/\omega$ of the cycling damping frequency) corresponding to half the frequency of the fundamental box mode, the frequency of the fundamental mode, and first overtone mode, respectively for each simulation. The horizontally averaged radiative flux was rescaled to the stellar surface following \citet{Trampedach98} and \citet{Ludwig06} assuming the stellar radius of $\epsilon$ Indi A (Lundkvist et al. in prep) for each simulation. The power spectra were then calculated as in \citet{Handberg11}.

We applied the original framework of \citet{Larsen2025b} to the power spectra of the simulations and recovered coherent results within the uncertainties for all three, suggesting that the choice of timescale for the artificial damping hardly affects our inferred parameters. The obtained timescales of the primary granulation component were then inspected and overplotted in Fig.~\ref{fig:TimescalePlateau} for the simulation with the longest damping timescale. Crucially, we find that the plateau in the granulation timescale is reproduced by the 3D simulations. The physical drivers in the stellar exteriors of K-dwarfs which result in longer timescales are thus present in our 3D simulations. This means that we can examine, in detail, what contributing factors play an important role in the convective behaviour of the K-dwarfs to produce the observed plateau.

\begin{figure}
    \centering
    \resizebox{\hsize}{!}{\includegraphics[width=\linewidth]{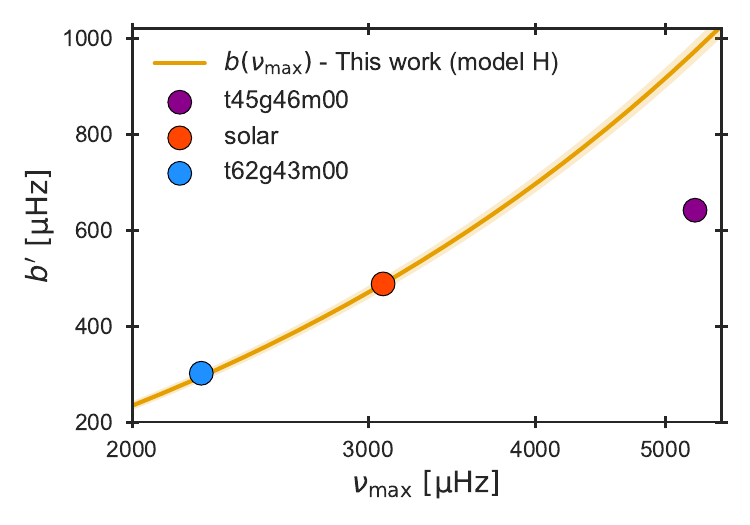}}
    \vspace*{-8mm}
    \caption{The inverse of the granulation timescale, $b'$, plotted against the oscillation timescale $\numax$, estimated from the three 3D simulations listed in Table \ref{tb:simu-info}. \numax was calculated using the asteroseismic scaling relation based on the fundamental parameters of the simulations. The gold solid line shows the fitted power law for model H from Fig.~\ref{fig:GranParams} with uncertainties indicated by the shaded bands, but scaled to pass the solar simulation.}
    \label{fig:tgran}
    \vspace*{-3mm}
\end{figure}

To this end, we introduce two additional 3D simulations, one solar and one corresponding to an F-type dwarf. Their basic parameters are listed in Table \ref{tb:simu-info}, together with those of the simulation replicating $\epsilon$-Indi. Recall that in a single-component model of granulation with exponent $l=2$, the parameter $b$ is inversely proportional to the characteristic timescale of granulation, that is $b = 1/(2\pi t_{\mathrm{gran}})$. The timescale, $b$, is tightly related to the typical size of the granules and the horizontal velocity at the optical surface. We estimated the granulation size similarly to \citet{Trampedach13}, where the 2D spatial power spectrum of bolometric intensity is computed for a series of simulation snapshots, followed by a radial average at different wave numbers. The peak of the time-averaged spatial spectrum indicates the typical size of granules. Next, we calculated the time-averaged distribution of horizontal velocity at the optical surface and selected the value with the highest probability to represent the characteristic surface horizontal velocity. The typical granulation timescale, $t_\mathrm{gran}$, is then crudely approximated as $t_\mathrm{gran}\approx d_{\mathrm{gran}} / (2\hat{v}_{\mathrm{h}})$. All quantities for the three 3D simulations are tabulated in Table \ref{tb:simu-info}.

The typical granulation size quantified from the three 3D simulations is not far from the approximate scaling relation $d_{\mathrm{gran}} \propto \teff / g$, due to its strong correlation with pressure scale height. However, when comparing horizontal velocities in the three simulations, we find a more pronounced decrease from the solar to the K-dwarf simulation than from the F-dwarf to solar, which implies that the granulation timescale increases more rapidly when moving from G- to F-type dwarfs. This is illustrated in Fig.~\ref{fig:tgran}, where the inverse of our estimated granulation timescale -- which has a similar physical meaning as the parameter $b$ -- is plotted against $\numax$. Although the `plateau' of $1/(2\pi t_{\mathrm{gran}})$ from the solar position onwards is not as apparent as for the observations in Fig.~\ref{fig:TimescalePlateau}, it is clear that the characteristic timescale of the granulation for the K-dwarf simulation \texttt{t45g46m00} does not follow the expected functional form of the correlation between granulation and oscillation timescales.  

The underlying reason for horizontal velocities decreasing more rapidly in K-dwarfs is likely associated with the fact that surface convection is much more efficient for cool dwarfs, as discussed in Sect.~\ref{sec:K-conv-1D}. 3D surface convection simulations suggest a moderate decrease of $\nabla - \nabla_{\mathrm{ad}}$ from F to G-type dwarfs, whereas the decrease of the gradient is notable for simulations with effective temperature less than ${\sim}5500$ K (cf.~Fig.~25 of \citealt{Magic2013}), implying convection is much more efficient in K-dwarfs. A smaller super-adiabatic temperature gradient translates to weaker vertical and horizontal velocity fields. The change in the efficiency of convection across spectral types is well-known from stellar models and manifests itself observationally in our detailed measurement of granulation timescales, as shown in Fig.~\ref{fig:TimescalePlateau}.

Lastly, we note that stellar oscillations are excited predominantly by the vertical convective motions, in particular by the strong convective downdrafts near the stellar surface \citep{Stein01}. While we find a reduction in horizontal velocities for K-dwarfs affecting the granulation timescales, the location of the oscillation envelope and \numax remain consistent with expectations from asteroseismic scaling relations. This is consistent with vertical convective velocities relevant for mode excitation, i.e. the downdrafs, being comparatively unaffected (see e.g. Sect.~3.2.2 in \citealt{Magic13}). As a result, the frequency separation between oscillation and granulation signals increases, as seen in the bottom panel of Fig.~\ref{fig:PDSComp}.

\section{Conclusion}\label{sec:Conclusion}
In this work we have further developed and applied the framework of \citet{Larsen2025b} to the largest sample of short-cadence \kepler\ stars analysed to date, drawing on the catalogue by \citet{Sayeed25}. Our aim was to characterise the surface granulation signatures and assess the robustness and applicability of the background modelling framework across a diverse stellar population. We tested an alternative model description (`peakbogging') for the power excess due to the presence of stellar oscillations, and while it shows promise, it requires additional work to remedy certain inconsistencies. Moving forwards with the traditional Gaussian envelope description, the following conclusions were reached:
\begin{itemize}
    \item Considering the background model preferences using the Bayesian evidences clearly shows no justification for a priori selection of background models. Rather, by applying a selection of models and choosing the one which best describes the given dataset, one allows for the optimal description of the background profile on a star-by-star basis.
    \item The inferred value of \numax is sensitive to model misspecification. The formal uncertainties on \numax are often comparable to or dominated by the systematic uncertainty stemming from the choice of background description. This underlines that reliable oscillation-based inferences require careful treatment of the granulation signal.
    \item The width of the Gaussian oscillation excess scales with \numax and indicates a temperature dependence. The dataset contributed by this work enables future studies of the correlations between envelope widths and various stellar parameters. 
    \item Total granulation amplitudes are consistent across background prescriptions and broadly align with the scaling relations of \citet{Kallinger2014} for evolved stars.
    \item The observed characteristic frequency of the primary granulation component shows a decoupling with the oscillation timescale \numax for MS stars cooler than the Sun ($\numax\gtrsim3000$ \muHz) -- indicating a plateau of prolonged granulation timescales for K-dwarfs. 
    \item The efficiency of the convective energy transport of K-dwarfs was examined by comparing 1D stellar models of Kepler-444 and the Sun. Due to the larger density in K-dwarf envelopes, and aided by a reduced luminosity, a lower velocity of the convective elements is necessary to transport the required flux. This leads to a smaller super-adiabatic temperature gradient and prolonged convective turnover times.
    \item Using 3D hydrodynamical simulations of convection, we replicated the K-dwarf $\epsilon$ Indi A with $\numax\approx 5000$ \muHz and applied the background inference framework of \citet{Larsen2025b} to the resulting power spectra. The 3D simulations reproduce the observed decoupling, reinforcing the physical picture supplied by the 1D stellar models. 
    \item The explanation for the decoupling likely stems from the much more efficient convective energy transport in K-dwarfs. The reduced super-adiabatic temperature gradients leads to weaker velocity fields, which from G- to K-dwarfs shows a significant decrease in the horizontal velocities, manifesting itself as prolonged granulation timescales in observations. 
\end{itemize}

As a direct consequence of this decoupling, the scaling relations often adopted from \citet{Kallinger2014} for describing the granulation signal of dwarfs are shown not to hold for MS stars with $\numax \gtrsim 3000$ \muHz. A fact which one should be aware of when, for example, simulations of power spectra for the upcoming ESA PLATO mission are made \citep{Samadi19,PLATO24}. Furthermore, this decoupling between granulation and oscillation timescales may offer a positive prospect for future detectability of asteroseismic signals in K-dwarfs, as it implies an increased frequency separation between the primary granulation signal and the oscillation excess. Moreover, it is noteworthy that the onset of the decoupling and the associated plateau in granulation timescales roughly coincides with the point where \citet{Campante24} and \citet{LiY25} observed a deviation from the expected oscillation amplitude scaling -- finding an apparent transition with decreasing $L/M$ from a scaling of $\sim\left(L/M\right)^{0.7}$ for the mode amplitudes to a steeper $\sim\left(L/M\right)^{1.5}$ for K-dwarfs. As both the granulation and oscillations are fundamentally tied to the convective motions of the star, a decrease in the convective velocities would influence the excitation mechanism of the oscillations, suggesting that the two observed effects may share a common origin. However, other factors, such as mode damping, is also an important contribution, and it will require dedicated future studies to disentangle their roles and pinpoint the exact causes. Such efforts will represent an important step towards refining asteroseismic inferences and general stellar characterisation of K-dwarfs on the lower MS. 

Alongside these physical insights, this work provides a comprehensive dataset: complete posterior distributions, parameter covariances, and Bayesian evidences for three background descriptions applied to \Nstars\ stars. When combined with the catalogue of \citet{Sayeed25}, the majority of them are also asteroseismically characterised, which provides estimates for the fundamental stellar parameters. This combination enables extensive future studies into how the surface signatures of convection ties to the stellar parameters across a diverse sample of stars spanning the MS, SGB and lower RGB.

\begin{acknowledgements}
    JRL wishes to thank the members of SAC in Aarhus and the Sun, Stars and Exoplanets group in Birmingham for comments and discussions regarding the paper -- in particular Jørgen Christensen-Dalsgaard and Hans Kjeldsen for helpful insights concerning stellar convection and thermodynamics. Additionally, the authors thank Maryum Sayeed for allowing us preliminary access to the catalogue data. This work was supported by a research grant (42101) from VILLUM FONDEN. MSL acknowledges support from The Independent Research Fund Denmark's Inge Lehmann  program (grant  agreement  no.:  1131-00014B). MNL acknowledges support from the ESA PRODEX programme (PEA 4000142995). YZ acknowledges support from the European Union's Horizon 2020 research and innovation programme under the Marie Skłodowska–Curie grant agreement No 101150921. Funding for the Stellar Astrophysics Centre was provided by The Danish National Research Foundation (grant agreement no.: DNRF106). The numerical results presented in this work were partly obtained at the Centre for Scientific Computing, Aarhus \url{https://phys.au.dk/forskning/faciliteter/cscaa/}. This paper received funding from the European Research Council (ERC) under the European Union’s Horizon 2020 research and innovation programme (CartographY GA. 804752). This paper includes data collected by the \kepler and TESS missions and obtained from the MAST data archive at the Space Telescope Science Institute (STScI). Funding for the Kepler mission was provided by the NASA Science Mission Directorate. Funding for the TESS mission is provided by the NASA Explorer Program. STScI is operated by the Association of Universities for Research in Astronomy, Inc., under NASA contract NAS 5–26555. This work presents results from the European Space Agency (ESA) space mission \gaia. \gaia data are being processed by the \gaia Data Processing and Analysis Consortium (DPAC). Funding for the DPAC is provided by national institutions, in particular the institutions participating in the \gaia MultiLateral Agreement (MLA). The \gaia mission website is \url{https://www.cosmos.esa.int/gaia}. The \gaia archive website is \url{https://archives.esac.esa.int/gaia}.
\end{acknowledgements}

\section*{Data availability}
All data products for the \Ninit \kepler short-cadence stars studied are available in this online repository: \url{https://www.erda.au.dk/archives/7edd1841b91dec2afd027a4bb7be3598/published-archive.html}. The framework for the background inference is accessible on Github upon reasonable request to the first author, as it is currently under further development.

\bibliographystyle{aa.bst} 
\bibliography{bibliography.bib}

@ARTICLE{Anderson1990,
       author = {{Anderson}, Edwin R. and {Duvall}, Thomas L., Jr. and {Jefferies}, Stuart M.},
        title = "{Modeling of Solar Oscillation Power Spectra}",
      journal = {\apj},
         year = 1990,
        month = dec,
       volume = {364},
        pages = {699},
          doi = {10.1086/169452},
       adsurl = {https://ui.adsabs.harvard.edu/abs/1990ApJ...364..699A}
}

@ARTICLE{Lund2017,
       author = {{Lund}, Mikkel N. and {Silva Aguirre}, V{\'\i}ctor and {Davies}, Guy R. and {Chaplin}, William J. and {Christensen-Dalsgaard}, J{\o}rgen and {Houdek}, G{\"u}nter and {White}, Timothy R. and {Bedding}, Timothy R. and {Ball}, Warrick H. and {Huber}, Daniel and {Antia}, H.~M. and {Lebreton}, Yveline and {Latham}, David W. and {Handberg}, Rasmus and {Verma}, Kuldeep and {Basu}, Sarbani and {Casagrande}, Luca and {Justesen}, Anders B. and {Kjeldsen}, Hans and {Mosumgaard}, Jakob R.},
        title = "{Standing on the Shoulders of Dwarfs: the Kepler Asteroseismic LEGACY Sample. I. Oscillation Mode Parameters}",
      journal = {\apj},
         year = 2017,
        month = feb,
       volume = {835},
       number = {2},
          eid = {172},
        pages = {172},
          doi = {10.3847/1538-4357/835/2/172},
archivePrefix = {arXiv},
       eprint = {1612.00436},
 primaryClass = {astro-ph.SR},
       adsurl = {https://ui.adsabs.harvard.edu/abs/2017ApJ...835..172L}
}

@ARTICLE{Borucki10,
       author = {{Borucki}, William J. and {Koch}, David and {Basri}, Gibor and {Batalha}, Natalie and {Brown}, Timothy and {Caldwell}, Douglas and {Caldwell}, John and {Christensen-Dalsgaard}, J{\o}rgen and {Cochran}, William D. and {DeVore}, Edna and {Dunham}, Edward W. and {Dupree}, Andrea K. and {Gautier}, Thomas N. and {Geary}, John C. and {Gilliland}, Ronald and {Gould}, Alan and {Howell}, Steve B. and {Jenkins}, Jon M. and {Kondo}, Yoji and {Latham}, David W. and {Marcy}, Geoffrey W. and {Meibom}, S{\o}ren and {Kjeldsen}, Hans and {Lissauer}, Jack J. and {Monet}, David G. and {Morrison}, David and {Sasselov}, Dimitar and {Tarter}, Jill and {Boss}, Alan and {Brownlee}, Don and {Owen}, Toby and {Buzasi}, Derek and {Charbonneau}, David and {Doyle}, Laurance and {Fortney}, Jonathan and {Ford}, Eric B. and {Holman}, Matthew J. and {Seager}, Sara and {Steffen}, Jason H. and {Welsh}, William F. and {Rowe}, Jason and {Anderson}, Howard and {Buchhave}, Lars and {Ciardi}, David and {Walkowicz}, Lucianne and {Sherry}, William and {Horch}, Elliott and {Isaacson}, Howard and {Everett}, Mark E. and {Fischer}, Debra and {Torres}, Guillermo and {Johnson}, John Asher and {Endl}, Michael and {MacQueen}, Phillip and {Bryson}, Stephen T. and {Dotson}, Jessie and {Haas}, Michael and {Kolodziejczak}, Jeffrey and {Van Cleve}, Jeffrey and {Chandrasekaran}, Hema and {Twicken}, Joseph D. and {Quintana}, Elisa V. and {Clarke}, Bruce D. and {Allen}, Christopher and {Li}, Jie and {Wu}, Haley and {Tenenbaum}, Peter and {Verner}, Ekaterina and {Bruhweiler}, Frederick and {Barnes}, Jason and {Prsa}, Andrej},
        title = "{Kepler Planet-Detection Mission: Introduction and First Results}",
      journal = {Science},
         year = 2010,
        month = feb,
       volume = {327},
       number = {5968},
        pages = {977},
          doi = {10.1126/science.1185402},
       adsurl = {https://ui.adsabs.harvard.edu/abs/2010Sci...327..977B}
}

@ARTICLE{Kjeldsen95,
       author = {{Kjeldsen}, H. and {Bedding}, T.~R.},
        title = "{Amplitudes of stellar oscillations: the implications for asteroseismology.}",
      journal = {\aap},
         year = 1995,
        month = jan,
       volume = {293},
        pages = {87-106},
          doi = {10.48550/arXiv.astro-ph/9403015},
archivePrefix = {arXiv},
       eprint = {astro-ph/9403015},
 primaryClass = {astro-ph},
       adsurl = {https://ui.adsabs.harvard.edu/abs/1995A&A...293...87K}
}

@ARTICLE{Achim08,
       author = {{Weiss}, Achim and {Schlattl}, Helmut},
        title = "{GARSTEC{\textemdash}the Garching Stellar Evolution Code. The direct descendant of the legendary Kippenhahn code}",
      journal = {\apss},
         year = 2008,
        month = aug,
       volume = {316},
       number = {1-4},
        pages = {99-106},
          doi = {10.1007/s10509-007-9606-5},
       adsurl = {https://ui.adsabs.harvard.edu/abs/2008Ap&SS.316...99W}
}

@ARTICLE{Winther23,
       author = {{Winther}, Mark Lykke and {Aguirre B{\o}rsen-Koch}, V{\'\i}ctor and {R{\o}rsted}, Jakob Lysgaard and {Stokholm}, Amalie and {Verma}, Kuldeep},
        title = "{Did Kepler-444 have a long-lived convective core?}",
      journal = {\mnras},
         year = 2023,
        month = oct,
       volume = {525},
       number = {1},
        pages = {1416-1430},
          doi = {10.1093/mnras/stad1802},
archivePrefix = {arXiv},
       eprint = {2306.08430},
 primaryClass = {astro-ph.SR},
       adsurl = {https://ui.adsabs.harvard.edu/abs/2023MNRAS.525.1416W}
}

@ARTICLE{Yu18,
       author = {{Yu}, Jie and {Huber}, Daniel and {Bedding}, Timothy R. and {Stello}, Dennis and {Hon}, Marc and {Murphy}, Simon J. and {Khanna}, Shourya},
        title = "{Asteroseismology of 16,000 Kepler Red Giants: Global Oscillation Parameters, Masses, and Radii}",
      journal = {\apjs},
         year = 2018,
        month = jun,
       volume = {236},
       number = {2},
          eid = {42},
        pages = {42},
          doi = {10.3847/1538-4365/aaaf74},
archivePrefix = {arXiv},
       eprint = {1802.04455},
 primaryClass = {astro-ph.SR},
       adsurl = {https://ui.adsabs.harvard.edu/abs/2018ApJS..236...42Y}
}

@ARTICLE{Asplund09,
       author = {{Asplund}, Martin and {Grevesse}, Nicolas and {Sauval}, A. Jacques and {Scott}, Pat},
        title = "{The Chemical Composition of the Sun}",
      journal = {\araa},
         year = 2009,
        month = sep,
       volume = {47},
       number = {1},
        pages = {481-522},
          doi = {10.1146/annurev.astro.46.060407.145222},
archivePrefix = {arXiv},
       eprint = {0909.0948},
 primaryClass = {astro-ph.SR},
       adsurl = {https://ui.adsabs.harvard.edu/abs/2009ARA&A..47..481A}
}

@BOOK{Kippenhahn13,
       author = {{Kippenhahn}, Rudolf and {Weigert}, Alfred and {Weiss}, Achim},
        title = "{Stellar Structure and Evolution}",
         year = 2013,
          doi = {10.1007/978-3-642-30304-3},
       adsurl = {https://ui.adsabs.harvard.edu/abs/2013sse..book.....K}
}

@ARTICLE{GaiaDR3,
       author = {{Gaia Collaboration} and {Vallenari}, A. and {Brown}, A.~G.~A. and {Prusti}, T. and {de Bruijne}, J.~H.~J. and {Arenou}, F. and {Babusiaux}, C. and {Biermann}, M. and {Creevey}, O.~L. and {Ducourant}, C. and {Evans}, D.~W. and {Eyer}, L. and {Guerra}, R. and {Hutton}, A. and {Jordi}, C. and {Klioner}, S.~A. and {Lammers}, U.~L. and {Lindegren}, L. and {Luri}, X. and {Mignard}, F. and {Panem}, C. and {Pourbaix}, D. and {Randich}, S. and {Sartoretti}, P. and {Soubiran}, C. and {Tanga}, P. and {Walton}, N.~A. and {Bailer-Jones}, C.~A.~L. and {Bastian}, U. and {Drimmel}, R. and {Jansen}, F. and {Katz}, D. and {Lattanzi}, M.~G. and {van Leeuwen}, F. and {Bakker}, J. and {Cacciari}, C. and {Casta{\~n}eda}, J. and {De Angeli}, F. and {Fabricius}, C. and {Fouesneau}, M. and {Fr{\'e}mat}, Y. and {Galluccio}, L. and {Guerrier}, A. and {Heiter}, U. and {Masana}, E. and {Messineo}, R. and {Mowlavi}, N. and {Nicolas}, C. and {Nienartowicz}, K. and {Pailler}, F. and {Panuzzo}, P. and {Riclet}, F. and {Roux}, W. and {Seabroke}, G.~M. and {Sordo}, R. and {Th{\'e}venin}, F. and {Gracia-Abril}, G. and {Portell}, J. and {Teyssier}, D. and {Altmann}, M. and {Andrae}, R. and {Audard}, M. and {Bellas-Velidis}, I. and {Benson}, K. and {Berthier}, J. and {Blomme}, R. and {Burgess}, P.~W. and {Busonero}, D. and {Busso}, G. and {C{\'a}novas}, H. and {Carry}, B. and {Cellino}, A. and {Cheek}, N. and {Clementini}, G. and {Damerdji}, Y. and {Davidson}, M. and {de Teodoro}, P. and {Nu{\~n}ez Campos}, M. and {Delchambre}, L. and {Dell'Oro}, A. and {Esquej}, P. and {Fern{\'a}ndez-Hern{\'a}ndez}, J. and {Fraile}, E. and {Garabato}, D. and {Garc{\'\i}a-Lario}, P. and {Gosset}, E. and {Haigron}, R. and {Halbwachs}, J. -L. and {Hambly}, N.~C. and {Harrison}, D.~L. and {Hern{\'a}ndez}, J. and {Hestroffer}, D. and {Hodgkin}, S.~T. and {Holl}, B. and {Jan{\ss}en}, K. and {Jevardat de Fombelle}, G. and {Jordan}, S. and {Krone-Martins}, A. and {Lanzafame}, A.~C. and {L{\"o}ffler}, W. and {Marchal}, O. and {Marrese}, P.~M. and {Moitinho}, A. and {Muinonen}, K. and {Osborne}, P. and {Pancino}, E. and {Pauwels}, T. and {Recio-Blanco}, A. and {Reyl{\'e}}, C. and {Riello}, M. and {Rimoldini}, L. and {Roegiers}, T. and {Rybizki}, J. and {Sarro}, L.~M. and {Siopis}, C. and {Smith}, M. and {Sozzetti}, A. and {Utrilla}, E. and {van Leeuwen}, M. and {Abbas}, U. and {{\'A}brah{\'a}m}, P. and {Abreu Aramburu}, A. and {Aerts}, C. and {Aguado}, J.~J. and {Ajaj}, M. and {Aldea-Montero}, F. and {Altavilla}, G. and {{\'A}lvarez}, M.~A. and {Alves}, J. and {Anders}, F. and {Anderson}, R.~I. and {Anglada Varela}, E. and {Antoja}, T. and {Baines}, D. and {Baker}, S.~G. and {Balaguer-N{\'u}{\~n}ez}, L. and {Balbinot}, E. and {Balog}, Z. and {Barache}, C. and {Barbato}, D. and {Barros}, M. and {Barstow}, M.~A. and {Bartolom{\'e}}, S. and {Bassilana}, J. -L. and {Bauchet}, N. and {Becciani}, U. and {Bellazzini}, M. and {Berihuete}, A. and {Bernet}, M. and {Bertone}, S. and {Bianchi}, L. and {Binnenfeld}, A. and {Blanco-Cuaresma}, S. and {Blazere}, A. and {Boch}, T. and {Bombrun}, A. and {Bossini}, D. and {Bouquillon}, S. and {Bragaglia}, A. and {Bramante}, L. and {Breedt}, E. and {Bressan}, A. and {Brouillet}, N. and {Brugaletta}, E. and {Bucciarelli}, B. and {Burlacu}, A. and {Butkevich}, A.~G. and {Buzzi}, R. and {Caffau}, E. and {Cancelliere}, R. and {Cantat-Gaudin}, T. and {Carballo}, R. and {Carlucci}, T. and {Carnerero}, M.~I. and {Carrasco}, J.~M. and {Casamiquela}, L. and {Castellani}, M. and {Castro-Ginard}, A. and {Chaoul}, L. and {Charlot}, P. and {Chemin}, L. and {Chiaramida}, V. and {Chiavassa}, A. and {Chornay}, N. and {Comoretto}, G. and {Contursi}, G. and {Cooper}, W.~J. and {Cornez}, T. and {Cowell}, S. and {Crifo}, F. and {Cropper}, M. and {Crosta}, M. and {Crowley}, C. and {Dafonte}, C. and {Dapergolas}, A. and {David}, M. and {David}, P. and {de Laverny}, P. and {De Luise}, F. and {De March}, R. and {De Ridder}, J. and {de Souza}, R. and {de Torres}, A. and {del Peloso}, E.~F. and {del Pozo}, E. and {Delbo}, M. and {Delgado}, A. and {Delisle}, J. -B. and {Demouchy}, C. and {Dharmawardena}, T.~E. and {Di Matteo}, P. and {Diakite}, S. and {Diener}, C. and {Distefano}, E. and {Dolding}, C. and {Edvardsson}, B. and {Enke}, H. and {Fabre}, C. and {Fabrizio}, M. and {Faigler}, S. and {Fedorets}, G. and {Fernique}, P. and {Fienga}, A. and {Figueras}, F. and {Fournier}, Y. and {Fouron}, C. and {Fragkoudi}, F. and {Gai}, M. and {Garcia-Gutierrez}, A. and {Garcia-Reinaldos}, M. and {Garc{\'\i}a-Torres}, M. and {Garofalo}, A. and {Gavel}, A. and {Gavras}, P. and {Gerlach}, E. and {Geyer}, R. and {Giacobbe}, P. and {Gilmore}, G. and {Girona}, S. and {Giuffrida}, G. and {Gomel}, R. and {Gomez}, A. and {Gonz{\'a}lez-N{\'u}{\~n}ez}, J. and {Gonz{\'a}lez-Santamar{\'\i}a}, I. and {Gonz{\'a}lez-Vidal}, J.~J. and {Granvik}, M. and {Guillout}, P. and {Guiraud}, J. and {Guti{\'e}rrez-S{\'a}nchez}, R. and {Guy}, L.~P. and {Hatzidimitriou}, D. and {Hauser}, M. and {Haywood}, M. and {Helmer}, A. and {Helmi}, A. and {Sarmiento}, M.~H. and {Hidalgo}, S.~L. and {Hilger}, T. and {H{\l}adczuk}, N. and {Hobbs}, D. and {Holland}, G. and {Huckle}, H.~E. and {Jardine}, K. and {Jasniewicz}, G. and {Jean-Antoine Piccolo}, A. and {Jim{\'e}nez-Arranz}, {\'O}. and {Jorissen}, A. and {Juaristi Campillo}, J. and {Julbe}, F. and {Karbevska}, L. and {Kervella}, P. and {Khanna}, S. and {Kontizas}, M. and {Kordopatis}, G. and {Korn}, A.~J. and {K{\'o}sp{\'a}l}, {\'A}. and {Kostrzewa-Rutkowska}, Z. and {Kruszy{\'n}ska}, K. and {Kun}, M. and {Laizeau}, P. and {Lambert}, S. and {Lanza}, A.~F. and {Lasne}, Y. and {Le Campion}, J. -F. and {Lebreton}, Y. and {Lebzelter}, T. and {Leccia}, S. and {Leclerc}, N. and {Lecoeur-Taibi}, I. and {Liao}, S. and {Licata}, E.~L. and {Lindstr{\o}m}, H.~E.~P. and {Lister}, T.~A. and {Livanou}, E. and {Lobel}, A. and {Lorca}, A. and {Loup}, C. and {Madrero Pardo}, P. and {Magdaleno Romeo}, A. and {Managau}, S. and {Mann}, R.~G. and {Manteiga}, M. and {Marchant}, J.~M. and {Marconi}, M. and {Marcos}, J. and {Marcos Santos}, M.~M.~S. and {Mar{\'\i}n Pina}, D. and {Marinoni}, S. and {Marocco}, F. and {Marshall}, D.~J. and {Martin Polo}, L. and {Mart{\'\i}n-Fleitas}, J.~M. and {Marton}, G. and {Mary}, N. and {Masip}, A. and {Massari}, D. and {Mastrobuono-Battisti}, A. and {Mazeh}, T. and {McMillan}, P.~J. and {Messina}, S. and {Michalik}, D. and {Millar}, N.~R. and {Mints}, A. and {Molina}, D. and {Molinaro}, R. and {Moln{\'a}r}, L. and {Monari}, G. and {Mongui{\'o}}, M. and {Montegriffo}, P. and {Montero}, A. and {Mor}, R. and {Mora}, A. and {Morbidelli}, R. and {Morel}, T. and {Morris}, D. and {Muraveva}, T. and {Murphy}, C.~P. and {Musella}, I. and {Nagy}, Z. and {Noval}, L. and {Oca{\~n}a}, F. and {Ogden}, A. and {Ordenovic}, C. and {Osinde}, J.~O. and {Pagani}, C. and {Pagano}, I. and {Palaversa}, L. and {Palicio}, P.~A. and {Pallas-Quintela}, L. and {Panahi}, A. and {Payne-Wardenaar}, S. and {Pe{\~n}alosa Esteller}, X. and {Penttil{\"a}}, A. and {Pichon}, B. and {Piersimoni}, A.~M. and {Pineau}, F. -X. and {Plachy}, E. and {Plum}, G. and {Poggio}, E. and {Pr{\v{s}}a}, A. and {Pulone}, L. and {Racero}, E. and {Ragaini}, S. and {Rainer}, M. and {Raiteri}, C.~M. and {Rambaux}, N. and {Ramos}, P. and {Ramos-Lerate}, M. and {Re Fiorentin}, P. and {Regibo}, S. and {Richards}, P.~J. and {Rios Diaz}, C. and {Ripepi}, V. and {Riva}, A. and {Rix}, H. -W. and {Rixon}, G. and {Robichon}, N. and {Robin}, A.~C. and {Robin}, C. and {Roelens}, M. and {Rogues}, H.~R.~O. and {Rohrbasser}, L. and {Romero-G{\'o}mez}, M. and {Rowell}, N. and {Royer}, F. and {Ruz Mieres}, D. and {Rybicki}, K.~A. and {Sadowski}, G. and {S{\'a}ez N{\'u}{\~n}ez}, A. and {Sagrist{\`a} Sell{\'e}s}, A. and {Sahlmann}, J. and {Salguero}, E. and {Samaras}, N. and {Sanchez Gimenez}, V. and {Sanna}, N. and {Santove{\~n}a}, R. and {Sarasso}, M. and {Schultheis}, M. and {Sciacca}, E. and {Segol}, M. and {Segovia}, J.~C. and {S{\'e}gransan}, D. and {Semeux}, D. and {Shahaf}, S. and {Siddiqui}, H.~I. and {Siebert}, A. and {Siltala}, L. and {Silvelo}, A. and {Slezak}, E. and {Slezak}, I. and {Smart}, R.~L. and {Snaith}, O.~N. and {Solano}, E. and {Solitro}, F. and {Souami}, D. and {Souchay}, J. and {Spagna}, A. and {Spina}, L. and {Spoto}, F. and {Steele}, I.~A. and {Steidelm{\"u}ller}, H. and {Stephenson}, C.~A. and {S{\"u}veges}, M. and {Surdej}, J. and {Szabados}, L. and {Szegedi-Elek}, E. and {Taris}, F. and {Taylor}, M.~B. and {Teixeira}, R. and {Tolomei}, L. and {Tonello}, N. and {Torra}, F. and {Torra}, J. and {Torralba Elipe}, G. and {Trabucchi}, M. and {Tsounis}, A.~T. and {Turon}, C. and {Ulla}, A. and {Unger}, N. and {Vaillant}, M.~V. and {van Dillen}, E. and {van Reeven}, W. and {Vanel}, O. and {Vecchiato}, A. and {Viala}, Y. and {Vicente}, D. and {Voutsinas}, S. and {Weiler}, M. and {Wevers}, T. and {Wyrzykowski}, {\L}. and {Yoldas}, A. and {Yvard}, P. and {Zhao}, H. and {Zorec}, J. and {Zucker}, S. and {Zwitter}, T.},
        title = "{Gaia Data Release 3. Summary of the content and survey properties}",
      journal = {\aap},
         year = 2023,
        month = jun,
       volume = {674},
          eid = {A1},
        pages = {A1},
          doi = {10.1051/0004-6361/202243940},
archivePrefix = {arXiv},
       eprint = {2208.00211},
 primaryClass = {astro-ph.GA},
       adsurl = {https://ui.adsabs.harvard.edu/abs/2023A&A...674A...1G}
}

@ARTICLE{Handberg11,
       author = {{Handberg}, R. and {Campante}, T.~L.},
        title = "{Bayesian peak-bagging of solar-like oscillators using MCMC: a comprehensive guide}",
      journal = {\aap},
         year = 2011,
        month = mar,
       volume = {527},
          eid = {A56},
        pages = {A56},
          doi = {10.1051/0004-6361/201015451},
archivePrefix = {arXiv},
       eprint = {1101.0084},
 primaryClass = {astro-ph.SR},
       adsurl = {https://ui.adsabs.harvard.edu/abs/2011A&A...527A..56H}
}

@INCOLLECTION{Bedding14,
       author = {{Bedding}, Timothy R.},
        title = "{Solar-like oscillations: An observational perspective}",
    booktitle = {Asteroseismology},
         year = 2014,
       editor = {{Pall{\'e}}, Pere L. and {Esteban}, Cesar},
        pages = {60},
          doi = {10.48550/arXiv.1107.1723},
       adsurl = {https://ui.adsabs.harvard.edu/abs/2014aste.book...60B}
}

@article{Corsaro14,
	adsurl = {http://adsabs.harvard.edu/abs/2014A%26A...571A..71C},
	archiveprefix = {arXiv},
	author = {{Corsaro}, E. and {De Ridder}, J.},
	doi = {10.1051/0004-6361/201424181},
	eid = {A71},
	eprint = {1408.2515},
	journal = {\aap},
	month = nov,
	pages = {A71},
	primaryclass = {astro-ph.IM},
	title = {{DIAMONDS: A new Bayesian nested sampling tool. Application to peak bagging of solar-like oscillations}},
	volume = 571,
	year = 2014
}

@article{Corsaro17,
	adsurl = {http://adsabs.harvard.edu/abs/2017A%26A...605A...3C},
	archiveprefix = {arXiv},
	author = {{Corsaro}, E. and {Mathur}, S. and {Garc{\'{\i}}a}, R.~A. and {Gaulme}, P. and {Pinsonneault}, M. and {Stassun}, K. and {Stello}, D. and {Tayar}, J. and {Trampedach}, R. and {Jiang}, C. and {Nitschelm}, C. and {Salabert}, D.},
	doi = {10.1051/0004-6361/201731094},
	eid = {A3},
	eprint = {1707.07474},
	journal = {\aap},
	month = aug,
	pages = {A3},
	primaryclass = {astro-ph.SR},
	title = {{Metallicity effect on stellar granulation detected from oscillating red giants in open clusters}},
	volume = 605,
	year = 2017
}

@INPROCEEDINGS{Ricker14,
       author = {{Ricker}, George R. and {Winn}, Joshua N. and {Vanderspek}, Roland and {Latham}, David W. and {Bakos}, G{\'a}sp{\'a}r. {\'A}. and {Bean}, Jacob L. and {Berta-Thompson}, Zachory K. and {Brown}, Timothy M. and {Buchhave}, Lars and {Butler}, Nathaniel R. and {Butler}, R. Paul and {Chaplin}, William J. and {Charbonneau}, David and {Christensen-Dalsgaard}, J{\o}rgen and {Clampin}, Mark and {Deming}, Drake and {Doty}, John and {De Lee}, Nathan and {Dressing}, Courtney and {Dunham}, E.~W. and {Endl}, Michael and {Fressin}, Francois and {Ge}, Jian and {Henning}, Thomas and {Holman}, Matthew J. and {Howard}, Andrew W. and {Ida}, Shigeru and {Jenkins}, Jon and {Jernigan}, Garrett and {Johnson}, John A. and {Kaltenegger}, Lisa and {Kawai}, Nobuyuki and {Kjeldsen}, Hans and {Laughlin}, Gregory and {Levine}, Alan M. and {Lin}, Douglas and {Lissauer}, Jack J. and {MacQueen}, Phillip and {Marcy}, Geoffrey and {McCullough}, P.~R. and {Morton}, Timothy D. and {Narita}, Norio and {Paegert}, Martin and {Palle}, Enric and {Pepe}, Francesco and {Pepper}, Joshua and {Quirrenbach}, Andreas and {Rinehart}, S.~A. and {Sasselov}, Dimitar and {Sato}, Bun'ei and {Seager}, Sara and {Sozzetti}, Alessandro and {Stassun}, Keivan G. and {Sullivan}, Peter and {Szentgyorgyi}, Andrew and {Torres}, Guillermo and {Udry}, Stephane and {Villasenor}, Joel},
        title = "{Transiting Exoplanet Survey Satellite (TESS)}",
    booktitle = {Space Telescopes and Instrumentation 2014: Optical, Infrared, and Millimeter Wave},
         year = 2014,
       editor = {{Oschmann}, Jacobus M., Jr. and {Clampin}, Mark and {Fazio}, Giovanni G. and {MacEwen}, Howard A.},
       series = {Society of Photo-Optical Instrumentation Engineers (SPIE) Conference Series},
       volume = {9143},
        month = aug,
          eid = {914320},
        pages = {914320},
          doi = {10.1117/12.2063489},
archivePrefix = {arXiv},
       eprint = {1406.0151},
 primaryClass = {astro-ph.EP},
       adsurl = {https://ui.adsabs.harvard.edu/abs/2014SPIE.9143E..20R}
}

@ARTICLE{PLATO24,
       author = {{Rauer}, Heike and {Aerts}, Conny and {Cabrera}, Juan and {Deleuil}, Magali and {Erikson}, Anders and {Gizon}, Laurent and {Goupil}, Mariejo and {Heras}, Ana and {Lorenzo-Alvarez}, Jose and {Marliani}, Filippo and {Martin-Garcia}, Cesar and {Mas-Hesse}, J. Miguel and {O'Rourke}, Laurence and {Osborn}, Hugh and {Pagano}, Isabella and {Piotto}, Giampaolo and {Pollacco}, Don and {Ragazzoni}, Roberto and {Ramsay}, Gavin and {Udry}, St{\'e}phane and {Appourchaux}, Thierry and {Benz}, Willy and {Brandeker}, Alexis and {G{\"u}del}, Manuel and {Janot-Pacheco}, Eduardo and {Kabath}, Petr and {Kjeldsen}, Hans and {Min}, Michiel and {Santos}, Nuno and {Smith}, Alan and {Suarez}, Juan-Carlos and {Werner}, Stephanie C. and {Aboudan}, Alessio and {Abreu}, Manuel and {Acu{\~n}a}, Lorena and {Adams}, Moritz and {Adibekyan}, Vardan and {Affer}, Laura and {Agneray}, Fran{\c{c}}ois and {Agnor}, Craig and {Aguirre B{\o}rsen-Koch}, Victor and {Ahmed}, Saad and {Aigrain}, Suzanne and {Al-Bahlawan}, Ashraf and {Alcacera Gil}, M de los Angeles and {Alei}, Eleonora and {Alencar}, Silvia and {Alexander}, Richard and {Alfonso-Garz{\'o}n}, Julia and {Alibert}, Yann and {Allende Prieto}, Carlos and {Almeida}, Leonardo and {Alonso Sobrino}, Roi and {Altavilla}, Giuseppe and {Althaus}, Christian and {Alonso Alvarez Trujillo}, Luis and {Amarsi}, Anish and {Ammler-von Eiff}, Matthias and {Am{\^o}res}, Eduardo and {Andrade}, Laerte and {Antoniadis-Karnavas}, Alexandros and {Ant{\'o}nio}, Carlos and {Aparicio del Moral}, Beatriz and {Appolloni}, Matteo and {Arena}, Claudio and {Armstrong}, David and {Aroca Aliaga}, Jose and {Asplund}, Martin and {Audenaert}, Jeroen and {Auricchio}, Natalia and {Avelino}, Pedro and {Baeke}, Ann and {Bailli{\'e}}, Kevin and {Balado}, Ana and {Balestra}, Andrea and {Ball}, Warrick and {Ballans}, Herve and {Ballot}, Jerome and {Barban}, Caroline and {Barbary}, Ga{\"e}le and {Barbieri}, Mauro and {Barcel{\'o} Forteza}, Sebasti{\`a} and {Barker}, Adrian and {Barklem}, Paul and {Barnes}, Sydney and {Barrado Navascues}, David and {Barragan}, Oscar and {Baruteau}, Cl{\'e}ment and {Basu}, Sarbani and {Baudin}, Frederic and {Baumeister}, Philipp and {Bayliss}, Daniel and {Bazot}, Michael and {Beck}, Paul G. and {Bedding}, Tim and {Belkacem}, Kevin and {Bellinger}, Earl and {Benatti}, Serena and {Benomar}, Othman and {B{\'e}rard}, Diane and {Bergemann}, Maria and {Bergomi}, Maria and {Bernardo}, Pierre and {Biazzo}, Katia and {Bignamini}, Andrea and {Bigot}, Lionel and {Billot}, Nicolas and {Binet}, Martin and {Biondi}, David and {Biondi}, Federico and {Birch}, Aaron C. and {Bitsch}, Bertram and {Bluhm Ceballos}, Paz Victoria and {B{\'o}di}, Attila and {Bogn{\'a}r}, Zs{\'o}fia and {Boisse}, Isabelle and {Bolmont}, Emeline and {Bonanno}, Alfio and {Bonavita}, Mariangela and {Bonfanti}, Andrea and {Bonfils}, Xavier and {Bonito}, Rosaria and {Bonomo}, Aldo Stefano and {B{\"o}rner}, Anko and {Boro Saikia}, Sudeshna and {Borreguero Mart{\'\i}n}, Elisa and {Borsa}, Francesco and {Borsato}, Luca and {Bossini}, Diego and {Bouchy}, Francois and {Bou{\'e}}, Gwena{\"e}l and {Boufleur}, Rodrigo and {Boumier}, Patrick and {Bourrier}, Vincent and {Bowman}, Dominic M. and {Bozzo}, Enrico and {Bradley}, Louisa and {Bray}, John and {Bressan}, Alessandro and {Breton}, Sylvain and {Brienza}, Daniele and {Brito}, Ana and {Brogi}, Matteo and {Brown}, Beverly and {Brown}, David and {Brun}, Allan Sacha and {Bruno}, Giovanni and {Bruns}, Michael and {Buchhave}, Lars A. and {Bugnet}, Lisa and {Buldgen}, Ga{\"e}l and {Burgess}, Patrick and {Busatta}, Andrea and {Busso}, Giorgia and {Buzasi}, Derek and {Caballero}, Jos{\'e} A. and {Cabral}, Alexandre and {Calderone}, Flavia and {Cameron}, Robert and {Cameron}, Andrew and {Campante}, Tiago and {Canto Martins}, Bruno Leonardo and {Cara}, Christophe and {Carone}, Ludmila and {Carrasco}, Josep Manel and {Casagrande}, Luca and {Casewell}, Sarah L. and {Cassisi}, Santi and {Castellani}, Marco and {Castro}, Matthieu and {Catala}, Claude and {Catal{\'a}n Fern{\'a}ndez}, Irene and {Catelan}, M{\'a}rcio and {Cegla}, Heather and {Cerruti}, Chiara and {Cessa}, Virginie and {Chadid}, Merieme and {Chaplin}, William and {Charpinet}, Stephane and {Chiappini}, Cristina and {Chiarucci}, Simone and {Chiavassa}, Andrea and {Chinellato}, Simonetta and {Chirulli}, Giovanni and {Christensen-Dalsgaard}, Jorgen and {Church}, Ross and {Claret}, Antonio and {Clarke}, Cathie and {Claudi}, Riccardo and {Clermont}, Lionel and {Coelho}, Hugo and {Coelho}, Joao and {Cogato}, Fabrizio and {Colom{\'e}}, Josep and {Condamin}, Mathieu and {Conseil}, Simon and {Corbard}, Thierry and {Correia}, Alexandre C.~M. and {Corsaro}, Enrico and {Cosentino}, Rosario and {Costes}, Jean and {Cottinelli}, Andrea and {Covone}, Giovanni and {Creevey}, Orlagh L. and {Crida}, Aurelien and {Csizmadia}, Szilard and {Cunha}, Margarida and {Curry}, Patrick and {da Costa}, Jefferson and {da Silva}, Francys and {Dalal}, Shweta and {Damasso}, Mario and {Damiani}, Cilia and {Damiani}, Francesco and {Liduina das Chagas}, Maria and {Davies}, Melvyn and {Davies}, Guy and {Davies}, Ben and {Davison}, Gary and {de Almeida}, Leandro and {de Angeli}, Francesca and {Cabral de Barros}, Susana Cristina and {de Castro Le{\~a}o}, Izan and {Brito de Freitas}, Daniel and {de Freitas}, Marcia Cristina and {De Martino}, Domitilla and {Renan de Medeiros}, Jos{\'e} and {de Paula}, Luiz Alberto and {de Plaa}, Jelle and {De Ridder}, Joris and {Deal}, Morgan and {Decin}, Leen and {Deeg}, Hans and {Degl'Innocenti}, Scilla and {Deheuvels}, Sebastien and {del Burgo}, Carlos and {Del Sordo}, Fabio and {Delgado-Mena}, Elisa and {Demangeon}, Olivier and {Denk}, Tilmann and {Derekas}, Aliz and {Desidera}, Silvano and {Dexet}, Marc and {Di Criscienzo}, Marcella and {Di Giorgio}, Anna Maria and {Di Mauro}, Maria Pia and {Diaz Rial}, Federico Jose and {D{\'\i}az-Garc{\'\i}a}, Jos{\'e}-Javier and {Dima}, Marco and {Dinuzzi}, Giacomo and {Dionatos}, Odysseas and {Distefano}, Elisa and {do Nascimento}, Jose-Dias, Jr. and {Domingo}, Albert and {D'Orazi}, Valentina and {Dorn}, Caroline and {Doyle}, Lauren and {Duarte}, Elena and {Ducellier}, Florent and {Dumaye}, Luc and {Dumusque}, Xavier and {Dupret}, Marc-Antoine and {Eggenberger}, Patrick and {Ehrenreich}, David and {Eigm{\"u}ller}, Philipp and {Eising}, Johannes and {Emilio}, Marcelo and {Eriksson}, Kjell and {Ermocida}, Marco and {Isidoro Escate Giribaldi}, Riano and {Eschen}, Yoshi and {Estrela}, In{\^e}s and {Evans}, Dafydd Wyn and {Fabbian}, Damian and {Fabrizio}, Michele and {Faria}, Jo{\~a}o Pedro and {Farina}, Maria and {Farinato}, Jacopo and {Feliz}, Dax and {Feltzing}, Sofia and {Fenouillet}, Thomas and {Ferrari}, Lorenza and {Ferraz-Mello}, Sylvio and {Fialho}, Fabio and {Fienga}, Agnes and {Figueira}, Pedro and {Fiori}, Laura and {Flaccomio}, Ettore and {Focardi}, Mauro and {Foley}, Steve and {Fontignie}, Jean and {Ford}, Dominic and {Fornazier}, Karin and {Forveille}, Thierry and {Fossati}, Luca and {de Marca Franca}, Rodrigo and {da Silva}, Lucas Franco and {Frasca}, Antonio and {Fridlund}, Malcolm and {Furlan}, Marco and {Gabler}, Sarah-Maria and {Gaido}, Marco and {Gallagher}, Andrew and {Galli}, Emanuele and {Garcia}, Rafael A. and {Garc{\'\i}a Hern{\'a}ndez}, Antonio and {Garcia Munoz}, Antonio and {Garc{\'\i}a-V{\'a}zquez}, Hugo and {Garrido Haba}, Rafael and {Gaulme}, Patrick and {Gauthier}, Nicolas and {Gehan}, Charlotte and {Gent}, Matthew and {Georgieva}, Iskra and {Ghigo}, Mauro and {Giana}, Edoardo and {Gill}, Samuel and {Girardi}, Leo and {Giuliatti Winter}, Silvia and {Giusi}, Giovanni and {Gomes da Silva}, Jo{\~a}o and {G{\'o}mez Zazo}, Luis Jorge and {Gomez-Lopez}, Juan Manuel and {Isai Gonz{\'a}lez Hern{\'a}ndez}, Jonay and {Gonzalez Murillo}, Kevin and {Gorius}, Nicolas and {Gouel}, Pierre-Vincent and {Goulty}, Duncan and {Granata}, Valentina and {Grenfell}, John Lee and {Grie{\ss}bach}, Denis and {Grolleau}, Emmanuel and {Grouffal}, Salom{\'e} and {Grziwa}, Sascha and {Guarcello}, Mario Giuseppe and {Gueguen}, Lo{\"\i}c and {Guenther}, Eike Wolf and {Guilhem}, Terrasa and {Guillerot}, Lucas and {Guiot}, Pierre and {Guterman}, Pascal and {Guti{\'e}rrez}, Antonio and {Guti{\'e}rrez-Canales}, Fernando and {Hagelberg}, Janis and {Haldemann}, Jonas and {Hall}, Cassandra and {Handberg}, Rasmus and {Harrison}, Ian and {Harrison}, Diana L. and {Hasiba}, Johann and {Haswell}, Carole A. and {Hatalova}, Petra and {Hatzes}, Artie and {Haywood}, Raphaelle and {H{\'e}brard}, Guillaume and {Heckes}, Frank and {Heiter}, Ulrike and {Hekker}, Saskia and {Heller}, Ren{\'e} and {Helling}, Christiane and {Helminiak}, Krzysztof and {Hemsley}, Simon and {Heng}, Kevin and {Hermans}, Aline and {Hermes}, JJ and {Hidalgo Torres}, Nadia and {Hinkel}, Natalie and {Hobbs}, David and {Hodgkin}, Simon and {Hofmann}, Karl and {Hojjatpanah}, Saeed and {Houdek}, G{\"u}nter and {Huber}, Daniel and {Huesler}, Joseph and {Hui-Bon-Hoa}, Alain and {Huygen}, Rik and {Huynh}, Duc-Dat and {Iro}, Nicolas and {Irwin}, Jonathan and {Irwin}, Mike and {Izidoro}, Andr{\'e} and {Jacquinod}, Sophie and {Emborg Jannsen}, Nicholas and {Janson}, Markus and {Jeszenszky}, Harald and {Jiang}, Chen and {Jos{\'e} Jimenez Mancebo}, Antonio and {Jofre}, Paula and {Johansen}, Anders and {Johnston}, Cole and {Jones}, Geraint and {Kallinger}, Thomas and {K{\'a}lm{\'a}n}, Szil{\'a}rd and {Kanitz}, Thomas and {Karjalainen}, Marie and {Karjalainen}, Raine and {Karoff}, Christoffer and {Kawaler}, Steven and {Kawata}, Daisuke and {Keereman}, Arnoud and {Keiderling}, David and {Kennedy}, Tom and {Kenworthy}, Matthew and {Kerschbaum}, Franz and {Kidger}, Mark and {Kiefer}, Flavien and {Kintziger}, Christian and {Kislyakova}, Kristina and {Kiss}, L{\'a}szl{\'o} and {Klagyivik}, Peter and {Klahr}, Hubert and {Klevas}, Jonas and {Kochukhov}, Oleg and {K{\"o}hler}, Ulrich and {Kolb}, Ulrich and {Koncz}, Alexander and {Korth}, Judith and {Kostogryz}, Nadiia and {Kov{\'a}cs}, G{\'a}bor and {Kov{\'a}cs}, J{\'o}zsef and {Kozhura}, Oleg and {Krivova}, Natalie and {Ku{\v{c}}inskas}, Arunas and {Kuhlemann}, Ilyas and {Kupka}, Friedrich and {Laauwen}, Wouter and {Labiano}, Alvaro and {Lagarde}, Nadege and {Laget}, Philippe and {Laky}, Gunter and {Lam}, Kristine Wai Fun and {Lambrechts}, Michiel and {Lammer}, Helmut and {Lanza}, Antonino Francesco and {Lanzafame}, Alessandro and {Lares Martiz}, Mariel and {Laskar}, Jacques and {Latter}, Henrik and {Lavanant}, Tony and {Lawrenson}, Alastair and {Lazzoni}, Cecilia and {Lebre}, Agnes and {Lebreton}, Yveline and {Lecavelier des Etangs}, Alain and {Leinhardt}, Zoe and {Leleu}, Adrien and {Lendl}, Monika and {Leto}, Giuseppe and {Levillain}, Yves and {Libert}, Anne-Sophie and {Lichtenberg}, Tim and {Ligi}, Roxanne and {Lignieres}, Francois and {Lillo-Box}, Jorge and {Linsky}, Jeffrey and {Scige Liu}, John and {Loidolt}, Dominik and {Longval}, Yuying and {Lopes}, Il{\'\i}dio and {Lorenzani}, Andrea and {Ludwig}, Hans-Guenter and {Lund}, Mikkel and {Sloth Lundkvist}, Mia and {Luri}, Xavier and {Maceroni}, Carla and {Madden}, Sean and {Madhusudhan}, Nikku and {Maggio}, Antonio and {Magliano}, Christian and {Magrin}, Demetrio and {Mahy}, Laurent and {Maibaum}, Olaf and {Malac-Allain}, LeeRoy and {Malapert}, Jean-Christophe and {Malavolta}, Luca and {Maldonado}, Jesus and {Mamonova}, Elena and {Manchon}, Louis and {Mann}, Andrew and {Mantovan}, Giacomo and {Marafatto}, Luca and {Marconi}, Marcella and {Mardling}, Rosemary and {Marigo}, Paola and {Marinoni}, Silvia and {Marques}, {\'E}rico and {Marques}, Joao Pedro and {Marrese}, Paola Maria and {Marshall}, Douglas and {Mart{\'\i}nez Perales}, Silvia and {Mary}, David and {Marzari}, Francesco and {Masana}, Eduard and {Mascher}, Andrina and {Mathis}, St{\'e}phane and {Mathur}, Savita and {Mattiuci Figueiredo}, Ana Carolina and {Maxted}, Pierre F.~L. and {Mazeh}, Tsevi and {Mazevet}, Stephane and {Mazzei}, Francesco and {McCormac}, James and {McMillan}, Paul and {Menou}, Lucas and {Merle}, Thibault and {Meru}, Farzana and {Mesa}, Dino and {Messina}, Sergio and {M{\'e}sz{\'a}ros}, Szabolcs and {Meunier}, Nad{\'e}ge and {Meunier}, Jean-Charles and {Micela}, Giuseppina and {Michaelis}, Harald and {Michel}, Eric and {Michielsen}, Mathias and {Michtchenko}, Tatiana and {Miglio}, Andrea and {Miguel}, Yamila and {Milligan}, David and {Mirouh}, Giovanni and {Mitchell}, Morgan A. and {Moedas}, Nuno and {Molendini}, Francesca and {Moln{\'a}r}, L{\'a}szl{\'o} and {Mombarg}, Joey and {Montalban}, Josefina and {Montalto}, Marco and {Monteiro}, M{\'a}rio J.~P.~F.~G. and {Morales}, Juan Carlos and {Morales-Calderon}, Maria and {Morbidelli}, Alessandro and {Mordasini}, Christoph and {Moreau}, Chrystel and {Morel}, Thierry and {Morello}, Guiseppe and {Morin}, Julien and {Mortier}, Annelies and {Mosser}, Beno{\^\i}t and {Mourard}, Denis and {Mousis}, Olivier and {Moutou}, Claire and {Mowlavi}, Nami and {Moya}, Andr{\'e}s and {Muehlmann}, Prisca and {Muirhead}, Philip and {Munari}, Matteo and {Musella}, Ilaria and {Mustill}, Alexander James and {Nardetto}, Nicolas and {Nardiello}, Domenico and {Narita}, Norio and {Nascimbeni}, Valerio and {Nash}, Anna and {Neiner}, Coralie and {Nelson}, Richard P. and {Nettelmann}, Nadine and {Nicolini}, Gianalfredo and {Nielsen}, Martin and {Niemi}, Sami-Matias and {Noack}, Lena and {Noels-Grotsch}, Arlette and {Noll}, Anthony and {Norazman}, Azib and {Norton}, Andrew J. and {Nsamba}, Benard and {Ofir}, Aviv and {Ogilvie}, Gordon and {Olander}, Terese and {Olivetto}, Christian and {Olofsson}, G{\"o}ran and {Ong}, Joel and {Ortolani}, Sergio and {Oshagh}, Mahmoudreza and {Ottacher}, Harald and {Ottensamer}, Roland and {Ouazzani}, Rhita-Maria and {Paardekooper}, Sijme-Jan and {Pace}, Emanuele and {Pajas}, Miriam and {Palacios}, Ana and {Palandri}, Gaelle and {Palle}, Enric and {Paproth}, Carsten and {Parro}, Vanderlei and {Parviainen}, Hannu and {Granado}, Javier Pascual and {Passegger}, Vera Maria and {Pastor-Morales}, Carmen and {P{\"a}tzold}, Martin and {Gade Pedersen}, May and {Pena Hidalgo}, David and {Pepe}, Francesco and {Pereira}, Filipe and {Persson}, Carina M. and {Pertenais}, Martin and {Peter}, Gisbert and {Petit}, Antoine C. and {Petit}, Pascal and {Pezzuto}, Stefano and {Pichierri}, Gabriele and {Pietrinferni}, Adriano and {Pinheiro}, Fernando and {Pinsonneault}, Marc and {Plachy}, Emese and {Plasson}, Philippe and {Plez}, Bertrand and {Poppenhaeger}, Katja and {Poretti}, Ennio and {Portaluri}, Elisa and {Portell}, Jordi and {Frederico Porto de Mello}, Gustavo and {Poyatos}, Julien and {Pozuelos}, Francisco J. and {Prada Moroni}, Pier Giorgio and {Pricopi}, Dumitru and {Prisinzano}, Loredana and {Quade}, Matthias and {Quirrenbach160}, ndreas and {Rabanal Reina6}, Julio Arturo and {Rabello Soares}, Maria Cristina and {Raimondo}, Gabriella and {Rainer}, Monica and {Ram{\'o}n Rod{\'o}n}, Jose and {Ram{\'o}n-Ballesta}, Alejandro and {Ramos Zapata}, Gonzalo and {R{\"a}tz}, Stefanie and {Rauterberg}, Christoph and {Redman}, Bob and {Redmer}, Ronald and {Reese}, Daniel and {Regibo}, Sara and {Reiners}, Ansgar and {Reinhold}, Timo and {Renie}, Christian and {Ribas}, Ignasi and {Ribeiro}, Sergio and {Pereira Ricciardi}, Thiago and {Rice}, Ken and {Richard}, Olivier and {Riello}, Marco and {Rieutord}, Michel and {Ripepi}, Vincenzo and {Rixon}, Guy and {Rockstein}, Steve and {Rodr{\'\i}guez}, Mar{\'\i}a Teresa Rodrigo and {Rodr{\'\i}guez D{\'\i}az}, Luisa Fernanda and {Rodriguez Garcia}, Juan Pablo and {Rodriguez-Gomez}, Julio and {Roehlly}, Yannick and {Roig}, Fernando and {Rojas-Ayala}, B{\'a}rbara and {Rolf}, Tobias and {Lysgaard R{\o}rsted}, Jakob and {Rosado}, Hugo and {Rosotti}, Giovanni and {Roth}, Olivier and {Roth}, Markus and {Rousseau}, Alex and {Roxburgh}, Ian and {Roy}, Fabrice and {Royer}, Pierre and {Ruane}, Kirk and {Rufini Mastropasqua}, Sergio and {Ruiz de Galarreta}, Claudia and {Russi}, Andrea and {Saar}, Steven and {Saillenfest}, Melaine and {Salaris}, Maurizio and {Salmon}, Sebastien and {Saltas}, Ippocratis and {Samadi}, R{\'e}za and {Samadi}, Aunia and {Samra}, Dominic and {Sanches da Silva}, Tiago and {Andr{\'e}s S{\'a}nchez Carrasco}, Miguel and {Santerne}, Alexandre and {Santoli}, Francesco and {Santos}, {\^A}ngela R.~G. and {Sanz Mesa}, Rosario and {Sarro}, Luis Manuel and {Scandariato}, Gaetano and {Sch{\"a}fer}, Martin and {Schlafly}, Edward and {Schmider}, Fran{\c{c}}ois-Xavier and {Schneider}, Jean and {Schou}, Jesper and {Schunker}, Hannah and {J{\"o}rg Schwarzkopf}, Gabriel and {Serenelli}, Aldo and {Seynaeve}, Dries and {Shan}, Yutong and {Shapiro}, Alexander and {Shipman}, Russel and {Sicilia}, Daniela and {Sierra Sanmartin}, Maria Angeles and {Sigot}, Axelle and {Silliman}, Kyle and {Silvotti}, Roberto and {Simon}, Attila E. and {Simoyama Napoli}, Ricardo and {Skarka}, Marek and {Smalley}, Barry and {Smiljanic}, Rodolfo and {Smit}, Samuel and {Smith}, Alexis and {Smith}, Leigh and {Snellen}, Ignas and {S{\'o}dor}, {\'A}d{\'a}m and {Sohl}, Frank and {Solanki}, Sami K. and {Sortino}, Francesca and {Sousa}, S{\'e}rgio and {Southworth}, John and {Souto}, Diogo and {Sozzetti}, Alessandro and {Stamatellos}, Dimitris and {Stassun}, Keivan and {Steller}, Manfred and {Stello}, Dennis and {Stelzer}, Beate and {Stiebeler}, Ulrike and {Stokholm}, Amalie and {Storelvmo}, Trude and {Strassmeier}, Klaus and {Str{\o}m}, Paul Anthony and {Strugarek}, Antoine and {Sulis}, Sophia and {{\v{S}}vanda}, Michal and {Szabados}, L{\'a}szl{\'o} and {Szab{\'o}}, R{\'o}bert and {Szab{\'o}}, Gyula M. and {Szuszkiewicz}, Ewa and {Talens}, Geert Jan and {Teti}, Daniele and {Theisen}, Tom and {Th{\'e}venin}, Fr{\'e}d{\'e}ric and {Thoul}, Anne and {Tiphene}, Didier and {Titz-Weider}, Ruth and {Tkachenko}, Andrew and {Tomecki}, Daniel and {Tonfat}, Jorge and {Tosi}, Nicola and {Trampedach}, Regner and {Traven}, Gregor and {Triaud}, Amaury and {Tr{\o}nnes}, Reidar and {Tsantaki}, Maria and {Tschentscher}, Matthias and {Turin}, Arnaud and {Tvaruzka}, Adam and {Ulmer}, Bernd and {Ulmer-Moll}, Sol{\`e}ne and {Ulusoy}, Ceren and {Umbriaco}, Gabriele and {Valencia}, Diana and {Valentini}, Marica and {Valio}, Adriana and {Valverde Guijarro}, {\'A}ngel Luis and {Van Eylen}, Vincent and {Van Grootel}, Valerie and {van Kempen}, Tim A. and {Van Reeth}, Timothy and {Van Zelst}, Iris and {Vandenbussche}, Bart and {Vasiliou}, Konstantinos and {Vasilyev}, Valeriy and {Vaz de Mascarenhas}, David and {Vazan}, Allona and {Vela Nunez}, Marina and {Nunes Velloso}, Eduardo and {Ventura}, Rita and {Ventura}, Paolo and {Venturini}, Julia and {Trallero}, Isabel Vera and {Veras}, Dimitri and {Verdugo}, Eva and {Verma}, Kuldeep and {Vibert}, Didier and {Vicanek Martinez}, Tobias and {Vida}, Kriszti{\'a}n and {Vigan}, Arthur and {Villacorta}, Antonio and {Villaver}, Eva and {Villaverde Aparicio}, Marcos and {Viotto}, Valentina and {Vorobyov}, Eduard and {Vorontsov}, Sergey and {Wagner}, Frank W. and {Walloschek}, Thomas and {Walton}, Nicholas and {Walton}, Dave and {Wang}, Haiyang and {Waters}, Rens and {Watson}, Christopher and {Wedemeyer}, Sven and {Weeks}, Angharad and {Weingril}, J{\"o}rg and {Weiss}, Annita and {Wendler}, Belinda and {West}, Richard and {Westerdorff}, Karsten and {Westphal}, Pierre-Amaury and {Wheatley}, Peter and {White}, Tim and {Whittaker}, Amadou and {Wickhusen}, Kai and {Wilson}, Thomas and {Windsor}, James and {Winter}, Othon and {Lykke Winther}, Mark and {Winton}, Alistair and {Witteck}, Ulrike and {Witzke}, Veronika and {Woitke}, Peter and {Wolter}, David and {Wuchterl}, G{\"u}nther and {Wyatt}, Mark and {Yang}, Dan and {Yu}, Jie and {Zanmar Sanchez}, Ricardo and {Rosa Zapatero Osorio}, Mar{\'\i}a and {Zechmeister}, Mathias and {Zhou}, Yixiao and {Ziemke}, Claas and {Zwintz}, Konstanze},
        title = "{The PLATO Mission}",
      journal = {arXiv e-prints},
         year = 2024,
        month = jun,
          eid = {arXiv:2406.05447},
        pages = {arXiv:2406.05447},
          doi = {10.48550/arXiv.2406.05447},
archivePrefix = {arXiv},
       eprint = {2406.05447},
 primaryClass = {astro-ph.IM},
       adsurl = {https://ui.adsabs.harvard.edu/abs/2024arXiv240605447R}
}

@ARTICLE{Handberg14,
       author = {{Handberg}, R. and {Lund}, M.~N.},
        title = "{Automated preparation of Kepler time series of planet hosts for asteroseismic analysis}",
      journal = {\mnras},
         year = 2014,
        month = dec,
       volume = {445},
       number = {3},
        pages = {2698-2709},
          doi = {10.1093/mnras/stu1823},
archivePrefix = {arXiv},
       eprint = {1409.1366},
 primaryClass = {astro-ph.IM},
       adsurl = {https://ui.adsabs.harvard.edu/abs/2014MNRAS.445.2698H}
}

@ARTICLE{Stello09,
       author = {{Stello}, Dennis and {Chaplin}, William J. and {Bruntt}, Hans and {Creevey}, Orlagh L. and {Garc{\'\i}a-Hern{\'a}ndez}, Antonio and {Monteiro}, Mario J.~P.~F.~G. and {Moya}, Andr{\'e}s and {Quirion}, Pierre-Olivier and {Sousa}, Sergio G. and {Su{\'a}rez}, Juan-Carlos and {Appourchaux}, Thierry and {Arentoft}, Torben and {Ballot}, Jerome and {Bedding}, Timothy R. and {Christensen-Dalsgaard}, J{\o}rgen and {Elsworth}, Yvonne and {Fletcher}, Stephen T. and {Garc{\'\i}a}, Rafael A. and {Houdek}, G{\"u}nter and {Jim{\'e}nez-Reyes}, Sebastian J. and {Kjeldsen}, Hans and {New}, Roger and {R{\'e}gulo}, Clara and {Salabert}, David and {Toutain}, Thierry},
        title = "{Radius Determination of Solar-type Stars Using Asteroseismology: What to Expect from the Kepler Mission}",
      journal = {\apj},
         year = 2009,
        month = aug,
       volume = {700},
       number = {2},
        pages = {1589-1602},
          doi = {10.1088/0004-637X/700/2/1589},
archivePrefix = {arXiv},
       eprint = {0906.0766},
 primaryClass = {astro-ph.SR},
       adsurl = {https://ui.adsabs.harvard.edu/abs/2009ApJ...700.1589S}
}

@ARTICLE{Kallinger2014,
       author = {{Kallinger}, T. and {De Ridder}, J. and {Hekker}, S. and {Mathur}, S. and {Mosser}, B. and {Gruberbauer}, M. and {Garc{\'\i}a}, R.~A. and {Karoff}, C. and {Ballot}, J.},
        title = "{The connection between stellar granulation and oscillation as seen by the Kepler mission}",
      journal = {\aap},
         year = 2014,
        month = oct,
       volume = {570},
          eid = {A41},
        pages = {A41},
          doi = {10.1051/0004-6361/201424313},
archivePrefix = {arXiv},
       eprint = {1408.0817},
 primaryClass = {astro-ph.SR},
       adsurl = {https://ui.adsabs.harvard.edu/abs/2014A&A...570A..41K}
}

@UNPUBLISHED{ruwe,
     author = {L.~Lindegren},
     title={{R}e-normalising the astrometric chi-square in {G}aia {D}{R}2},
     institution={Lund Observatory},
     year={2018},
     month={August},
     url={http://www.rssd.esa.int/doc_fetch.php?id=3757412},
     note={GAIA-C3-TN-LU-LL-124},
     type={Technical Note}
}

@ARTICLE{Serenelli17,
       author = {{Serenelli}, Aldo and {Johnson}, Jennifer and {Huber}, Daniel and {Pinsonneault}, Marc and {Ball}, Warrick H. and {Tayar}, Jamie and {Silva Aguirre}, Victor and {Basu}, Sarbani and {Troup}, Nicholas and {Hekker}, Saskia and {Kallinger}, Thomas and {Stello}, Dennis and {Davies}, Guy R. and {Lund}, Mikkel N. and {Mathur}, Savita and {Mosser}, Benoit and {Stassun}, Keivan G. and {Chaplin}, William J. and {Elsworth}, Yvonne and {Garc{\'\i}a}, Rafael A. and {Handberg}, Rasmus and {Holtzman}, Jon and {Hearty}, Fred and {Garc{\'\i}a-Hern{\'a}ndez}, D.~A. and {Gaulme}, Patrick and {Zamora}, Olga},
        title = "{The First APOKASC Catalog of Kepler Dwarf and Subgiant Stars}",
      journal = {\apjs},
         year = 2017,
        month = dec,
       volume = {233},
       number = {2},
          eid = {23},
        pages = {23},
          doi = {10.3847/1538-4365/aa97df},
archivePrefix = {arXiv},
       eprint = {1710.06858},
 primaryClass = {astro-ph.SR},
       adsurl = {https://ui.adsabs.harvard.edu/abs/2017ApJS..233...23S}
}

@ARTICLE{Handberg17,
       author = {{Handberg}, R. and {Brogaard}, K. and {Miglio}, A. and {Bossini}, D. and {Elsworth}, Y. and {Slumstrup}, D. and {Davies}, G.~R. and {Chaplin}, W.~J.},
        title = "{NGC 6819: testing the asteroseismic mass scale, mass loss and evidence for products of non-standard evolution}",
      journal = {\mnras},
         year = 2017,
        month = nov,
       volume = {472},
       number = {1},
        pages = {979-997},
          doi = {10.1093/mnras/stx1929},
archivePrefix = {arXiv},
       eprint = {1707.08223},
 primaryClass = {astro-ph.SR},
       adsurl = {https://ui.adsabs.harvard.edu/abs/2017MNRAS.472..979H}
}

@article{Diaz22,
    author = {Rodríguez Díaz, Luisa Fernanda and Bigot, Lionel and Aguirre Børsen-Koch, Víctor and Lund, Mikkel N and Rørsted, Jakob Lysgaard and Kallinger, Thomas and Sulis, Sophia and Mary, David},
    title = {Scaling relations of convective granulation noise across the HR diagram from 3D stellar atmosphere models},
    journal = {Monthly Notices of the Royal Astronomical Society},
    volume = {514},
    number = {2},
    pages = {1741-1756},
    year = {2022},
    month = {05},
    issn = {0035-8711},
    doi = {10.1093/mnras/stac1467},
    url = {https://doi.org/10.1093/mnras/stac1467},
    eprint = {https://academic.oup.com/mnras/article-pdf/514/2/1741/44055313/stac1467.pdf},
}

@ARTICLE{Ludwig06,
       author = {{Ludwig}, H. -G.},
        title = "{Hydrodynamical simulations of convection-related stellar micro-variability. I. Statistical relations for photometric and photocentric variability}",
      journal = {\aap},
         year = 2006,
        month = jan,
       volume = {445},
       number = {2},
        pages = {661-671},
          doi = {10.1051/0004-6361:20042102},
archivePrefix = {arXiv},
       eprint = {astro-ph/0509441},
 primaryClass = {astro-ph},
       adsurl = {https://ui.adsabs.harvard.edu/abs/2006A&A...445..661L}
}

@INPROCEEDINGS{Harvey85,
       author = {{Harvey}, J.},
        title = "{High-Resolution Helioseismology}",
    booktitle = {Future Missions in Solar, Heliospheric \& Space Plasma Physics},
         year = 1985,
       editor = {{Rolfe}, Erica and {Battrick}, Bruce},
       series = {ESA Special Publication},
       volume = {235},
        month = jun,
        pages = {199},
       adsurl = {https://ui.adsabs.harvard.edu/abs/1985ESASP.235..199H}
}

@ARTICLE{Chaplin11b,
       author = {{Chaplin}, W.~J. and {Kjeldsen}, H. and {Bedding}, T.~R. and {Christensen-Dalsgaard}, J. and {Gilliland}, R.~L. and {Kawaler}, S.~D. and {Appourchaux}, T. and {Elsworth}, Y. and {Garc{\'\i}a}, R.~A. and {Houdek}, G. and {Karoff}, C. and {Metcalfe}, T.~S. and {Molenda-{\.Z}akowicz}, J. and {Monteiro}, M.~J.~P.~F.~G. and {Thompson}, M.~J. and {Verner}, G.~A. and {Batalha}, N. and {Borucki}, W.~J. and {Brown}, T.~M. and {Bryson}, S.~T. and {Christiansen}, J.~L. and {Clarke}, B.~D. and {Jenkins}, J.~M. and {Klaus}, T.~C. and {Koch}, D. and {An}, D. and {Ballot}, J. and {Basu}, S. and {Benomar}, O. and {Bonanno}, A. and {Broomhall}, A. -M. and {Campante}, T.~L. and {Corsaro}, E. and {Creevey}, O.~L. and {Esch}, L. and {Gai}, N. and {Gaulme}, P. and {Hale}, S.~J. and {Handberg}, R. and {Hekker}, S. and {Huber}, D. and {Mathur}, S. and {Mosser}, B. and {New}, R. and {Pinsonneault}, M.~H. and {Pricopi}, D. and {Quirion}, P. -O. and {R{\'e}gulo}, C. and {Roxburgh}, I.~W. and {Salabert}, D. and {Stello}, D. and {Suran}, M.~D.},
        title = "{Predicting the Detectability of Oscillations in Solar-type Stars Observed by Kepler}",
      journal = {\apj},
         year = 2011,
        month = may,
       volume = {732},
       number = {1},
          eid = {54},
        pages = {54},
          doi = {10.1088/0004-637X/732/1/54},
archivePrefix = {arXiv},
       eprint = {1103.0702},
 primaryClass = {astro-ph.SR},
       adsurl = {https://ui.adsabs.harvard.edu/abs/2011ApJ...732...54C}
}

@ARTICLE{Karoff13,
       author = {{Karoff}, C. and {Campante}, T.~L. and {Ballot}, J. and {Kallinger}, T. and {Gruberbauer}, M. and {Garc{\'\i}a}, R.~A. and {Caldwell}, D.~A. and {Christiansen}, J.~L. and {Kinemuchi}, K.},
        title = "{Observations of Intensity Fluctuations Attributed to Granulation and Faculae on Sun-like Stars from the Kepler Mission}",
      journal = {\apj},
         year = 2013,
        month = apr,
       volume = {767},
       number = {1},
          eid = {34},
        pages = {34},
          doi = {10.1088/0004-637X/767/1/34},
archivePrefix = {arXiv},
       eprint = {1302.5563},
 primaryClass = {astro-ph.SR},
       adsurl = {https://ui.adsabs.harvard.edu/abs/2013ApJ...767...34K}
}

@ARTICLE{Lundkvist21,
       author = {{Lundkvist}, Mia S. and {Ludwig}, Hans-G{\"u}nter and {Collet}, Remo and {Straus}, Thomas},
        title = "{The signature of granulation in a solar power spectrum as seen with CO$^{5}$BOLD}",
      journal = {\mnras},
         year = 2021,
        month = feb,
       volume = {501},
       number = {2},
        pages = {2512-2521},
          doi = {10.1093/mnras/staa3656},
archivePrefix = {arXiv},
       eprint = {2011.10045},
 primaryClass = {astro-ph.SR},
       adsurl = {https://ui.adsabs.harvard.edu/abs/2021MNRAS.501.2512L}
}

@ARTICLE{Speagle20,
       author = {{Speagle}, Joshua S.},
        title = "{DYNESTY: a dynamic nested sampling package for estimating Bayesian posteriors and evidences}",
      journal = {\mnras},
         year = 2020,
        month = apr,
       volume = {493},
       number = {3},
        pages = {3132-3158},
          doi = {10.1093/mnras/staa278},
archivePrefix = {arXiv},
       eprint = {1904.02180},
 primaryClass = {astro-ph.IM},
       adsurl = {https://ui.adsabs.harvard.edu/abs/2020MNRAS.493.3132S}
}

@ARTICLE{Zhou21,
       author = {{Zhou}, Yixiao and {Nordlander}, Thomas and {Casagrande}, Luca and {Joyce}, Meridith and {Li}, Yaguang and {Amarsi}, Anish M. and {Reggiani}, Henrique and {Asplund}, Martin},
        title = "{The relationship between photometric and spectroscopic oscillation amplitudes from 3D stellar atmosphere simulations}",
      journal = {\mnras},
         year = 2021,
        month = may,
       volume = {503},
       number = {1},
        pages = {13-27},
          doi = {10.1093/mnras/stab337},
archivePrefix = {arXiv},
       eprint = {2102.02135},
 primaryClass = {astro-ph.SR},
       adsurl = {https://ui.adsabs.harvard.edu/abs/2021MNRAS.503...13Z}
}

@ARTICLE{Sreenivas24,
       author = {{Sreenivas}, K.~R. and {Bedding}, Timothy R. and {Li}, Yaguang and {Huber}, Daniel and {Crawford}, Courtney L. and {Stello}, Dennis and {Yu}, Jie},
        title = "{A simple method to measure {\ensuremath{\nu}}$_{max}$ for asteroseismology: application to 16 000 oscillating Kepler red giants}",
      journal = {\mnras},
         year = 2024,
        month = may,
       volume = {530},
       number = {3},
        pages = {3477-3487},
          doi = {10.1093/mnras/stae991},
archivePrefix = {arXiv},
       eprint = {2401.17557},
 primaryClass = {astro-ph.SR},
       adsurl = {https://ui.adsabs.harvard.edu/abs/2024MNRAS.530.3477S}
}

@ARTICLE{Stello09b,
       author = {{Stello}, D. and {Chaplin}, W.~J. and {Basu}, S. and {Elsworth}, Y. and {Bedding}, T.~R.},
        title = "{The relation between {\ensuremath{\Delta}}{\ensuremath{\nu}} and {\ensuremath{\nu}}$_{max}$ for solar-like oscillations}",
      journal = {\mnras},
         year = 2009,
        month = nov,
       volume = {400},
       number = {1},
        pages = {L80-L84},
          doi = {10.1111/j.1745-3933.2009.00767.x},
archivePrefix = {arXiv},
       eprint = {0909.5193},
 primaryClass = {astro-ph.SR},
       adsurl = {https://ui.adsabs.harvard.edu/abs/2009MNRAS.400L..80S}
}

@ARTICLE{Hekker09,
       author = {{Hekker}, S. and {Kallinger}, T. and {Baudin}, F. and {De Ridder}, J. and {Barban}, C. and {Carrier}, F. and {Hatzes}, A.~P. and {Weiss}, W.~W. and {Baglin}, A.},
        title = "{Characteristics of solar-like oscillations in red giants observed in the CoRoT exoplanet field}",
      journal = {\aap},
         year = 2009,
        month = oct,
       volume = {506},
       number = {1},
        pages = {465-469},
          doi = {10.1051/0004-6361/200911858},
archivePrefix = {arXiv},
       eprint = {0906.5002},
 primaryClass = {astro-ph.SR},
       adsurl = {https://ui.adsabs.harvard.edu/abs/2009A&A...506..465H}
}

@ARTICLE{Mathur11,
       author = {{Mathur}, S. and {Hekker}, S. and {Trampedach}, R. and {Ballot}, J. and {Kallinger}, T. and {Buzasi}, D. and {Garc{\'\i}a}, R.~A. and {Huber}, D. and {Jim{\'e}nez}, A. and {Mosser}, B. and {Bedding}, T.~R. and {Elsworth}, Y. and {R{\'e}gulo}, C. and {Stello}, D. and {Chaplin}, W.~J. and {De Ridder}, J. and {Hale}, S.~J. and {Kinemuchi}, K. and {Kjeldsen}, H. and {Mullally}, F. and {Thompson}, S.~E.},
        title = "{Granulation in Red Giants: Observations by the Kepler Mission and Three-dimensional Convection Simulations}",
      journal = {\apj},
         year = 2011,
        month = nov,
       volume = {741},
       number = {2},
          eid = {119},
        pages = {119},
          doi = {10.1088/0004-637X/741/2/119},
archivePrefix = {arXiv},
       eprint = {1109.1194},
 primaryClass = {astro-ph.SR},
       adsurl = {https://ui.adsabs.harvard.edu/abs/2011ApJ...741..119M}
}

@ARTICLE{Kjeldsen11,
       author = {{Kjeldsen}, H. and {Bedding}, T.~R.},
        title = "{Amplitudes of solar-like oscillations: a new scaling relation}",
      journal = {\aap},
         year = 2011,
        month = may,
       volume = {529},
          eid = {L8},
        pages = {L8},
          doi = {10.1051/0004-6361/201116789},
archivePrefix = {arXiv},
       eprint = {1104.1659},
 primaryClass = {astro-ph.SR},
       adsurl = {https://ui.adsabs.harvard.edu/abs/2011A&A...529L...8K}
}

@ARTICLE{Samadi13b,
       author = {{Samadi}, R. and {Belkacem}, K. and {Ludwig}, H. -G. and {Caffau}, E. and {Campante}, T.~L. and {Davies}, G.~R. and {Kallinger}, T. and {Lund}, M.~N. and {Mosser}, B. and {Baglin}, A. and {Mathur}, S. and {Garcia}, R.~A.},
        title = "{Stellar granulation as seen in disk-integrated intensity. II. Theoretical scaling relations compared with observations}",
      journal = {\aap},
         year = 2013,
        month = nov,
       volume = {559},
          eid = {A40},
        pages = {A40},
          doi = {10.1051/0004-6361/201220817},
archivePrefix = {arXiv},
       eprint = {1309.1488},
 primaryClass = {astro-ph.SR},
       adsurl = {https://ui.adsabs.harvard.edu/abs/2013A&A...559A..40S}
}

@ARTICLE{Magic2013,
       author = {{Magic}, Z. and {Collet}, R. and {Asplund}, M. and {Trampedach}, R. and {Hayek}, W. and {Chiavassa}, A. and {Stein}, R.~F. and {Nordlund}, {\r{A}}.},
        title = "{The Stagger-grid: A grid of 3D stellar atmosphere models. I. Methods and general properties}",
      journal = {\aap},
         year = 2013,
        month = sep,
       volume = {557},
          eid = {A26},
        pages = {A26},
          doi = {10.1051/0004-6361/201321274},
archivePrefix = {arXiv},
       eprint = {1302.2621},
 primaryClass = {astro-ph.SR},
       adsurl = {https://ui.adsabs.harvard.edu/abs/2013A&A...557A..26M}
}

@ARTICLE{Ballot11,
       author = {{Ballot}, J. and {Barban}, C. and {van't Veer-Menneret}, C.},
        title = "{Visibilities and bolometric corrections for stellar oscillation modes observed by Kepler}",
      journal = {\aap},
         year = 2011,
        month = jul,
       volume = {531},
          eid = {A124},
        pages = {A124},
          doi = {10.1051/0004-6361/201016230},
archivePrefix = {arXiv},
       eprint = {1105.4557},
 primaryClass = {astro-ph.SR},
       adsurl = {https://ui.adsabs.harvard.edu/abs/2011A&A...531A.124B}
}

@ARTICLE{Michel09,
       author = {{Michel}, E. and {Samadi}, R. and {Baudin}, F. and {Barban}, C. and {Appourchaux}, T. and {Auvergne}, M.},
        title = "{Intrinsic photometric characterisation of stellar oscillations and granulation. Solar reference values and CoRoT response functions}",
      journal = {\aap},
         year = 2009,
        month = mar,
       volume = {495},
       number = {3},
        pages = {979-987},
          doi = {10.1051/0004-6361:200810353},
archivePrefix = {arXiv},
       eprint = {0809.1078},
 primaryClass = {astro-ph},
       adsurl = {https://ui.adsabs.harvard.edu/abs/2009A&A...495..979M}
}

@INPROCEEDINGS{Trampedach98,
       author = {{Trampedach}, R. and {Christensen-Dalsgaard}, J. and {Nordlund}, A. and {Stein}, R.~F.},
        title = "{Stellar background power spectra from hydrodynamical simulations of stellar atmospheres}",
    booktitle = {The First MONS Workshop: Science with a Small Space Telescope},
         year = 1998,
       editor = {{Kjeldsen}, H. and {Bedding}, T.~R.},
        month = nov,
        pages = {59},
       adsurl = {https://ui.adsabs.harvard.edu/abs/1998mons.proc...59T}
}

@ARTICLE{Samadi19,
       author = {{Samadi}, R. and {Deru}, A. and {Reese}, D. and {Marchiori}, V. and {Grolleau}, E. and {Green}, J.~J. and {Pertenais}, M. and {Lebreton}, Y. and {Deheuvels}, S. and {Mosser}, B. and {Belkacem}, K. and {B{\"o}rner}, A. and {Smith}, A.~M.~S.},
        title = "{The PLATO Solar-like Light-curve Simulator. A tool to generate realistic stellar light-curves with instrumental effects representative of the PLATO mission}",
      journal = {\aap},
         year = 2019,
        month = apr,
       volume = {624},
          eid = {A117},
        pages = {A117},
          doi = {10.1051/0004-6361/201834822},
archivePrefix = {arXiv},
       eprint = {1903.02747},
 primaryClass = {astro-ph.IM},
       adsurl = {https://ui.adsabs.harvard.edu/abs/2019A&A...624A.117S}
}

@ARTICLE{Huber11,
       author = {{Huber}, D. and {Bedding}, T.~R. and {Stello}, D. and {Hekker}, S. and {Mathur}, S. and {Mosser}, B. and {Verner}, G.~A. and {Bonanno}, A. and {Buzasi}, D.~L. and {Campante}, T.~L. and {Elsworth}, Y.~P. and {Hale}, S.~J. and {Kallinger}, T. and {Silva Aguirre}, V. and {Chaplin}, W.~J. and {De Ridder}, J. and {Garc{\'\i}a}, R.~A. and {Appourchaux}, T. and {Frandsen}, S. and {Houdek}, G. and {Molenda-{\.Z}akowicz}, J. and {Monteiro}, M.~J.~P.~F.~G. and {Christensen-Dalsgaard}, J. and {Gilliland}, R.~L. and {Kawaler}, S.~D. and {Kjeldsen}, H. and {Broomhall}, A.~M. and {Corsaro}, E. and {Salabert}, D. and {Sanderfer}, D.~T. and {Seader}, S.~E. and {Smith}, J.~C.},
        title = "{Testing Scaling Relations for Solar-like Oscillations from the Main Sequence to Red Giants Using Kepler Data}",
      journal = {\apj},
         year = 2011,
        month = dec,
       volume = {743},
       number = {2},
          eid = {143},
        pages = {143},
          doi = {10.1088/0004-637X/743/2/143},
archivePrefix = {arXiv},
       eprint = {1109.3460},
 primaryClass = {astro-ph.SR},
       adsurl = {https://ui.adsabs.harvard.edu/abs/2011ApJ...743..143H}
}

@ARTICLE{VitenseMLT58,
       author = {{B{\"o}hm-Vitense}, E.},
        title = "{{\"U}ber die Wasserstoffkonvektionszone in Sternen verschiedener Effektivtemperaturen und Leuchtkr{\"a}fte. Mit 5 Textabbildungen}",
      journal = {\zap},
         year = 1958,
        month = jan,
       volume = {46},
        pages = {108},
       adsurl = {https://ui.adsabs.harvard.edu/abs/1958ZA.....46..108B}
}

@ARTICLE{Trampedach13,
       author = {{Trampedach}, Regner and {Asplund}, Martin and {Collet}, Remo and {Nordlund}, {\r{A}}ke and {Stein}, Robert F.},
        title = "{A Grid of Three-dimensional Stellar Atmosphere Models of Solar Metallicity. I. General Properties, Granulation, and Atmospheric Expansion}",
      journal = {\apj},
         year = 2013,
        month = may,
       volume = {769},
       number = {1},
          eid = {18},
        pages = {18},
          doi = {10.1088/0004-637X/769/1/18},
archivePrefix = {arXiv},
       eprint = {1303.1780},
 primaryClass = {astro-ph.SR},
       adsurl = {https://ui.adsabs.harvard.edu/abs/2013ApJ...769...18T}
}

@ARTICLE{Froehlich95,
       author = {{Fr{\"o}hlich}, Claus and {Romero}, Jos{\'e} and {Roth}, Hansj{\"o}rg and {Wehrli}, Christoph and {Andersen}, Bo N. and {Appourchaux}, Thierry and {Domingo}, Vicente and {Telljohann}, Udo and {Berthomieu}, Gabrielle and {Delache}, Philippe and {Provost}, Janine and {Toutain}, Thierry and {Crommelynck}, Dominique A. and {Chevalier}, Andr{\'e} and {Fichot}, Alain and {D{\"a}ppen}, Werner and {Gough}, Douglas and {Hoeksema}, Todd and {Jim{\'e}nez}, Antonio and {G{\'o}mez}, Maria F. and {Herreros}, Jos{\'e} M. and {Cort{\'e}s}, Teodoro Roca and {Jones}, Andrew R. and {Pap}, Judit M. and {Willson}, Richard C.},
        title = "{VIRGO: Experiment for Helioseismology and Solar Irradiance Monitoring}",
      journal = {\solphys},
         year = 1995,
        month = dec,
       volume = {162},
       number = {1-2},
        pages = {101-128},
          doi = {10.1007/BF00733428},
       adsurl = {https://ui.adsabs.harvard.edu/abs/1995SoPh..162..101F}
}

@ARTICLE{Larsen2025b,
       author = {{Larsen}, J.~R. and {Lundkvist}, M.~S. and {Davies}, G.~R. and {Nielsen}, M.~B. and {Ludwig}, H.-G. and {Zhou}, Y. and {Rodr{\'\i}guez D{\'\i}az}, L.~F. and {Kjeldsen}, H.},
        title = "{Granulation signatures in 3D hydrodynamical simulations: Evaluating background model performance using a Bayesian nested sampling framework}",
      journal = {\aap},
         year = 2025,
        month = sep,
       volume = {701},
          eid = {A92},
        pages = {A92},
          doi = {10.1051/0004-6361/202555661},
archivePrefix = {arXiv},
       eprint = {2507.11699},
 primaryClass = {astro-ph.SR},
       adsurl = {https://ui.adsabs.harvard.edu/abs/2025A&A...701A..92L}
}

@ARTICLE{Sayeed25,
       author = {{Sayeed}, Maryum and {Huber}, Daniel and {Chontos}, Ashley and {Li}, Yaguang},
        title = "{A Homogeneous Catalog of Oscillating Solar-Type Stars Observed by the Kepler Mission and a New Amplitude Scaling Relation Including Chromospheric Activity}",
      journal = {arXiv e-prints},
         year = 2025,
        month = mar,
          eid = {arXiv:2503.15599},
        pages = {arXiv:2503.15599},
          doi = {10.48550/arXiv.2503.15599},
archivePrefix = {arXiv},
       eprint = {2503.15599},
 primaryClass = {astro-ph.SR},
       adsurl = {https://ui.adsabs.harvard.edu/abs/2025arXiv250315599S}
}

@ARTICLE{Nielsen21,
       author = {{Nielsen}, M.~B. and {Davies}, G.~R. and {Ball}, W.~H. and {Lyttle}, A.~J. and {Li}, T. and {Hall}, O.~J. and {Chaplin}, W.~J. and {Gaulme}, P. and {Carboneau}, L. and {Ong}, J.~M.~J. and {Garc{\'\i}a}, R.~A. and {Mosser}, B. and {Roxburgh}, I.~W. and {Corsaro}, E. and {Benomar}, O. and {Moya}, A. and {Lund}, M.~N.},
        title = "{PBjam: A Python Package for Automating Asteroseismology of Solar-like Oscillators}",
      journal = {\aj},
         year = 2021,
        month = feb,
       volume = {161},
       number = {2},
          eid = {62},
        pages = {62},
          doi = {10.3847/1538-3881/abcd39},
archivePrefix = {arXiv},
       eprint = {2012.00580},
 primaryClass = {astro-ph.SR},
       adsurl = {https://ui.adsabs.harvard.edu/abs/2021AJ....161...62N}
}

@ARTICLE{Nielsen25,
       author = {{Nielsen}, M.~B. and {Ong}, J.~M.~J. and {Hatt}, E.~J. and {Davies}, G.~R. and {Chaplin}, W.~J. and {Hookway}, G.~T. and {Stokholm}, A. and {Scutt}, O.~J. and {Lund}, M.~N. and {Garc{\'\i}a}, R.~A.},
        title = "{Asteroseismology with PBjam 2.0: Measuring Dipole Mode Frequencies in Coupling Regimes from Main-sequence to Low-luminosity Red Giant Stars}",
      journal = {\aj},
         year = 2025,
        month = jun,
       volume = {169},
       number = {6},
          eid = {322},
        pages = {322},
          doi = {10.3847/1538-3881/adcb37},
archivePrefix = {arXiv},
       eprint = {2506.20382},
 primaryClass = {astro-ph.SR},
       adsurl = {https://ui.adsabs.harvard.edu/abs/2025AJ....169..322N}
}

@ARTICLE{Aguirre15_KAGES,
       author = {{Silva Aguirre}, V. and {Davies}, G.~R. and {Basu}, S. and {Christensen-Dalsgaard}, J. and {Creevey}, O. and {Metcalfe}, T.~S. and {Bedding}, T.~R. and {Casagrande}, L. and {Handberg}, R. and {Lund}, M.~N. and {Nissen}, P.~E. and {Chaplin}, W.~J. and {Huber}, D. and {Serenelli}, A.~M. and {Stello}, D. and {Van Eylen}, V. and {Campante}, T.~L. and {Elsworth}, Y. and {Gilliland}, R.~L. and {Hekker}, S. and {Karoff}, C. and {Kawaler}, S.~D. and {Kjeldsen}, H. and {Lundkvist}, M.~S.},
        title = "{Ages and fundamental properties of Kepler exoplanet host stars from asteroseismology}",
      journal = {\mnras},
         year = 2015,
        month = sep,
       volume = {452},
       number = {2},
        pages = {2127-2148},
          doi = {10.1093/mnras/stv1388},
archivePrefix = {arXiv},
       eprint = {1504.07992},
 primaryClass = {astro-ph.SR},
       adsurl = {https://ui.adsabs.harvard.edu/abs/2015MNRAS.452.2127S}
}

@ARTICLE{Mosser12,
       author = {{Mosser}, B. and {Elsworth}, Y. and {Hekker}, S. and {Huber}, D. and {Kallinger}, T. and {Mathur}, S. and {Belkacem}, K. and {Goupil}, M.~J. and {Samadi}, R. and {Barban}, C. and {Bedding}, T.~R. and {Chaplin}, W.~J. and {Garc{\'\i}a}, R.~A. and {Stello}, D. and {De Ridder}, J. and {Middour}, C.~K. and {Morris}, R.~L. and {Quintana}, E.~V.},
        title = "{Characterization of the power excess of solar-like oscillations in red giants with Kepler}",
      journal = {\aap},
         year = 2012,
        month = jan,
       volume = {537},
          eid = {A30},
        pages = {A30},
          doi = {10.1051/0004-6361/20111735210.1086/141952},
archivePrefix = {arXiv},
       eprint = {1110.0980},
 primaryClass = {astro-ph.SR},
       adsurl = {https://ui.adsabs.harvard.edu/abs/2012A&A...537A..30M}
}

@INPROCEEDINGS{TACOPrelim,
       author = {{Theme{\ss}l}, N. and {Kuszlewicz}, J.~S. and {Garc{\'\i}a Saravia Ortiz de Montellano}, A. and {Hekker}, S.},
        title = "{From light-curves to frequencies of oscillation modes using TACO}",
    booktitle = {Stars and their Variability Observed from Space},
         year = 2020,
       editor = {{Neiner}, C. and {Weiss}, W.~W. and {Baade}, D. and {Griffin}, R.~E. and {Lovekin}, C.~C. and {Moffat}, A.~F.~J.},
        month = jan,
        pages = {287-291},
       adsurl = {https://ui.adsabs.harvard.edu/abs/2020svos.conf..287T}
}

@ARTICLE{Llorente22,
       author = {{Llorente}, F. and {Martino}, L. and {Curbelo}, E. and {Lopez-Santiago}, J. and {Delgado}, D.},
        title = "{On the safe use of prior densities for Bayesian model selection}",
      journal = {arXiv e-prints},
         year = 2022,
        month = jun,
          eid = {arXiv:2206.05210},
        pages = {arXiv:2206.05210},
          doi = {10.48550/arXiv.2206.05210},
archivePrefix = {arXiv},
       eprint = {2206.05210},
 primaryClass = {stat.ME},
       adsurl = {https://ui.adsabs.harvard.edu/abs/2022arXiv220605210L}
}

@INPROCEEDINGS{Chontos21,
       author = {{Chontos}, Ashley and {Sayeed}, Maryum and {Huber}, Daniel},
        title = "{pySYD: Automated Measurements of Global Asteroseismic Parameters}",
    booktitle = {Posters from the TESS Science Conference II (TSC2)},
         year = 2021,
        month = jul,
          eid = {189},
        pages = {189},
          doi = {10.5281/zenodo.5140574},
       adsurl = {https://ui.adsabs.harvard.edu/abs/2021tsc2.confE.189C}
}

@ARTICLE{Huber09,
       author = {{Huber}, D. and {Stello}, D. and {Bedding}, T.~R. and {Chaplin}, W.~J. and {Arentoft}, T. and {Quirion}, P. -O. and {Kjeldsen}, H.},
        title = "{Automated extraction of oscillation parameters for Kepler observations of solar-type stars}",
      journal = {Communications in Asteroseismology},
         year = 2009,
        month = oct,
       volume = {160},
        pages = {74},
          doi = {10.48550/arXiv.0910.2764},
archivePrefix = {arXiv},
       eprint = {0910.2764},
 primaryClass = {astro-ph.SR},
       adsurl = {https://ui.adsabs.harvard.edu/abs/2009CoAst.160...74H}
}

@MISC{KeplerHandbook16,
       author = {{Van Cleve}, Jeffrey E. and {Christiansen}, Jessie L. and {Jenkins}, Jon M. and {Caldwell}, Douglas A. and {Barclay}, Thomas and {Bryson}, Stephen T. and {Burke}, Christopher J. and {Cambell}, Jennifer and {Catanzarite}, Joseph and {Clarke}, Bruce D. and {Coughlin}, Jeffrey L. and {Girouard}, F. and {Haas}, Michael R. and {Klaus}, Todd C. and {Kolodziejczak}, Jeffrey J. and {Li}, Jie and {McCauliff}, Sean D. and {Morris}, Robert L. and {Mullally}, Fergal and {Quintana}, Elisa V. and {Rowe}, Jason and {Sabale}, Anima and {Seader}, Shawn and {Smith}, Jeffrey C. and {Still}, Martin D. and {Tenenbaum}, Peter G. and {Thompson}, Susan E. and {Twicken}, Joesph D. and {Kamal Uddin}, Akm and {Zamudio}, Khadeejah},
        title = "{Kepler Data Characteristics Handbook}",
 howpublished = {Kepler Science Document KSCI-19040-005, id. 2. Edited by Doug Caldwell, Jon M. Jenkins, Michael R. Haas and Natalie Batalha},
         year = 2016,
        month = dec,
          eid = {2},
        pages = {2},
       adsurl = {https://ui.adsabs.harvard.edu/abs/2016ksci.rept....2V}
}

@BOOK{Johnson95,
  author = {{Johnson}, N. L. and {Kotz}, S. and {Balakrishnan}, N.},
  title = {Continuous Univariate Distributions, Volume 2 (2nd ed.)},
  year = {1995},
  publisher = {John Wiley \& Sons},
  address = {New York},
  isbn = {978‑0‑471‑58494‑0},
  series = {Wiley Series in Probability and Statistics},
  pages = {752}
}

@ARTICLE{Hon24,
       author = {{Hon}, Marc and {Huber}, Daniel and {Li}, Yaguang and {Metcalfe}, Travis S. and {Bedding}, Timothy R. and {Ong}, Joel and {Chontos}, Ashley and {Rubenzahl}, Ryan and {Halverson}, Samuel and {Garc{\'\i}a}, Rafael A. and {Kjeldsen}, Hans and {Stello}, Dennis and {Hey}, Daniel R. and {Campante}, Tiago and {Howard}, Andrew W. and {Gibson}, Steven R. and {Rider}, Kodi and {Roy}, Arpita and {Baker}, Ashley D. and {Edelstein}, Jerry and {Smith}, Chris and {Fulton}, Benjamin J. and {Walawender}, Josh and {Brodheim}, Max and {Brown}, Matt and {Chan}, Dwight and {Dai}, Fei and {Deich}, William and {Gottschalk}, Colby and {Grillo}, Jason and {Hale}, Dave and {Hill}, Grant M. and {Holden}, Bradford and {Householder}, Aaron and {Isaacson}, Howard and {Ishikawa}, Yuzo and {Jelinsky}, Sharon R. and {Kassis}, Marc and {Kaye}, Stephen and {Laher}, Russ and {Lanclos}, Kyle and {Lee}, Chien-Hsiu and {Lilley}, Scott and {McCarney}, Ben and {Miller}, Timothy N. and {Payne}, Joel and {Petigura}, Erik A. and {Poppett}, Claire and {Raffanti}, Michael and {Rockosi}, Constance and {Sanford}, Dale and {Schwab}, Christian and {Shaum}, Abby P. and {Sirk}, Martin M. and {Smith}, Roger and {Thorne}, Jim and {Valliant}, John and {Vandenberg}, Adam and {Wang}, Shin Ywan and {Wishnow}, Edward and {Wold}, Truman and {Yeh}, Sherry and {Baker}, Ashley and {Basu}, Sarbani and {Bedell}, Megan and {Cegla}, Heather M. and {Crossfield}, Ian and {Dressing}, Courtney and {Dumusque}, Xavier and {Knutson}, Heather and {Mawet}, Dimitri and {O'Meara}, John and {Stef{\'a}nsson}, Gu{\dj}mundur and {Teske}, Johanna and {Vasisht}, Gautam and {Wang}, Sharon Xuesong and {Weiss}, Lauren M. and {Winn}, Joshua N. and {Wright}, Jason T.},
        title = "{Asteroseismology of the Nearby K Dwarf {\ensuremath{\sigma}} Draconis Using the Keck Planet Finder and TESS}",
      journal = {\apj},
         year = 2024,
        month = nov,
       volume = {975},
       number = {1},
          eid = {147},
        pages = {147},
          doi = {10.3847/1538-4357/ad76a9},
archivePrefix = {arXiv},
       eprint = {2407.21234},
 primaryClass = {astro-ph.SR},
       adsurl = {https://ui.adsabs.harvard.edu/abs/2024ApJ...975..147H}
}

@ARTICLE{Lundkvist24,
       author = {{Lundkvist}, Mia S. and {Kjeldsen}, Hans and {Bedding}, Timothy R. and {McCaughrean}, Mark J. and {Butler}, R. Paul and {Slumstrup}, Ditte and {Campante}, Tiago L. and {Aerts}, Conny and {Arentoft}, Torben and {Bruntt}, Hans and {Cardoso}, C{\'a}tia V. and {Carrier}, Fabien and {Close}, Laird M. and {Gomes da Silva}, Jo{\~a}o and {Kallinger}, Thomas and {King}, Robert R. and {Li}, Yaguang and {Murphy}, Simon J. and {R{\o}rsted}, Jakob L. and {Stello}, Dennis},
        title = "{Low-amplitude Solar-like Oscillations in the K5 V Star ϵ Indi A}",
      journal = {\apj},
         year = 2024,
        month = apr,
       volume = {964},
       number = {2},
          eid = {110},
        pages = {110},
          doi = {10.3847/1538-4357/ad25f2},
archivePrefix = {arXiv},
       eprint = {2403.04509},
 primaryClass = {astro-ph.SR},
       adsurl = {https://ui.adsabs.harvard.edu/abs/2024ApJ...964..110L}
}

@MISC{LightCurve18,
   author = {{Lightkurve Collaboration} and {Cardoso}, J.~V.~d.~M. and
             {Hedges}, C. and {Gully-Santiago}, M. and {Saunders}, N. and
             {Cody}, A.~M. and {Barclay}, T. and {Hall}, O. and
             {Sagear}, S. and {Turtelboom}, E. and {Zhang}, J. and
             {Tzanidakis}, A. and {Mighell}, K. and {Coughlin}, J. and
             {Bell}, K. and {Berta-Thompson}, Z. and {Williams}, P. and
             {Dotson}, J. and {Barentsen}, G.},
    title = "{Lightkurve: Kepler and TESS time series analysis in Python}",
howpublished = {Astrophysics Source Code Library},
     year = 2018,
    month = dec,
archivePrefix = "ascl",
   eprint = {1812.013},
   adsurl = {http://adsabs.harvard.edu/abs/2018ascl.soft12013L},
}

@ARTICLE{GaiaDR3_AstParamRelease,
       author = {{Creevey}, O.~L. and {Sordo}, R. and {Pailler}, F. and {Fr{\'e}mat}, Y. and {Heiter}, U. and {Th{\'e}venin}, F. and {Andrae}, R. and {Fouesneau}, M. and {Lobel}, A. and {Bailer-Jones}, C.~A.~L. and {Garabato}, D. and {Bellas-Velidis}, I. and {Brugaletta}, E. and {Lorca}, A. and {Ordenovic}, C. and {Palicio}, P.~A. and {Sarro}, L.~M. and {Delchambre}, L. and {Drimmel}, R. and {Rybizki}, J. and {Torralba Elipe}, G. and {Korn}, A.~J. and {Recio-Blanco}, A. and {Schultheis}, M.~S. and {De Angeli}, F. and {Montegriffo}, P. and {Abreu Aramburu}, A. and {Accart}, S. and {{\'A}lvarez}, M.~A. and {Bakker}, J. and {Brouillet}, N. and {Burlacu}, A. and {Carballo}, R. and {Casamiquela}, L. and {Chiavassa}, A. and {Contursi}, G. and {Cooper}, W.~J. and {Dafonte}, C. and {Dapergolas}, A. and {de Laverny}, P. and {Dharmawardena}, T.~E. and {Edvardsson}, B. and {Le Fustec}, Y. and {Garc{\'\i}a-Lario}, P. and {Garc{\'\i}a-Torres}, M. and {Gomez}, A. and {Gonz{\'a}lez-Santamar{\'\i}a}, I. and {Hatzidimitriou}, D. and {Jean-Antoine Piccolo}, A. and {Kontiza}, M. and {Kordopatis}, G. and {Lanzafame}, A.~C. and {Lebreton}, Y. and {Licata}, E.~L. and {Lindstr{\o}m}, H.~E.~P. and {Livanou}, E. and {Magdaleno Romeo}, A. and {Manteiga}, M. and {Marocco}, F. and {Marshall}, D.~J. and {Mary}, N. and {Nicolas}, C. and {Pallas-Quintela}, L. and {Panem}, C. and {Pichon}, B. and {Poggio}, E. and {Riclet}, F. and {Robin}, C. and {Santove{\~n}a}, R. and {Silvelo}, A. and {Slezak}, I. and {Smart}, R.~L. and {Soubiran}, C. and {S{\"u}veges}, M. and {Ulla}, A. and {Utrilla}, E. and {Vallenari}, A. and {Zhao}, H. and {Zorec}, J. and {Barrado}, D. and {Bijaoui}, A. and {Bouret}, J.-C. and {Blomme}, R. and {Brott}, I. and {Cassisi}, S. and {Kochukhov}, O. and {Martayan}, C. and {Shulyak}, D. and {Silvester}, J.},
        title = "{Gaia Data Release 3. Astrophysical parameters inference system (Apsis). I. Methods and content overview}",
      journal = {\aap},
         year = 2023,
        month = jun,
       volume = {674},
          eid = {A26},
        pages = {A26},
          doi = {10.1051/0004-6361/202243688},
archivePrefix = {arXiv},
       eprint = {2206.05864},
 primaryClass = {astro-ph.GA},
       adsurl = {https://ui.adsabs.harvard.edu/abs/2023A&A...674A..26C}
}

@ARTICLE{Campante24,
       author = {{Campante}, T.~L. and {Kjeldsen}, H. and {Li}, Y. and {Lund}, M.~N. and {Silva}, A.~M. and {Corsaro}, E. and {Gomes da Silva}, J. and {Martins}, J.~H.~C. and {Adibekyan}, V. and {Azevedo Silva}, T. and {Bedding}, T.~R. and {Bossini}, D. and {Buzasi}, D.~L. and {Chaplin}, W.~J. and {Costa}, R.~R. and {Cunha}, M.~S. and {Cristo}, E. and {Faria}, J.~P. and {Garc{\'\i}a}, R.~A. and {Huber}, D. and {Lundkvist}, M.~S. and {Metcalfe}, T.~S. and {Monteiro}, M.~J.~P.~F.~G. and {Neitzel}, A.~W. and {Nielsen}, M.~B. and {Poretti}, E. and {Santos}, N.~C. and {Sousa}, S.~G.},
        title = "{Expanding the frontiers of cool-dwarf asteroseismology with ESPRESSO. Detection of solar-like oscillations in the K5 dwarf ϵ Indi}",
      journal = {\aap},
         year = 2024,
        month = mar,
       volume = {683},
          eid = {L16},
        pages = {L16},
          doi = {10.1051/0004-6361/202449197},
archivePrefix = {arXiv},
       eprint = {2403.16333},
 primaryClass = {astro-ph.SR},
       adsurl = {https://ui.adsabs.harvard.edu/abs/2024A&A...683L..16C}
}

@ARTICLE{Ramirez13,
       author = {{Ram{\'\i}rez}, I. and {Allende Prieto}, C. and {Lambert}, D.~L.},
        title = "{Oxygen Abundances in Nearby FGK Stars and the Galactic Chemical Evolution of the Local Disk and Halo}",
      journal = {\apj},
         year = 2013,
        month = feb,
       volume = {764},
       number = {1},
          eid = {78},
        pages = {78},
          doi = {10.1088/0004-637X/764/1/78},
archivePrefix = {arXiv},
       eprint = {1301.1582},
 primaryClass = {astro-ph.SR},
       adsurl = {https://ui.adsabs.harvard.edu/abs/2013ApJ...764...78R}
}

@ARTICLE{Stein24,
       author = {{Stein}, Robert F. and {Nordlund}, {\r{A}}ke and {Collet}, Remo and {Trampedach}, Regner},
        title = "{The Stagger Code for Accurate and Efficient, Radiation-coupled Magnetohydrodynamic Simulations}",
      journal = {\apj},
         year = 2024,
        month = jul,
       volume = {970},
       number = {1},
          eid = {24},
        pages = {24},
          doi = {10.3847/1538-4357/ad4706},
archivePrefix = {arXiv},
       eprint = {2405.02483},
 primaryClass = {astro-ph.IM},
       adsurl = {https://ui.adsabs.harvard.edu/abs/2024ApJ...970...24S}
}

@ARTICLE{Lund15,
       author = {{Lund}, Mikkel N. and {Handberg}, Rasmus and {Davies}, Guy R. and {Chaplin}, William J. and {Jones}, Caitlin D.},
        title = "{K2P$^{2}${\textemdash} A Photometry Pipeline for the K2 Mission}",
      journal = {\apj},
         year = 2015,
        month = jun,
       volume = {806},
       number = {1},
          eid = {30},
        pages = {30},
          doi = {10.1088/0004-637X/806/1/30},
archivePrefix = {arXiv},
       eprint = {1504.05199},
 primaryClass = {astro-ph.SR},
       adsurl = {https://ui.adsabs.harvard.edu/abs/2015ApJ...806...30L}
}

@ARTICLE{LiY25,
       author = {{Li}, Yaguang and {Huber}, Daniel and {Ong}, J.~M. Joel and {van Saders}, Jennifer and {Costa}, R.~R. and {Larsen}, Jens Reersted and {Basu}, Sarbani and {Bedding}, Timothy R. and {Dai}, Fei and {Chontos}, Ashley and {Carmichael}, Theron W. and {Hey}, Daniel and {Kjeldsen}, Hans and {Hon}, Marc and {Campante}, Tiago L. and {Monteiro}, M{\'a}rio J.~P.~F.~G. and {Lundkvist}, Mia Sloth and {Saunders}, Nicholas and {Isaacson}, Howard and {Howard}, Andrew W. and {Gibson}, Steven R. and {Halverson}, Samuel and {Rider}, Kodi and {Roy}, Arpita and {Baker}, Ashley D. and {Edelstein}, Jerry and {Smith}, Chris and {Fulton}, Benjamin J. and {Walawender}, Josh},
        title = "{K Dwarf Radius Inflation and a 10 Gyr Spin-down Clock Unveiled through Asteroseismology of HD 219134 from the Keck Planet Finder}",
      journal = {\apj},
         year = 2025,
        month = may,
       volume = {984},
       number = {2},
          eid = {125},
        pages = {125},
          doi = {10.3847/1538-4357/adc737},
archivePrefix = {arXiv},
       eprint = {2502.00971},
 primaryClass = {astro-ph.SR},
       adsurl = {https://ui.adsabs.harvard.edu/abs/2025ApJ...984..125L}
}

@ARTICLE{Gizon03,
       author = {{Gizon}, L. and {Solanki}, S.~K.},
        title = "{Determining the Inclination of the Rotation Axis of a Sun-like Star}",
      journal = {\apj},
         year = 2003,
        month = jun,
       volume = {589},
       number = {2},
        pages = {1009-1019},
          doi = {10.1086/374715},
       adsurl = {https://ui.adsabs.harvard.edu/abs/2003ApJ...589.1009G}
}

@ARTICLE{Santos24,
       author = {{Santos}, {\^A}ngela R.~G. and {Godoy-Rivera}, Diego and {Finley}, Adam J. and {Mathur}, Savita and {Garc{\'\i}a}, Rafael A. and {Breton}, Sylvain N. and {Broomhall}, Anne-Marie},
        title = "{Kepler main-sequence solar-like stars: surface rotation and magnetic-activity evolution}",
      journal = {Frontiers in Astronomy and Space Sciences},
         year = 2024,
        month = mar,
       volume = {11},
          eid = {1356379},
        pages = {1356379},
          doi = {10.3389/fspas.2024.1356379},
archivePrefix = {arXiv},
       eprint = {2404.15911},
 primaryClass = {astro-ph.SR},
       adsurl = {https://ui.adsabs.harvard.edu/abs/2024FrASS..1156379S}
}

@book{laplace_thorie_1812,
  address = {Paris},
  author = {Laplace, {Pierre-Simon}},
  publisher = {Courcier},
  title = {Théorie analytique des probabilités},
  url = {http://gallica.bnf.fr/ark:/12148/bpt6k88764q},
  year = 1812
}

@phdthesis{Schofield19thesis,
    title        = {The Asteroseismic Potential of the NASA TESS satellite},
    author       = {Mathew Schofield},
    year         = 2019,
    month        = {February},
    school       = {University of Birmingham},
    type         = {PhD thesis}
}

@ARTICLE{Lund25,
       author = {{Lund}, Mikkel N. and {Chontos}, Ashley and {Grundahl}, Frank and {Mathur}, Savita and {Garc{\'\i}a}, Rafael A. and {Huber}, Daniel and {Buzasi}, Derek and {Bedding}, Timothy R. and {Hon}, Marc and {Li}, Yaguang},
        title = "{Luminaries in the sky: The TESS legacy sample of bright stars: I. Asteroseismic detections in naked-eye main-sequence and subgiant solar-like oscillators}",
      journal = {\aap},
         year = 2025,
        month = sep,
       volume = {701},
          eid = {A285},
        pages = {A285},
          doi = {10.1051/0004-6361/202555485},
archivePrefix = {arXiv},
       eprint = {2508.08699},
 primaryClass = {astro-ph.SR},
       adsurl = {https://ui.adsabs.harvard.edu/abs/2025A&A...701A.285L}
}

@ARTICLE{Campante15,
       author = {{Campante}, T.~L. and {Barclay}, T. and {Swift}, J.~J. and {Huber}, D. and {Adibekyan}, V. Zh. and {Cochran}, W. and {Burke}, C.~J. and {Isaacson}, H. and {Quintana}, E.~V. and {Davies}, G.~R. and et al.},
        title = "{An Ancient Extrasolar System with Five Sub-Earth-size Planets}",
      journal = {\apj},
         year = 2015,
        month = feb,
       volume = {799},
       number = {2},
          eid = {170},
        pages = {170},
          doi = {10.1088/0004-637X/799/2/170},
archivePrefix = {arXiv},
       eprint = {1501.06227},
 primaryClass = {astro-ph.EP},
       adsurl = {https://ui.adsabs.harvard.edu/abs/2015ApJ...799..170C}
}

@ARTICLE{Appourchaux04,
       author = {{Appourchaux}, T.},
        title = "{On detecting short-lived p modes in a stellar oscillation spectrum}",
      journal = {\aap},
         year = 2004,
        month = dec,
       volume = {428},
        pages = {1039-1042},
          doi = {10.1051/0004-6361:20041682},
       adsurl = {https://ui.adsabs.harvard.edu/abs/2004A&A...428.1039A}
}

@ARTICLE{Stein01,
       author = {{Stein}, R.~F. and {Nordlund}, {\r{A}}.},
        title = "{Solar Oscillations and Convection. II. Excitation of Radial Oscillations}",
      journal = {\apj},
         year = 2001,
        month = jan,
       volume = {546},
       number = {1},
        pages = {585-603},
          doi = {10.1086/318218},
archivePrefix = {arXiv},
       eprint = {astro-ph/0008048},
 primaryClass = {astro-ph},
       adsurl = {https://ui.adsabs.harvard.edu/abs/2001ApJ...546..585S}
}

@ARTICLE{Kim21,
       author = {{Kim}, Ki-Beom and {Chang}, Heon-Young},
        title = "{Scaling relations for width of the power excess of stellar oscillations}",
      journal = {\na},
         year = 2021,
        month = apr,
       volume = {84},
          eid = {101522},
        pages = {101522},
          doi = {10.1016/j.newast.2020.101522},
       adsurl = {https://ui.adsabs.harvard.edu/abs/2021NewA...8401522K}
}

@ARTICLE{Magic13,
       author = {{Magic}, Z. and {Collet}, R. and {Asplund}, M. and {Trampedach}, R. and {Hayek}, W. and {Chiavassa}, A. and {Stein}, R.~F. and {Nordlund}, {\r{A}}.},
        title = "{The Stagger-grid: A grid of 3D stellar atmosphere models. I. Methods and general properties}",
      journal = {\aap},
         year = 2013,
        month = sep,
       volume = {557},
          eid = {A26},
        pages = {A26},
          doi = {10.1051/0004-6361/201321274},
archivePrefix = {arXiv},
       eprint = {1302.2621},
 primaryClass = {astro-ph.SR},
       adsurl = {https://ui.adsabs.harvard.edu/abs/2013A&A...557A..26M}
}

\newpage
\begin{appendix}
\section{Specific removals from sample}\label{App:Removals}
\begin{table}
\renewcommand{\arraystretch}{1.3}
\centering
\caption{Specific stars removed from the \citet{Sayeed25} catalogue with the reasons noted}
\begin{tabular}{lcccc}
\hline
\multicolumn{1}{c}{\thead{Star}} & \multicolumn{1}{c}{\thead{Reason}} \\
\hline 
KIC12069424 & \thead{Seismic binary} \\
KIC12069449 & \thead{Seismic binary} \\
KIC3427720 & \thead{Seismic binary} \\
KIC10295224 & \thead{Artefacts in the PDS at low frequency, \\ low granulation amplitudes} \\
KIC11456910 & \thead{Strong blue noise, weak oscillations, \\ strong contamination peals} \\
KIC6106120 & \thead{Strong contamination, \\ low granulation amplitudes, blue noise} \\
KIC9912680 & \thead{Strong contamination near granulation level} \\
KIC3430893 & \thead{Strong contamination near granulation level, \\ high white noise} \\
KIC5514383 & \thead{Strong contamination near granulation level, \\ weak oscillations} \\
KIC12068975 & \thead{PDS is dominated by white noise} \\
KIC8715511 & \thead{\numax estimate of $6.79$ \muHz \\ below adopted cut-off} \\
KIC11759838 & \thead{\numax estimate of $5.10$ \muHz \\ below adopted cut-off} \\
\hline
\end{tabular}
\label{tab:removals}
\end{table}
This appendix contains Table~\ref{tab:removals} which specifies the 12 stars that were removed from the \citet{Sayeed25} catalogue, along with the reasons why.

\section{Prior densities in model selection}\label{App:ModSelectDiscuss}
This appendix briefly discusses the considerations underlying our choice of prior densities. A comprehensive treatment of the associated challenges and caveats can be found in \citet{Llorente22}. In this work, we are concerned with addressing two related but distinct inference problems. The first is to determine the optimal parameters $\theta$ of a given model $M$ that best describe a set of data $D$ -- what \citet{Llorente22} refer to as a level-1 inference. In such cases, if no strong prior knowledge is available, non-informative priors (e.g. uniform distributions spanning the entire parameter space) provide an objective starting point. These allow the nested sampler to explore the full parameter space and identify the optimal set of $\theta$ values, albeit often at the cost of slower convergence. The second problem is model selection, based on the Bayesian evidence (or marginal likelihood), which \citet{Llorente22} classify as a level-2 inference. Here, the choice of priors directly influences the result and thereby our later interpretations. The evidence is given by
\begin{equation}\label{eq:Evidence}
    p(D|M) = \mathcal{Z}= \int_\theta p(\theta|M)p(D|\theta,M) \ \textup{d}\theta \ .
\end{equation}
From this expression, it is evident that the volume of the prior $p(\theta|M)$ affects the evidence $\mathcal{Z}$. Broad or non-informative priors with large volumes reduce the evidence and thus penalise the corresponding model. We must be astutely aware of this fact when defining our priors, at the same time as ensuring that we define priors which allow for the optimal set of parameters $\theta$ to be recovered. 

This penalisation of large prior volumes is, however, an intended feature of the Bayesian framework. Models with a greater number of free parameters are naturally penalised unless those parameters substantially improve the likelihood $p(D|\theta,M)$, thereby providing genuine explanatory power. Even so, it is essential to ensure that the priors are defined consistently across models to prevent unintended biases. For parameters common to multiple models -- such as the first two Harvey components in models H and T -- we therefore adopt priors with identical functional forms and boundaries broad enough to include the plausible ranges for all models. For parameters that are directly data-driven (e.g. the white-noise level), the priors are defined solely from the data and applied identically to all models, ensuring a fair contribution to Eq.~\ref{eq:Evidence}. The only exception is the amplitude parameter of model J, whose functional form differs fundamentally from those of models H and T. In this case, the prior’s central value is adjusted accordingly, while its functional form and scale remains identical (see Table~\ref{tab:PriorCompModparams}).

\section{Overview of priors}\label{App:Priors}
\begin{table*}
\renewcommand{\arraystretch}{1.1}
\centering
\caption{Compilation of the granulation background model specific priors used for the three prescriptions in Table~\ref{tab:Models}.}
\begin{threeparttable}
\begin{tabular}{cccc}
\hline
\multicolumn{1}{c}{\thead{\large\textbf{Model specific} \\ \large\textbf{fit parameters}}} & \multicolumn{1}{c}{\thead{\large\textbf{Model J}}} & \multicolumn{1}{c}{\thead{\large\textbf{Model H}}} & \multicolumn{1}{c}{\thead{\large\textbf{Model T}}}\\ \hline 
\multicolumn{4}{c}{\thead{Correlated-inference setup (Sect.~\ref{subsec:corrinf})}} \\
\hline
$\sigma_a$    & 
\thead{-- Samples $a$ freely as below --} & 
\thead{lognormal \\ $\mu=0$ \\ $\sigma = 0.1$ } & 
\thead{lognormal \\ $\mu=0$ \\ $\sigma = 0.1$ }    \\

$\sigma_b$    & 
\thead{beta \\ $a, b = 6, 6$ \\ loc, scale $=0.6, 0.8$ } & 
\thead{beta \\ $a, b = 6, 6$ \\ loc, scale $=0.6, 0.8$ } & 
\thead{beta \\ $a, b = 6, 6$ \\ loc, scale $=0.6, 0.8$ } \\

$\sigma_d$    & 
\thead{beta \\ $a, b = 6, 6$ \\ loc, scale $=0.6, 0.8$ } & 
\thead{beta \\ $a, b = 6, 6$ \\ loc, scale $=0.6, 0.8$ } & 
\thead{beta \\ $a, b = 6, 6$ \\ loc, scale $=0.6, 0.8$ }  \\

\hline
\multicolumn{4}{c}{\thead{Free variable inference}} \\
\hline
$a$ & 
\thead{lognormal \\ $\mu=f(a_c=3.555, a_e=-1.006,\numax)$\tnote{§1} \\ $\sigma = 0.1$} & 
\thead{lognormal \\ $\mu=f(a_c=3.530, a_e=-0.609,\numax)$ \\ $\sigma = 0.1$} & 
\thead{lognormal \\ $\mu=f(a_c=3.530, a_e=-0.609,\numax)$ \\ $\sigma = 0.1$} \\

$b$    & 
\thead{lognormal \\ $\mu=f(b_c=-0.499, b_e=0.970,\numax)$ \\ $\sigma = 0.1$} & 
\thead{lognormal \\ $\mu=f(b_c=-0.499, b_e=0.970,\numax)$ \\ $\sigma = 0.1$} & 
\thead{lognormal \\ $\mu=f(b_c=-0.499, b_e=0.970,\numax)$ \\ $\sigma = 0.1$} \\

$d$    & 
\thead{lognormal \\ $\mu=f(d_c=-0.020, d_e=0.992,\numax)$ \\ $\sigma = 0.1$} & 
\thead{lognormal \\ $\mu=f(d_c=-0.020, d_e=0.992,\numax)$ \\ $\sigma = 0.1$} & 
\thead{lognormal \\ $\mu=f(d_c=-0.020, d_e=0.992,\numax)$ \\ $\sigma = 0.1$} \\ 

\hline
\multicolumn{4}{c}{\thead{Universal parameter priors}} \\
\hline
$c$    & 
\thead{--} & 
\thead{lognormal \\ $\mu=f(c_c=3.477, c_e=-0.609,\numax)$ \\ $\sigma = 0.1$ } & 
\thead{lognormal \\ $\mu=f(c_c=3.477, c_e=-0.609,\numax)$ \\ $\sigma = 0.1$ }     \\

$e$    & 
\thead{--} & 
\thead{--} & 
\thead{lognormal \\ $\mu=\frac{1}{2}f(c_c=3.477, c_e=-0.609,\numax)$ \\ $\sigma = 0.1$ } \\

$f$    & 
\thead{--} &
\thead{--} & 
\thead{beta \\
$a,b =$ data-driven \\
loc $= 0.9\numax$ \\
scale = $\min(4\numax,0.9\nu_{\rm Nyq})-$ loc \\
mode $= 1.1\numax$\tnote{§2}}  \\

$l$    &
\thead{beta \\ $a, b = 1.8, 3.0$ \\ loc, scale $=1, 8$ } & 
\thead{beta \\ $a, b = 1.8, 3.0$ \\ loc, scale $=1, 8$ } &
\thead{beta \\ $a, b = 1.8, 3.0$ \\ loc, scale $=1, 8$ }      \\

$k$    & 
\thead{beta \\ $a, b = 2.0, 5.0$ \\ loc, scale $=1, 9$ } &
\thead{beta \\ $a, b = 2.0, 5.0$ \\ loc, scale $=1, 9$ } & 
\thead{beta \\ $a, b = 2.0, 5.0$ \\ loc, scale $=1, 9$ }       \\

$m$    & 
\thead{--} & 
\thead{--} & 
\thead{beta \\ $a, b = 2.0, 5.0$ \\ loc, scale $=1, 10$ }       \\
\hline
\end{tabular}
\begin{tablenotes}\footnotesize
\item[§1] Amplitude prior for model J follows the prescription in \citet{Larsen2025b}.
\item[§2] Shape parameters a(m),b(m) set according to bounds such that mode lies just above \numax.
\end{tablenotes}
\end{threeparttable}
\tablefoot{The log-space coefficients for the power law $f(\textup{constant}, \textup{exponent},\numax) = \textup{constant} + \log_{10}(\numax)^{\textup{exponent}}$ are specified directly when used and originate from \citet{Kallinger2014}, except for the amplitude $a$ of model J which is from \citet{Larsen2025b}. When the prior uses \numax as input, it is the observed value obtained as specified in Sect.~\ref{sec:Sample}.}
\vspace*{-3mm}
\label{tab:PriorCompModparams}
\end{table*}

\begin{table*}
\renewcommand{\arraystretch}{2.2}
\centering
\caption{Compilation of the general priors for the complete background model terms describing stellar activity, the oscillation excess, and white noise.}
\begin{threeparttable}
\begin{tabular}{ccc}
\hline
\multicolumn{1}{c}{\thead{\large\textbf{Fit parameters}}} & \multicolumn{1}{c}{\thead{\large\textbf{Prior distribution}}} & \multicolumn{1}{c}{\thead{\large\textbf{Prior setup}}} \\ 
\hline 
\multicolumn{3}{c}{\thead{Activity component priors}} \\
\hline
\multicolumn{3}{c}{\large $\mathcal{M}_\mathrm{act} = \frac{a_2^2/b_2}{1+\left(\frac{\nu}{b_2}\right)^2}$} \\
$a_2$ &
\thead{beta} &
\thead{
a,b = data-driven \\
loc $= 0.0$, scale $= 3 \times a_2^\mathrm{est}$ \\
mode $= a_2^\mathrm{est} = \sqrt{2\, P_\mathrm{peak}\, b_2^\mathrm{est}}$ \\
$P_\mathrm{peak} = \max(\text{Power}[\nu < 0.15\,\numax]$)
} \\

$b_2$ &
\thead{beta} &
\thead{
a,b = data-driven \\
loc $= 1\times10^{-6}$, scale $= (1/8) \, b_\mathrm{gran}(\numax) - 1\times10^{-6}$ \\
mode $= b_2^\mathrm{est} = \arg\max(\text{Power}[\nu < 0.15\,\numax]$)
} \\
\hline 
\multicolumn{3}{c}{\thead{Gaussian excess specific priors}} \\
\hline
\multicolumn{3}{c}{\large $\mathcal{M}_\mathrm{osc} = P_\mathrm{osc}\exp{\frac{-(\nu-\numax)^2}{2\sigma^2}}$} \\
$\sigma_\mathrm{osc}$ & 
\thead{beta} & 
\thead{
a,b = data-driven \\
loc $= 1\mathrm{e}{-4}$, scale $= 3 \times \text{mode} - 1\mathrm{e}{-4}$ \\
mode $= \Delta\nu \cdot \max(1,4(\nu_\text{max}/3090)^{0.2}) / (2\sqrt{2\ln 2})$\tnote{§1}
} \\

$P_\mathrm{osc}$ & 
\thead{beta} & 
\thead{
a,b = data-driven \\
loc $= 1\mathrm{e}{-4}$, scale $= 15 \times \text{mode} - 1\mathrm{e}{-4}$ \\
mode $= \alpha \cdot \text{median(Power[$\nu_\text{max} \pm 2\sigma_\mathrm{osc}$]))}$ \\
$\alpha = 0.4$ ($\numax < 1000\mu$Hz), $0.1$ ($\numax > 1000\mu$Hz)
} \\

$\numax$ & 
\thead{beta} & 
\thead{
a,b = 11, 11 \\
loc $= 0.75 \, \numax$, scale $= 1.25 \, \numax - 0.75\,\numax$
} \\
\hline
\multicolumn{3}{c}{\thead{Peakbogging specific priors}} \\
\hline
$H$  & 
\thead{beta} & 
\thead{
a,b = 1, 5 \\
loc, scale $=0,1$
} \\  

$\beta$  & 
\thead{beta} & 
\thead{
a,b = 1, 10 \\
loc, scale $=1,19$
} \\  
\hline
\multicolumn{3}{c}{\thead{White noise prior}} \\
\hline
$W$ &
\thead{beta} &
\thead{
a,b = 1.5, 4 \\
loc, scale $= 0.5\,W_\mathrm{est},\; 2.5\,W_\mathrm{est}$ \\
$W_\mathrm{est}=\textup{median(Power[-100:])}$
} \\
\hline
\end{tabular}
\begin{tablenotes}\footnotesize
\item[§1] Oscillation excess width prior set as in the SYD pipeline \citep{Huber09} and its python implementation pySYD \citep{Chontos21}, using an approximation of \dnu \citep{Stello09b, Hekker09, Huber11}. 
\end{tablenotes}
\end{threeparttable}
\tablefoot{The functional forms of the stellar activity (standard Harvey profile, \citealt{Harvey85}) and Gaussian oscillation excess components are provided for clarity of the parameter definitions. When the prior uses \numax as input, it is the observed value obtained as specified in Sect.~\ref{sec:Sample}.}
\vspace*{-3mm}
\label{tab:PriorCompGeneral}
\end{table*}

This appendix defines the priors adopted in this work. In Table~\ref{tab:PriorCompModparams} the priors for the parameters of the granulation background models from Table~\ref{tab:Models} are presented. The priors for the general term parameters identical across granulation background models -- which describe stellar activity, the oscillation excess, and white noise -- are found in Table~\ref{tab:PriorCompGeneral}, and are primarily defined based on the PDS data for the given star.

The model-specific priors in Table~\ref{tab:PriorCompModparams} are presented for both the correlated-inference setup described in Sect.~\ref{subsec:corrinf}, which introduces the scatter parameters $\sigma_{a,b,d}$, and additionally also the case when all parameters are treated as free variables. For model~J, whose functional form differs from that of models H and T, the amplitude parameter $a$ is always inferred directly, and its prior follows the prescription of \citet{Larsen2025b}. In the correlated inference, a broader prior is used for the amplitude scatter parameter, implemented as a log-normal distribution, while the characteristic frequencies are constrained to lie within $\pm40\%$ of the values predicted by the scaling relations of \citet{Kallinger2014}. This choice permits substantial deviation from the scaling relations, reflecting the fact that our sample consists primarily of short-cadence MS and SGB stars, whereas the relations of \citet{Kallinger2014} were derived for RGB stars. Thus, we use those relations only as general guidelines rather than strict prescriptions. Furthermore, especially when using the peakbogging approach described in Appendix.~\ref{App:peakbogging}, the framework is by construction entirely different to that of \citet{Kallinger2014}. Thereby, the choices made for the priors of $\sigma_{a,b,d}$ ensure that we do not unintentionally bias our results. Consistent with the discussion in Appendix~\ref{App:ModSelectDiscuss}, we note that many of the priors listed in Table~\ref{tab:PriorCompModparams} are identical across the granulation background models, ensuring fair and consistent treatment in the model comparison.

For completeness, we provide the functional forms of the beta and log-normal distributions used throughout this work. For parameters bounded within finite intervals, we employ beta distributions with probability density function:
\begin{equation}
\begin{split}
    p(x \mid a, b, \text{loc}, \text{scale}) = & \frac{1}{B(a, b) \cdot \text{scale}} \left(\frac{x - \text{loc}}{\text{scale}}\right)^{a-1} \\ &\left(1 - \frac{x - \text{loc}}{\text{scale}}\right)^{b-1},
\end{split}
\end{equation}
where $x \in [\text{loc}, \text{loc} + \text{scale}]$, $B(a, b) = \Gamma(a)\Gamma(b)/\Gamma(a+b)$ is the beta function, and the shape parameters $a, b > 0$ control the distribution's concentration and skewness. The parameters loc and scale define the lower bound and width of the distribution, respectively. When data-driven priors are used, the shape parameters are determined via the mode-concentration relation $a(m) = m(k-2) + 1$ and $b(m) = (1-m)(k-2) + 1$ with concentration parameter $k=10$ and mode value $m$ as specified from the data, ensuring the distribution peaks at the desired location \citep{Johnson95}.

For strictly positive parameters varying over orders of magnitude, we may adopt log-normal priors implemented by placing a normal distribution on the base-10 logarithm of the parameter:
\begin{equation}
\log_{10}(\theta) \sim \mathcal{N}(\mu, \sigma),
\end{equation}
which corresponds to the probability density:
\begin{equation}
p(\theta \mid \mu, \sigma) = \frac{1}{\theta \ln(10) \sqrt{2\pi} \sigma} \exp\left[-\frac{(\log_{10}\theta - \mu)^2}{2\sigma^2}\right],
\end{equation}
for $\theta > 0$. Here, $\mu$ represents the centre of the prior on the logarithmic scale (corresponding to a median of $10^{\mu}$ in linear space), and $\sigma$ controls the width, with typical values of $\sigma= 0.1$ providing mild regularization.

\clearpage
\section{A TESS K-dwarf sample}\label{App:TESSKs}
In Sect.~\ref{subsec:grantimescales} we noted that the primary granulation timescale appears to plateau for cool MS stars (K dwarfs; Fig.~\ref{fig:GranParams}), although this suggestion is based on only a modest number of objects. To reinforce this indication we sought additional K dwarfs beyond those observed by \kepler, turning to photometry from TESS \citep{Ricker14}.

We did not expect the stars recovered to display clear oscillatory signatures. However, we required approximate \numax values both to guide our priors and to compare against the resulting granulation timescales. To obtain such an estimate, we used to the asteroseismic scaling relation \citep{Kjeldsen95}:
\begin{equation}\label{Eq:numaxscal}
    \numax = \nu_\mathrm{max,\odot} 
             \left(\frac{g}{g_\odot}\right)
             \left(\frac{\teff}{T_\mathrm{eff,\odot}}\right)^{1/2},
\end{equation}
with $\nu_\mathrm{max,\odot}=3090$\,\muHz, $g_\odot=10^{4.44}\,\mathrm{g\,cm^{-2}}$, and $T_\mathrm{eff,\odot}=5777$\,K. Thus we needed estimates for the surface gravities $g$ and effective temperatures $\teff$ for the sample stars. In order to form a sample of K-dwarfs most likely to display granulation in the TESS photometry, while also obtaining the above parameters, we turned to \gaia \citep{GaiaDR3}. 

We queried the \href{https://gea.esac.esa.int/archive/}{\gaia archive} for bright, high-quality single stars consistent with K-dwarf temperatures and luminosities, imposing the following criteria:
\begin{enumerate}
    \item $m_G < 11$
    \item \texttt{parallax} $>1$\,mas ($d<1$\,kpc)
    \item \texttt{ruwe} $<1.4$ 
    \item \texttt{non\_single\_star} $=0$
    \item \texttt{phot\_proc\_mode} $=0$
    \item \texttt{phot\_bp\_rp\_excess\_factor} $\in [0.5,\,1.5]$
    \item $\texttt{bp\_rp} \in [0.9,\,1.5]$
    \item $M_G \in [4.5,\,7.0]$
    \item \texttt{logg\_gspphot} $>4.4$\,dex
\end{enumerate}
Conditions 1–6 ensure bright, well-behaved single stars; conditions 7-9 restrict the selection to a broad K-dwarf region in colour–magnitude space and suitable surface gravities. We retrieved the top 200 stars sorted by apparent $G$ magnitude. For each star we obtained the corresponding TIC ID via \href{https://exofop.ipac.caltech.edu/tess/find_ticid.php}{ExoFOP}, and checked for available 20\,s cadence TESS data through \href{https://archive.stsci.edu}{MAST} using \textsc{LightCurve} \citep{LightCurve18}. When such data were available, we constructed a preliminary power density spectrum from the SPOC light curves for visual inspection of any tentative granulation signal present. This yielded an initial list of 72 candidates.

\begin{figure}[t]
    \resizebox{\hsize}{!}{\includegraphics[width=\linewidth]{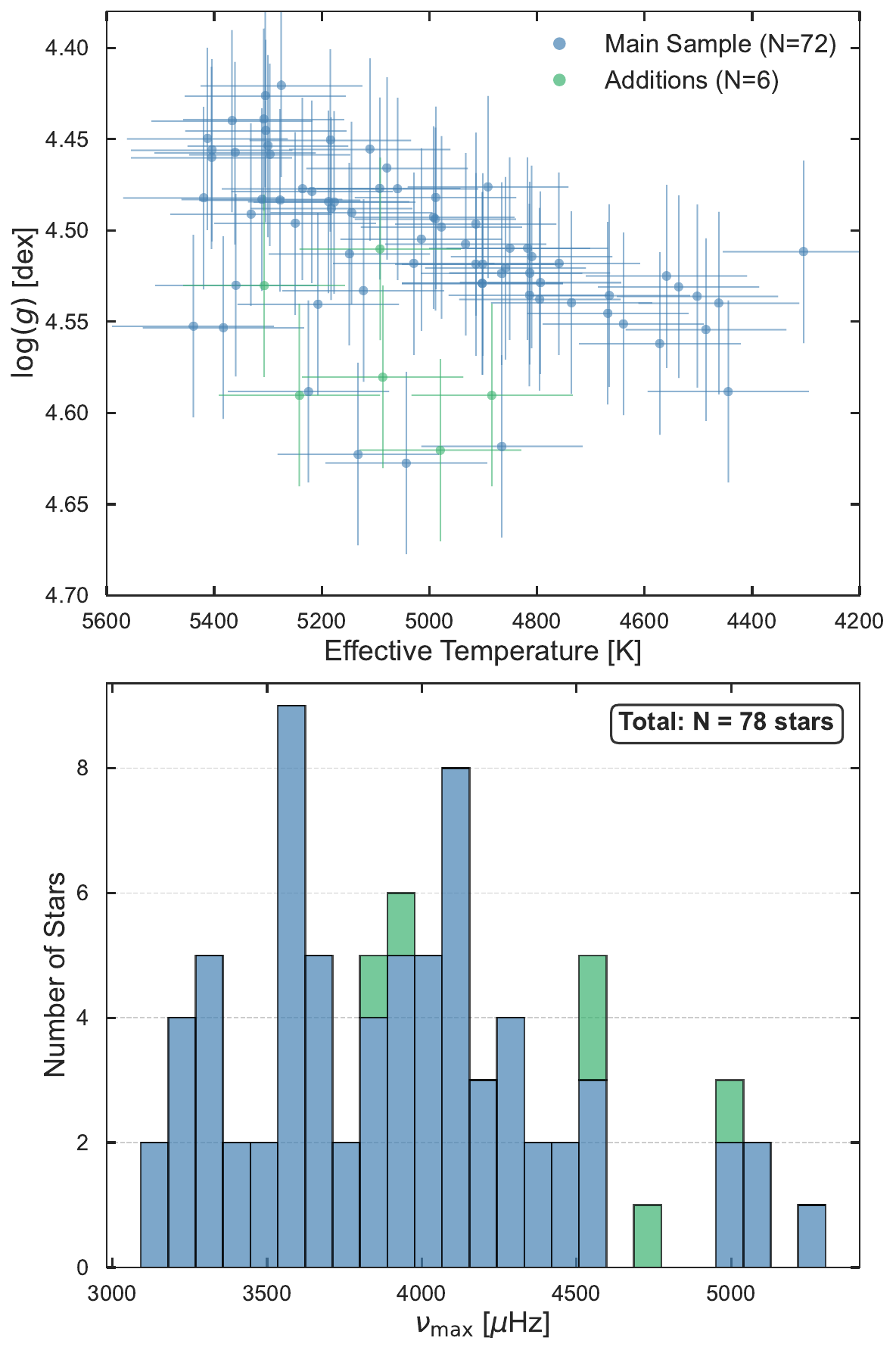}}
    \vspace*{-6mm}
    \centering
    \caption{Kiel diagram and \numax distribution of the TESS K-dwarf sample. The initial list obtained through the Gaia query and TESS 20 second cadence overlap is shown in blue, while the manual additions are indicated in green.}
    \label{fig:TESSKdwarfs}    
\end{figure}

We supplemented this list with several known TESS K-dwarf candidate oscillators observed at 20 s cadence:
TIC\,283722336 (HD\,219134),  
TIC\,398120047 (70\,Oph\,A),  
TIC\,79454735 (36\,Oph),  
TIC\,389198736 (HD\,191408),  
Furthermore, we also added the confirmed oscillators TIC\,259237827 ($\sigma$~Dra; \citealt{Hon24}) and TIC\,67772871 (40\,Eri\,A; \citealt{Lund25}), moreover noting that another known oscillator TIC\,231698181 ($\epsilon$~Indi ; \citealt{Lundkvist24, Campante24}) was already included in the \gaia-selected sample. For these additions we adopted $T_\mathrm{eff}$ and $\log g$ from the TESS Input Catalog v8.2, except for TIC\,79454735 and TIC\,389198736 where the estimates were recovered from \href{https://exofop.ipac.caltech.edu/tess/find_ticid.php}{ExoFOP} and \citet{Ramirez13}, respectively.

Altogether, we obtained 78 targets. For each, the TESS target pixel files were subsequently retrieved and custom aperture photometry was performed following \citet{Lund15} and \citet{Lund25} to construct the light curves. Power density spectra were then computed following \citet{Handberg11}.

\subsection{The resulting sample}
The \gaia \texttt{gspphot} parameters for the selected stars pass stringent quality filters and are suitable for estimating \numax\ for our purposes \citep{GaiaDR3_AstParamRelease}. Because the formal (internal) \gaia uncertainties are unrealistically small, we adopted conservative values of $\sigma_{T_\mathrm{eff}}=150$\,K and $\sigma_{\log g}=0.05$\,dex for all stars, which propagate to uncertainties of roughly $360$–$600$\,\muHz in \numax. For the additional targets drawn from the TIC, the quoted $\log g$ uncertainties can approach $0.1$\,dex, yielding \numax\ errors of order $1000$\,\muHz. Since \numax\ serves only to guide prior placement and to indicate the approximate position of each star on the potential timescale plateau, these uncertainties are acceptable. For $\sigma$~Dra and $\epsilon$~Indi we directly adopt the published \numax\ values and uncertainties from \citet{Hon24} and \citet{Campante24}, respectively.

The final TESS K-dwarf sample is shown in Fig.~\ref{fig:TESSKdwarfs}. Despite the sizeable uncertainties -- particularly in the surface gravity -- the stars broadly trace the expected lower MS trend in the Kiel diagram. The resulting \numax\ estimates span the desired range, from slightly above the solar value up to just beyond $5000$\,\muHz.

\clearpage
\section{Peakbogging}\label{App:peakbogging}
The assumption of a Gaussian oscillation excess means overlaying an entire envelope to describe the power arising from the stellar oscillations. Yet, as described previously, the power contained in the narrow Lorentzian peaks is not well represented by a smooth Gaussian excess. The most realistic description would be to individually model the observed frequencies as each their own Lorentzian excess, colloquially known as `peakbagging', which is used extensively when performing mode identification of seismic targets (see e.g., \citealt{Corsaro14}; \citealt{TACOPrelim}; \citealt{Nielsen21}; \citealt{Nielsen25}). This is, however, a very complex process with many nuances in the chosen treatment of the stellar pulsations as a function of evolution, furthermore affected by phenomena such as rotation, binarity, and magnetic fields. In the present work we aim for a middle ground: to go beyond a Gaussian envelope without having to individually model the stellar oscillations. This is because our interest is in the underlying stellar granulation background and not the oscillations themselves. We wish to essentially treat the power contained in the peaks as noise standing on top of the background. For this purpose we present `peakbogging', which is a methodology that uses a mixture model for the likelihood during the inference of the background model.

\subsection{A mixture-model likelihood setup}\label{App:subsec:mixlike}
The likelihood function $\mathcal{L}$ is identical to that used for the Gaussian approach as specified in Eq.~\ref{eq:loglike}. Peakbogging uses a mixture-model likelihood function for the inference, which is the sum of a foreground and background contribution,
\begin{equation}\label{eq:mixedmod}
    \ln{\mathcal{L}_\mathrm{tot}} = \left(1 - G(\nu,\theta) \right)\mathcal{L}_1(M) + G(\nu,\theta) \mathcal{L}_2(M\cdot\beta)
\end{equation}
Here, $M$ denotes the given background model, $\beta$ is a free scaling parameter, and $G(\nu,\theta)$ is the mixture coefficient assumed to be Gaussian with parameters $\theta$. The definition of the background and foreground contributions are in essence quite simple, while it is the interplay between them during sampling that becomes complex. The first term in Eq.~\ref{eq:mixedmod} containing $\mathcal{L}_1$ is the background contribution and contains the pure model from Table~\ref{tab:Models} assumed to describe the granulation background signal. The latter term containing $\mathcal{L}_2$ is the foreground contribution which serves the purpose of absorbing any excess in power not predicted by the assumed background model, which it may do by amplifying the predicted power of the background model by the scaling factor $\beta$. 

Notably, the assumed background model $M$ for the granulation signal is present in both the foreground and background contributions. This means that even in regions where the foreground plays a significant role, there is still some sensitivity to the background model and the parameters it contains (such as granulation amplitudes and timescales). In summary, the aim of peakbogging is thus to allow the background model parameters to be determined given the possibility of excess power being present in a certain frequency bin, which is absorbed to avoid unwanted perturbation of the background model inference.

\subsection{The mixture coefficient for control}\label{App:subsec:mixcorr}
To construct the mixture model, we introduce the Gaussian mixture coefficient $G(\nu,\theta)$, which regulates the relative contribution of the foreground and background components as a function of frequency. Its formulation allows explicit control over the frequency range where the foreground model is permitted to influence the likelihood. The coefficient is defined as a Gaussian function normalised between 0 and 1,
\begin{equation}\label{eq:Gaussmixture}
    G(\nu,\theta) = H \exp{\left[-\frac{(\nu-\nu_\mathrm{max})^2}{2\sigma^2}\right]} \ ,
\end{equation}
where $H$ denotes the height of the Gaussian and thus the strength of the foreground contribution. In practice, $H$ may be interpreted as a measure of the probability that excess power is present beyond what is predicted by the background model (i.e. `peakbog detection probability') -- hence, $H>0$ indicates the detection of such an excess. The centroid $\nu_\mathrm{max}$ identifies the frequency around which the foreground component becomes active, corresponding to the region where most of the excess power resides. When no significant artefacts affect the PDS and the oscillations are well resolved, this centroid is roughly expected to coincide with \numax\ from the traditional Gaussian envelope description. Furthermore, we anticipate that the determination of \numax in the peakbogging approach may be less sensitive to model misspecification; a suspicion that we will be examined further in Appendix~\ref{app:D_numaxres}

Lastly, we have the standard deviation of the Gaussian $\sigma$ which controls the size of the region where the foreground model is significant. Importantly, $\sigma$ is not allowed to vary freely during the inference. We reinterpret the mixture coefficient width $\sigma$ not as a parameter to be inferred, but as a prior-informed localization term that reflects the expected region of the oscillation excess. This approach prevents the mixture model from becoming sensitive to spurious features across the spectrum and ensures that the foreground model is only applied in a physically meaningful domain. For a given star, we use the estimate for the FWHM of the oscillation excess from \citet{Mosser12} to inform the width, 
\begin{equation}
    \sigma =\frac{1.5}{2\sqrt{2\ln(2)}} \ 0.66 \ \nu_\mathrm{max,obs}^{0.88} \ .
\end{equation}
We make the conservative choice to enhance the width of the mixture coefficient to be larger than expected by a factor of $1.5$, while the factor $1/2\sqrt{2\ln(2)}$ converts the FWHM to the width $\sigma$. Crucially, as we use a Gaussian mixture coefficient, non-zero tails exist throughout the spectrum. Hence, if there, for example, exists a strong systematic peak in the PDS -- which is not described by the background model -- the large difference in the log-likelihood probability between the foreground and background contributions of Eq.~\ref{eq:mixedmod} will still allow for peakbogging to take effect.

\begin{figure*}[t]
    \includegraphics[width=\linewidth]{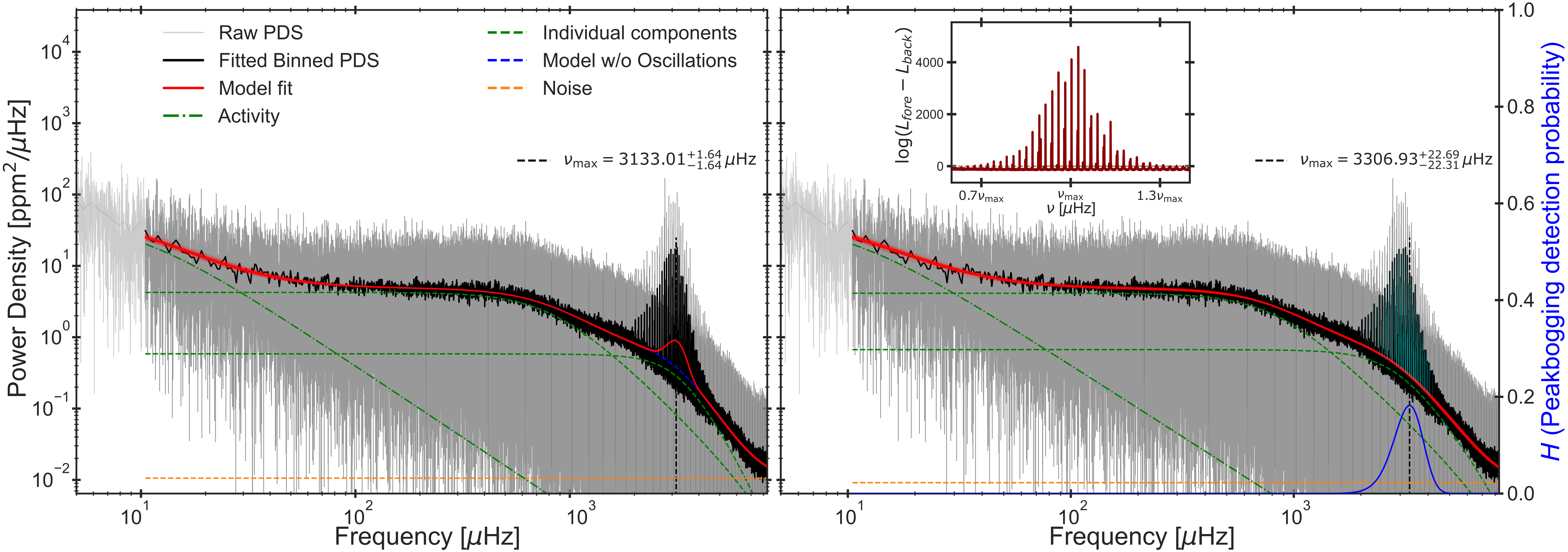}
    \centering
    \vspace*{-4mm}
    \caption{Background fits for the Sun using a Gaussian envelope (left) and peakbogging (right). A two-component Harvey model (H) is used for the inference on the PDS of a $\sim$3.15 year time series of VIRGO \citep{Froehlich95} blue band data taken during the solar minimum between cycle 23 and 24. \textit{Left}: the unbinned PDS shown in grey and the binned version is overplotted in black. The model is plotted in red using the median of the obtained posteriors for each fit parameter. Additionally, 50 randomly drawn samples from the posteriors are plotted to indicate the scatter. The individual granulation components are plotted as dashed green profiles. The fitted value of \numax is given and indicated by the vertical dashed black line, while the noise is shown by the horizontal dashed orange line. The activity component is the dash-dotted green line. The model without the influence of the Gaussian oscillation excess is plotted as the dashed blue profile, visible underneath the oscillation excess. \textit{Right}: same as left, but the Gaussian mixture coefficient controlling the peakbog mixture-model likelihood in Eq.~\ref{eq:mixedmod} is plotted as the blue profile, with the secondary y-axis showing the amplitude, $H$. Whenever $\log(\mathcal{L}_\mathrm{foreground}-\mathcal{L}_\mathrm{background})>2$, the data is coloured cyan to indicate the dominance of the foreground contribution in the peakbogging. The insert shows the log-likelihood ratio between the foreground and background contributions of the mixture model in a region around \numax.}
    \label{fig:SunBog}
    \vspace*{-2.5mm}
\end{figure*}

\subsection{Binning for peakbogging}
In Sect.~\ref{subsec:InfSampBinning} it was discussed how moderately binning the data was beneficial and did not affect the inference for the traditional Gaussian approach. For peakbogging, binning is essential as it enhances the contrast between the granulation background and contaminating signals, whether from stellar oscillations or otherwise. Since peakbogging models the oscillation contributions through a flexible likelihood mixture component able to absorb signal not described by the assumed background, unbinned data can lead the sampler to confuse stochastic noise peaks with genuine pulsation power, causing the mixture model to overestimate the foreground contribution. When this happens, the granulation parameters are biased downwards, i.e. the model component(s) accounts for less overall power, which in turn makes it even easier for the mixture model to continue absorbing the true signal during subsequent sampling steps. Moderate binning mitigates this instability by reducing the prominence of spurious peaks. 

For the three stars considered in the main article -- KIC 6679371, KIC 8866102, and KIC 8006161 -- no change was found as a function of binning for the Gaussian approach. In contrast, for the peakbogging approach, nuances occur at low degrees of binning (i.e. high frequency resolution) which vary from star to star, but common to all is that they converge as the binning becomes moderate. When the bin size reflects a frequency resolution larger than $\sim0.3$ \muHz (conservative boundary), further increasing the bin size causes negligible changes. As was argued in Sect.~\ref{subsec:InfSampBinning} we bin to a resolution of $0.5$ \muHz if the time series duration allows, which is above this boundary. The nuances of how peakbogging is sensitive to the binning is displeasing and points towards the challenges of constraining a model with significantly more freedom during the statistical inference.

\subsection{Peakbogging the Sun}
In order to visualise the nature of peakbogging and its different aspects, Fig.~\ref{fig:SunBog} shows the background model inference with a two-component Harvey model (see model H in Table~\ref{tab:Models}) for the Sun when employing a Gaussian envelope vs. peakbogging. Given the insert of Fig.~\ref{fig:SunBog} we can inspect the log-likelihood ratio between the foreground and background contributions of Eq.~\ref{eq:mixedmod}, and visually see which peaks have been absorbed by the peakbogging. We see that for the Sun, the foreground contribution dominates in a wide region around the observed \numax as desired. Contrary to the traditional method with a Gaussian envelope for the oscillation excess, which presents as a `bump' on top of the background-only profile, we see the peakbogged data standing in discrete peaks on top of the background model. 

The \numax obtained from the two approaches have some conceptual differences to be aware of. The Gaussian envelope provides an estimate of \numax which connects to literature \citep{Bedding14} as the centroid of the distribution. The peakbogging does not, however, estimate its corresponding $\numax$ as the centre of the stellar oscillations in frequency, but rather as the centre of a Gaussian mixture coefficient describing the probability for excess signal. The resulting estimate of \numax ought to be closely related and both suffer from generalities, i.e. they are both generated from the data but lack rigour, both suffer from model misspecification, and both will need effort concerning systematic corrections when referenced to models of stellar evolution. For this specific case, this difference means a shift from $\numax = 3133.0^{+1.6}_{-1.6}$ \muHz for the Gaussian case to $\numax=3306.9^{+22.7}_{-22.3}$ \muHz when peakbogging is used. 

In Appendix~\ref{App:subsec:mixcorr} the setup of peakbogging using a pre-defined and locked width of the mixture coefficient was discussed, and how despite this choice, a clear preference for the foreground contribution may lead to it dominating at a location displaced from \numax. This situation, however, is not showcased in Fig.~\ref{fig:SunBog}. However, although the log-likelihood ratio in the insert is below 0, reflecting preference for the pure background, peakboggin was allowed to asses the entire PDS for potential excess signal. Peakbogging should thereby be widely applicable when dealing with excess signal not described by the assumed model, be it instrumental peaks or potentially rotational peaks at low frequencies. If and how such peaks affect the \numax determination, however, is uncertain and has not been studied in this context. A proper approach may be to implement three terms into the mixture-model, such that you have an oscillation absorber and instrumental and/or rotation absorber, acting in isolation from each other in different regions of the PDS.

For the case of the Sun presented here, the difference between using the traditional Gaussian envelope and peakbogging is most clearly seen in the second granulation component. For the Gaussian case the characteristic frequency, $d=3258.2^{+20.4}_{-20.3}$ \muHz, which roughly indicates the position of the knee in the component, lies above the determined \numax. It is thus shifted towards higher frequency than would be expected \citep{Karoff13}, possibly because the sampler absorbs some of the power of the oscillation excess into the granulation component. In contrary, when using the peakbogging setup $d=2917.0^{+23.2}_{-24.9}$ \muHz lies below the determined \numax, and the corresponding component amplitude $c$ is lowered accordingly -- which follows the expectations from \citet{Karoff13} and \citet{Kallinger2014} almost exactly. Note that these nuances may differ when considering alternate background models, and in this context simply indicates the difference in behaviour and potential benefits of using the peakbogging approach. Yet, the presented example is for the unique case of the Sun, and we will now discuss the application and limitations of peakbogging when applied to the entire sample of stars studied in this work.

\subsection{Background model preferences and sensitivities for peakbogging}\label{App:subsec:bogprefs}
\begin{figure}[t]
    \resizebox{\hsize}{!}{\includegraphics[width=\linewidth]{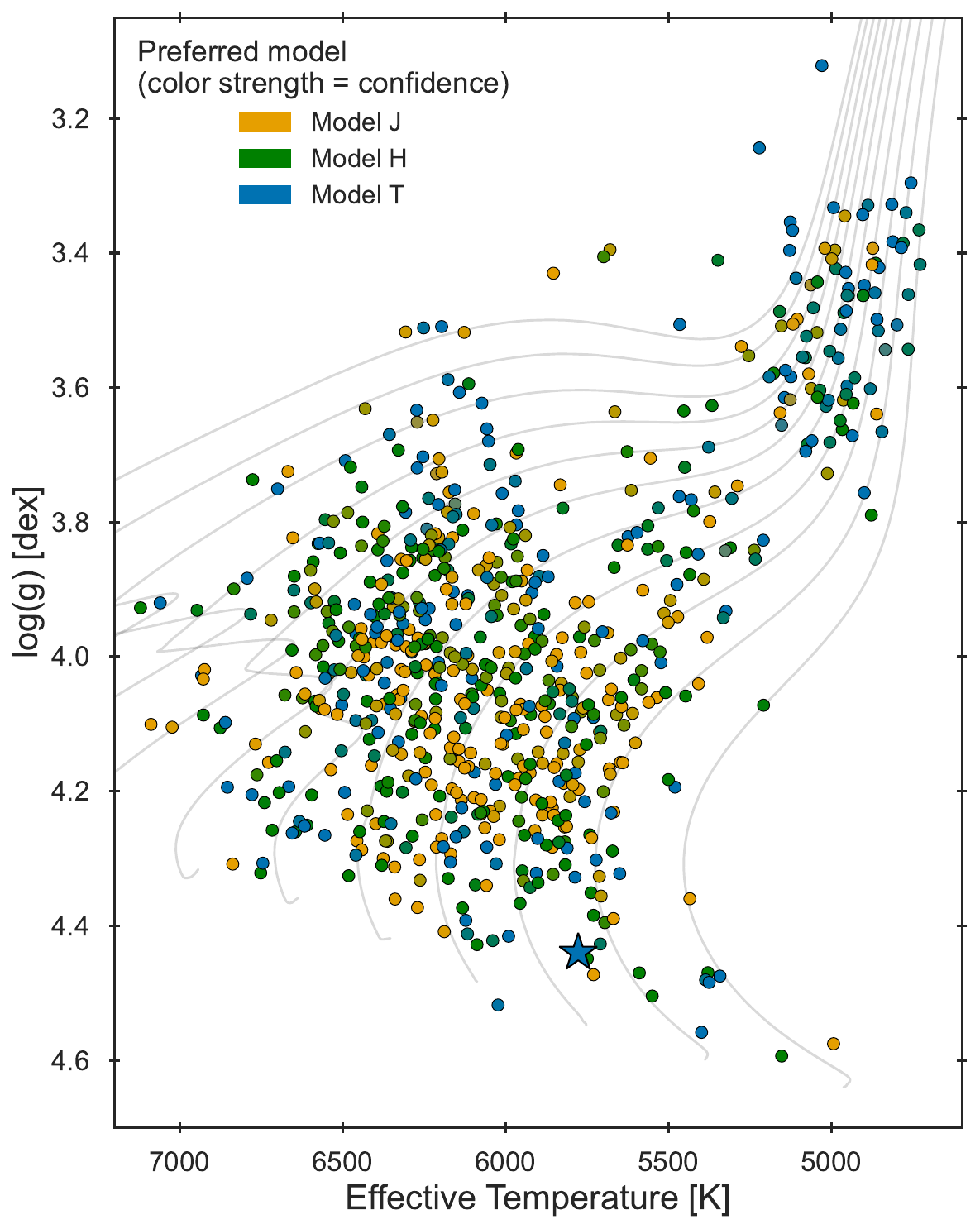}}
    \centering
    \vspace*{-5.5mm}
    \caption{Kiel diagram with colouring according to normalised evidence ratios, with model preferences as indicated by the legend. When models are comparable in their evidences the colour is blended between the two competing models. The Sun is overplotted as the enlarged star symbol at the solar location and significantly prefers model T.}
    \label{fig:Evidences_Peakbog}    
    \vspace*{-2.7mm}
\end{figure}

For the remainder of this appendix, we now largely follow the approach of the main article. In this section we thus evaluate the background model preferences and sensitivities, before considering the granulation parameters themselves in Appendix~\ref{app:bog_granscal}. For peakbogging a total of 7 stars failed to produce meaningful posteriors during the inference and were therefore discarded, resulting in a sample of 746 stars.

\subsubsection{Background model preferences}
In Fig.~\ref{fig:Evidences_Peakbog} we see the evidence ratios obtained when using the peakbogging approach. There is a clear difference in the model preferences, namely that the hybrid model J is preferred in far more cases than for the Gaussian approach (see Sect.~\ref{sec:ModPref} and Fig.~\ref{fig:Evidences_Gauss}). For peakbogging, the number of stars preferring each background model of Table~\ref{tab:Models} is split roughly into thirds, showing no overall preference or trend in the Kiel diagram. Due to the clear differences between what was obtained with the Gaussian approach, we see how the implemented model for the oscillation excess directly affects the model preferences. This was our initially worry as well as motivation for testing an alternative approach to the traditional Gaussian envelope approach. 

Thus, when considering the changes between Figs.~\ref{fig:Evidences_Gauss} and \ref{fig:Evidences_Peakbog}, we see how the optimal model for describing the stellar data is sensitive to our setup and associated assumptions. This provides further emphasis for the arguments presented by \citet{Handberg17}, namely that one should not a priori fix the background model, but rather assess which model best describes the given dataset.

\subsubsection{Background model sensitivity of \numax}\label{app:D_numaxres}
\begin{figure}[t]
    \resizebox{\hsize}{!}{\includegraphics[width=\linewidth]{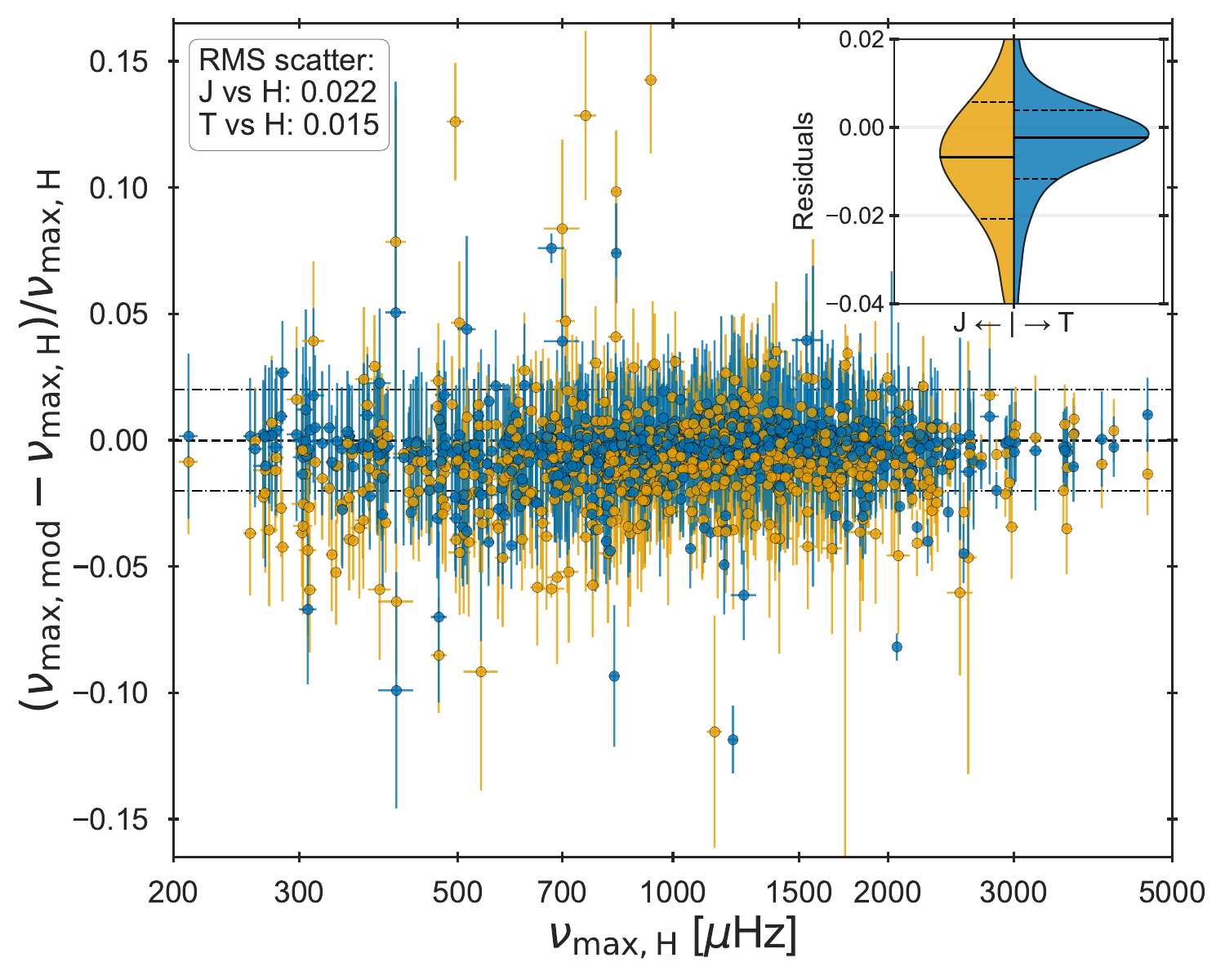}}
    \vspace*{-6.5mm}
    \centering
    \caption{Comparison of the \numax determination across the different models. The \numax fractional residuals of models J and T to those obtained by model H are plotted in yellow and blue, respectively. The horizontal dashed line indicates perfect agreement in \numax determinations, while the dot-dashed show the $2\%$ bounds. The RMS scatter was calculated for both cases and is provided in the inserted box \textbf{in the top left}. The insert shows a split violin plot of the \numax residual distributions for model J (left) and model T (right) versus model H, with medians and 16th/84th percentiles overplotted as full and dashed horizontal lines, respectively.}
    \label{fig:numaxresid_peakbog}
    \vspace*{-2.7mm}
\end{figure}

In Appendix~\ref{App:subsec:mixcorr} we discussed how peakbogging may be less sensitive to model misspecification. Figure~\ref{fig:numaxresid_peakbog} shows the \numax residuals akin to those shown in Fig.~\ref{fig:numaxresid} for the Gaussian approach. There are two things to note: 1) the RMS is actually slightly larger than that obtained by the Gaussian approach, meaning we do not see a reduction in sensitivity to the choice of background model, and 2), the formal uncertainties of \numax obtained through peakbogging are larger. 

\begin{figure*}[t]
    \includegraphics[width=\linewidth]{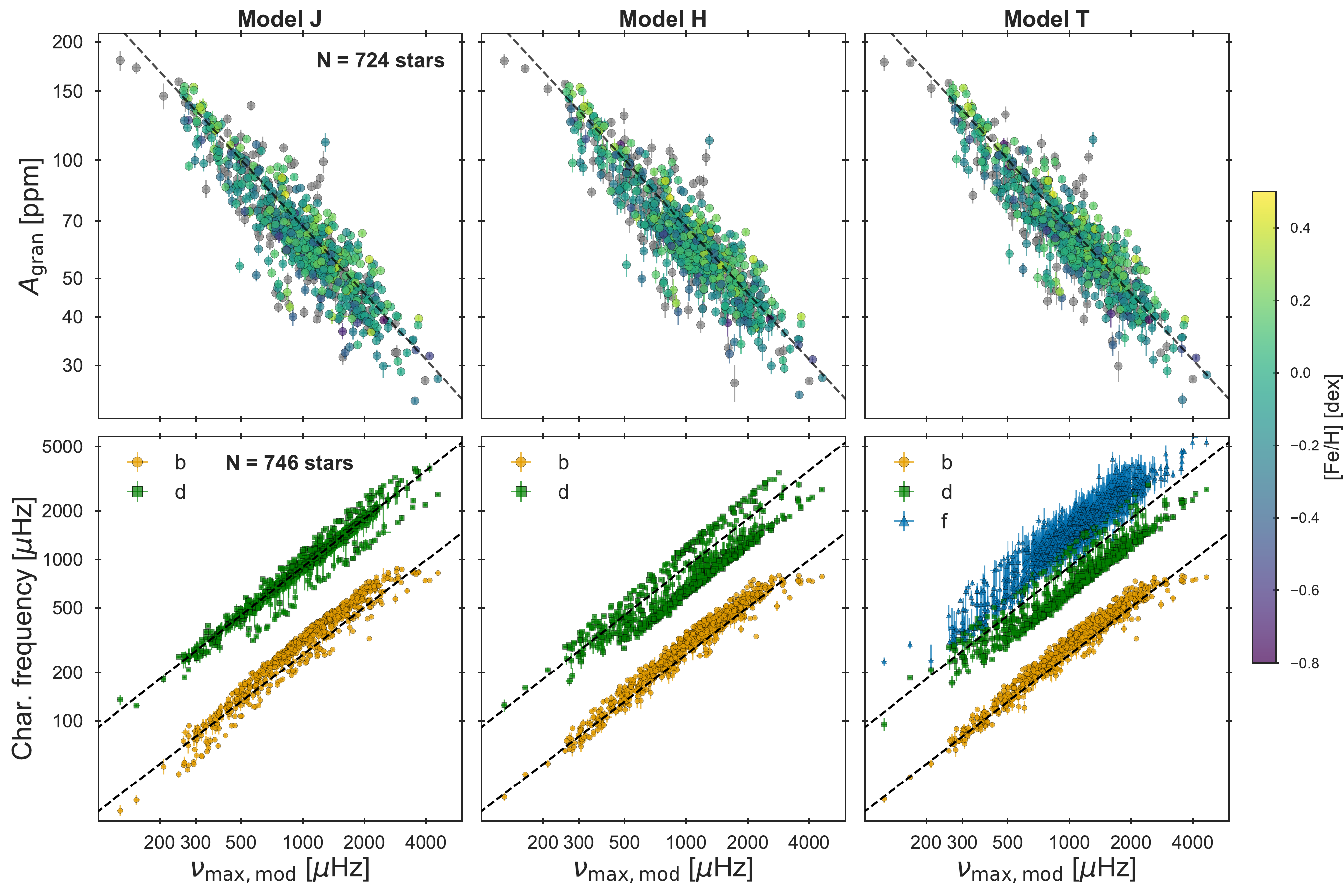}
    \centering
    \vspace*{-6mm}
    \caption{Total granulation amplitudes and characteristic frequencies as a function of \numax obtained by the three background models of Table~\ref{tab:Models}. In all panels, the dashed lines represent the corresponding scaling relation for the parameter from \citet{Kallinger2014}. \textit{Top row:} total granulation amplitudes colour-coded by the stellar metallicity \FeH for the 724 stars with available temperatures. \textit{Bottom row:} the characteristic frequencies of the individual granulation components for the 746 stars with consistent posteriors, with colours as indicated by the legend.}
    \label{fig:GranParams_Peakbog}
    \vspace*{-2mm}
\end{figure*}

Contrary to the Gaussian approach, we find somewhat less clear systematics in the obtained residuals, and since the formal uncertainties are larger they better explain the observed scatter -- for both comparisons with model H the uncertainties are consistent with zero $96\%$ of the time. Generally it thus seems that the peakbogging approach returns more realistic uncertainties on \numax when implemented into our framework, in comparison to what the simplistic and constrained Gaussian model obtains.

\subsection{Scaling of granulation parameters for peakbogging}\label{app:bog_granscal}
Figure~\ref{fig:GranParams_Peakbog} presents the total granulation amplitudes and characteristic frequencies obtained for the sample using the peakbogging approach. We have chosen not to fit power laws to determine scaling relations in the case of peakbogging, as certain pathologies which plague peakbogging will become evident in the subsequent discussion. 

The total granulation amplitudes are largely identical to those predicted by the Gaussian approach in Fig.~\ref{fig:GranParams}. This is a redeeming result for peakbogging, since the approaches vary significantly but the total power attributed to the granulation remains consistent. Moreover, the amplitudes also follow those predicted by \citet{Kallinger2014}. The primary granulation timescale across the three background models behave similarly for peakbogging as it did for the Gaussian approach. We still notice some systematic differences to the scaling relation of \citet{Kallinger2014}. Crucially, the departure from the strict correlation between the granulation and oscillation timescale is also recovered by peakbogging, seen as the apparent formation of the plateau in the timescale above \numax$\sim 3000$ \muHz. 

However, this is where the consistency of the results ends and the method begins to exhibit pathologies. When examining the characteristic frequency of the second granulation component, the measurements split into two sequences (most clearly for model~H). This behaviour was not present when using the Gaussian approach and originates from the additional freedom in the peakbogging mixture model. In some cases, peakbogging increases the likelihood by adopting an unphysically high exponent, $k$, for the characteristic frequency of the second granulation component, $d$. This rapidly depletes the power assigned to that component, after which peakbogging reallocates the missing power to the foreground component -- that is, to excess power not captured by the assumed background model. 

This pathology arises from the flexibility of the mixture model and is highly sensitive to the contrast between the background and the oscillation signal. Consequently, when the oscillations are weak, or the contrast is reduced by shorter time series or fainter targets, the model is prone to this behaviour. This also explains the observed sensitivity of peakbogging to the binning. In summary, this unresolved pathology motivated us to exclude peakbogging from the main results of this work.

\subsection{Use cases and unresolved pathologies}\label{subsec:PeakbogLimits}
During the development and testing of peakbogging, both its strengths and its limitations became clear. The method is broadly applicable and, in favourable cases -- such as the Sun or other stars with high-quality data -- it performs as intended and can serve as a viable alternative to the Gaussian approach. Applied across the full stellar sample, peakbogging is generally stable. The total granulation amplitudes it yields, dominated by the primary low-frequency granulation component that is well separated from oscillatory power, are consistent with those from the Gaussian method. Both approaches also broadly agree with the expected scaling relations of \citet{Kallinger2014}. Finally, as shown in Appendix~\ref{app:D_numaxres}, the \numax\ values inferred with peakbogging, though similarly affected by model misspecification as for the Gaussian case, provide more realistic formal uncertainties that reflect the scatter of the residuals.

However, the difficulties encountered with peakbogging primarily arise from the freedom afforded by the mixture model during nested sampling. Its flexibility makes the model harder to constrain, leading to several unresolved pathologies when applied to stars of varying data quality. The most prominent issue is the physical implausibility of the extreme exponents adopted for the second granulation component. In problematic cases, the sampler pushes the exponent $k$ to very large values, forcing the component into an artificially sharp decay and thereby distorting the inferred characteristic frequency $d$. Such exponents or correspondingly shifted timescales have no physical motivation in current descriptions of convection or surface granulation, and instead reflect a degeneracy in how the mixture model redistributes power among components.

This degeneracy has two main consequences. First, collapsing the second component into an unphysical regime reduces its background contribution, prompting the model to reassign the missing power to the foreground (excess-power) component. If taken at face value, this redistribution could be misinterpreted as unresolved oscillatory signal or as a deficiency in the background model rather than a modelling artefact. Second, the resulting parameter instability causes the characteristic frequencies of the second component to split into distinct sequences. These sequences have no astrophysical interpretation and never appear in results obtained with the Gaussian approach, making them clear indicators of model pathologies.

These issues highlight a broader challenge: while the mixture-model formulation underlying peakbogging is powerful, its freedom demands stronger constraints to ensure physically plausible behaviour across heterogeneous data. In high-quality observations the method behaves well, but for fainter targets, shorter light curves, or intrinsically low-contrast oscillators, this flexibility leads to pathological solutions that obscure rather than reveal the underlying stellar signals. Addressing this will likely require further modifications to the mixture-model structures, additional priors on granulation exponents, or more physically motivated granulation prescriptions. Until such developments are implemented, the occurrence of these pathologies warrants caution, and for this reason we have excluded peakbogging from the primary analysis of this work.

\end{appendix}

\end{document}